\documentclass[a4paper,11pt]{article}
\usepackage[toc,page]{appendix}
\usepackage{graphicx}
\usepackage{colortbl}
\usepackage{amsmath}
\usepackage{subfigure}
\usepackage{mathtools}

\DeclarePairedDelimiterX\braket[2]{\langle}{\rangle}{#1 \delimsize\vert #2}
\usepackage[utf8]{inputenc}
\usepackage{float}
\usepackage[normalem]{ulem}
\usepackage{epstopdf}
\usepackage{amsfonts,amsmath,amssymb}
\usepackage{hyperref}

\usepackage{epsfig,multicol,bbm}
\usepackage{color}
\usepackage[dvipsnames]{xcolor}

\definecolor{darkblue}{rgb}{0.0, 0.0, 0.55}
\definecolor{grey}{rgb}{0.57, 0.64, 0.69}
\definecolor{lightbrown}{rgb}{0.71, 0.4, 0.11}

\newcommand{\be}{\begin{equation}}
\newcommand{\ee}{\end{equation}}

\date{}

\newcommand\fverb{\setbox\pippobox=\hbox\bgroup\verb}
\newcommand\fverbit{\egroup\item[\fbox{\unhbox\pippobox}]}

\newbox\pippobox
\begin{document}
\title{\bf Holographic Anisotropic Background in 5D Einstien-Gauss-Bonnet Gravity}
\author{S. N. Sajadi\thanks{Electronic address: naseh.sajadi@gmail.com}
\\
\small School of Physics, Institute for Research in Fundamental Sciences (IPM), \\P. O. Box 19395-5531, Tehran, Iran\\
}
\maketitle
\begin{abstract}
In this paper, we extend the work on the AdS/QCD model to quadratic gravity to gain insight into the influence of gravity. We obtain an anisotropic black brane solution to a $5D$ Einstein-Gauss-Bonnet-two Maxwell-dilaton system. 
The background is specified by an arbitrary exponent, a dilaton field, a time component of the first Maxwell field, and a magnetic component of the second Maxwell field. 
The system in three cases has been investigated and in each case the effect of the parameter of theory, the anisotropic parameter has been considered. The blackening function supports the thermodynamical phase transition between small/large and AdS/large black brane for a suitable chemical potential and other parameters. 
\end{abstract}

\maketitle
\section{Introduction}
Quantum chromodynamics (QCD) is a non-abelian gauge theory that describes the strong interaction between quarks and gluons. QCD at low temperatures exhibits confinement whereas at high temperature undergoes a phase transition to a chiral symmetry. The investigation and understanding of the phase diagram of QCD and the search for new phases of matter are of attracting attention in the theoretical and experimental communities.
The gauge/gravity duality provided another way to further understand the dynamics of the strong-couple system, where standard methods do not work \cite{Maldacena:1997re}, \cite{Witten:1998qj}. The quark-gluon plasma (QGP) is one such system created in a short time in heavy ion collisions, it is believed to be anisotropic during this time \cite{Casalderrey-Solana:2011dxg}, \cite{Arefeva:2014kyw}. Therefore, various properties of QCD have been investigated in an anisotropic background \cite{Strickland:2013uga}. In \cite{Andreev:2006ct}, \cite{He:2013qq} the confinement-deconfinement phase transition in the framework of the Einstein-dilaton-Maxwell theory for the isotropic case has been studied. 
In \cite{Arefeva:2018hyo}, the confinement-deconfinement phase transition in the framework of $5D$ Einstein-dilaton two-Maxwell theory with an anisotropic background has been studied. In \cite{Bohra:2019ebj}, the authors extended the work of \cite{Arefeva:2018hyo}, by introducing a background magnetic field to gain insight into the influence of such field on QCD observables.\\
Higher-order gravitational models have recently received attention \cite{Fan:2014ala}-\cite{Sajadi:2022pcz}, in part because string theory predicts that at low energies Einstein's equations are subject to first-order corrections \cite{Ferrara:1996hh}.
In AdS/CFT context, higher-order gravities have been used as tools to characterize numerous properties of strongly coupled conformal field theories \cite{Hofman:2009ug}-\cite{deBoer:2009pn}. 
From quantum gravity viewpoint, in order to unify quantum mechanics and gravitational interactions, going beyond the Einstein gravity is necessary \cite{Camanho:2009hu}. 
The first correction of Lovelock gravity to the Einstein-Hilbert action appears in five and higher dimensions and is given by a precise combination of quadratic curvature terms yields the second-order field equations known as the Gauss-Bonnet density \cite{Garraffo:2008hu}, \cite{Charmousis:2008kc}, \cite{Canfora:2021ttl}.
Cosmological models, including in the inflation, and in the framework of Brane cosmology have been well studied in this theory \cite{Deruelle:2003ck}. {Black hole solutions of the theory have been studied in \cite{Zwiebach:1985uq}-\cite{Kanti:1997br}.} The thermodynamics of black holes has also been studied in the framework of this theory \cite{Padmanabhan:2013xyr}. The Gauss-Bonnet term in 4D gives a non-zero contribution to the field equations in the presence of the dilatonic scalar field $\phi$ \cite{Horndeski:1974wa}, \cite{Kobayashi:2019hrl}, \cite{Fernandes:2022zrq}.
In this paper, we extend the work of \cite{Arefeva:2018hyo} to the Einstein Quadratic Gravity, which is general relativity extended by quadratic curvature invariants in the action to find the effect of higher derivative terms on QCD.\\
The paper is organized as follows. In section \ref{sec2} we construct the anisotropic $5$-dimensional solution with an arbitrary dynamical exponent, an exponential quadratic warp function, a non-zero time component of the first Maxwell field and a non-zero magnetic
component of the second Maxwell field in the framework of EGB gravity. In section \ref{sext2.1} first we consider zero warp function and obtain the exact solution for blackening function and other unknown quantities. We have shown the behavior of the quantities with plots and we discuss the thermodynamics of the constructed background. In section \ref{sect2.2}, we consider exponential quadratic warp function and zero chemical potential and solved the differential equations approximatly and show that for negative exponential warp function the dilaton field is real. Then, we discuss the thermodynamics of the constructed background and find out the small/large phase transition. 
In section \ref{sect2.3} we consider the non-zero warp function and non-zero chemical potential and obtained the approximatly solution for the unknown functions. In this case we study the thermodynamics of the black brane and find out the small/large and AdS/large phase transitions. We finish the paper with some concluding remarks in section \ref{conclud}.

\section{Basic Formalism}\label{sec2}
We consider a $5$-dimensional Einstein-quadratic-dilaton-two-Maxwell system. The action of the system in the Einstein frame is specified as \cite{Arefeva:2018hyo}
\begin{equation}\label{action1}
S=\dfrac{1}{16\pi G_{5}}\int d^{5}x \sqrt{-g}L,
\end{equation}
where the Lagrangian is
\begin{equation}
L=R+\gamma R_{a b c d}R^{a b c d}+\beta R_{a b}R^{a b}+\alpha R^{2}-\dfrac{1}{4}f_{1}(\phi)F_{(1)}^{2}-\dfrac{1}{4}f_{2}(\phi)F_{(2)}^{2}-\dfrac{1}{2}\partial_{\mu}\phi\partial^{\mu}\phi - V(\phi),
\end{equation}
and $F_{(i)}^{2}=F_{\mu \nu}F^{\mu \nu}$, $\phi$ is the dilaton field, $f_{1}(\phi)$ and $f_{2}(\phi)$ are the gauge functions representing the coupling between the two $U(1)$ gauge fields on one hand and the dilaton on the other hand. $V(\phi)$ is the potential of the dilaton field, and $G_{5}$ is the Newton constant in five dimensions. ($\alpha,\beta,\gamma$) are coupling constants of theory. We use the metric ansatz $g_{\mu \nu}$, dilaton field $\phi$ and field strength tensor $F_{(i)}^{\mu \nu}$ in the following form:
\begin{equation}
ds^{2}=\dfrac{l^{2}b(z)}{z^{2}}\left(-g(z)dt^{2}+\dfrac{dz^2}{g(z)}+dx^{2}+P(z)(dy_{1}^{2}+dy_{2}^{2})\right),
\end{equation}
with
\begin{equation}
A^{(1)}_{\mu}=A_{t}(z)\delta^{0}_{\mu},\;\;\;\;\; F_{(2)}=q dy^{1}\wedge dy^{2},\;\;\;\; \phi=\phi(z),
\end{equation}
where $b(z)$ is the warp function, $g(z)$ is the metric function and $l$ is the AdS length scale. $z = 0$ corresponds to the boundary of the $5d$ spacetime. {The first gauge field ($F^{(1)}$) is the electric part of the Maxwell tensor which causes the black hole to become electrically charged. In relation \eqref{eqqcond10}, we relate the charge of the black hole to the chemical potential of the dual quantum field system. The second gauge field ($F^{(2)}$) is the magnetic part of the Maxwell tensor on a plane $y_{1}y_{2}$ and causes the anisotropy of the metric spatial part.} The variation of the action (\ref{action1}) over metric $g_{\mu \nu}$, the scalar field $\phi$ and $A_{t}$ gives the field equations as follows
\begin{small}
\begin{eqnarray}\label{eqq5q}
&&E_{\mu\nu}=G_{\mu \nu}+\alpha\left[2R(R_{\mu\nu}-\frac{1}{4}g_{\mu\nu}R)+2(g_{\mu\nu}\square-\nabla_{\mu}\nabla_{\nu})R\right]+\beta\left[(g_{\mu\nu}\square-\nabla_{\mu}\nabla_{\nu})R+\square G_{\mu\nu}+2R^{\lambda \rho}(R_{\mu\lambda\nu\rho}\right.  \nonumber\\
&&\left.-\frac{1}{4}g_{\mu\nu}R_{\lambda\rho})\right]
+\gamma\left[-\frac{1}{2}g_{\mu\nu}R_{\alpha\beta\gamma\eta}R^{\alpha\beta\gamma\eta}+
2R_{\mu\lambda\rho\sigma}R_{\nu}{}^{\lambda\rho\sigma}+4R_{\mu\lambda\nu\rho}R^{\lambda\rho}-4R_{\mu\sigma}
R^{\sigma}_{\nu}+4\square R_{\mu\nu}-2\nabla_{\mu}\nabla_{\nu}R\right]
\nonumber\\
&&=\dfrac{1}{2}f_{(i)}\left(F^{(i)}_{\mu \rho}F_{\nu}^{(i)\rho}-\dfrac{1}{4}g_{\mu\nu}F^{2(i)}\right)+\dfrac{1}{2}\left(\partial_{\mu}\phi\partial_{\nu}\phi-\dfrac{1}{2}g_{\mu\nu}(\partial \phi)^{2}-g_{\mu\nu}V\right),\;\;\;\;\;\;\;
\nabla^{2}\phi=\dfrac{\partial V}{\partial\phi}+\dfrac{1}{4}\dfrac{\partial f_{(i)}}{\partial\phi}(F^{2(i)}),\nonumber\\
&&\nabla_{\mu}\left(f_{(i)}F^{\mu\nu(i)}\right)=0,\;\;\;\;\;\;\;\;\;(i=1,2)
\end{eqnarray}
\end{small}where $G_{\mu\nu}$ is the Einstien tensor. Using the ansatz of the metric, the Maxwell fields and the dilaton field \eqref{eqq5q}, it is easy to obtain the equations of motion for the background fields. The explicit components of the field equation are large and bulky and we have not included them here. The field equations for $\phi$ and $A_{t}$ are given by:
\begin{align}
&-P^{2}z^{4}f_{1}^{\prime}h^{\prime 2}+q^{2}z^{4}\dfrac{df_{2}}{d\phi}+2b^{2}l^{4}P^{2}\dfrac{dV}{d\phi}-3l^{2}P^{2}gz^{2}\phi^{\prime}b^{\prime}-2l^{2}bP^{2}
z^{2}g\phi^{\prime\prime}+6l^{2}zbP^{2}g\phi^{\prime}
\nonumber\\
&-2l^{2}z^{2}bP^{2}\phi^{\prime}g^{\prime}=0,\\
&f_{1}A_{t}^{\prime}b^{\prime}zP-2f_{1}A_{t}^{\prime}bP+2f_{1}bPA_{t}^{\prime\prime}z+2bPzf^{\prime}_{1}A_{t}^{\prime}=0\label{eqq8},
\end{align}
where prime is differential with respect to $z$. One can check that the equation of motion for the second Maxwell field will not give any additional equation.
To find the solution for the field equations, we assume \cite{Arefeva:2018hyo}
\begin{equation}
b(z)=e^{-\frac{cz^{2}}{2}},\;\;\;\;\; f_{1}=z^{-2+\frac{2}{\nu}},\;\;\;\;P(z)=z^{2-\frac{2}{\nu}},
\end{equation}
where $\nu$ is a parameter that specified the anisotropic backgrounds.
To solve the background, we also impose the boundary conditions in the form
\begin{equation}\label{eqqcond}
{b(0)=1},\;\;\;\; g(0)=1,\;\;\;\; g(z_{h})=0,\;\;\;\; A_{t}(0)=\mu ,\;\;\;\; A_{t}(z_{h})=0
\end{equation}
where $z_{h}$ is the horizon and $\mu$ is the chemical potential of the boundary theory. The boundary conditions are used to fix the integration constants.
Now, we are going to solve the field equations.
First, by solving the differential equation \eqref{eqq8}, one can get
\begin{equation}\label{eqqAt}
A_{t}(z)=\mathfrak{c}_{1}+\mathfrak{c}_{2}e^{\frac{cz^{2}}{4}},
\end{equation}
where
\begin{align}\label{eqqcond10}
\mathfrak{c}_{1}=\dfrac{\mu e^{\frac{c z_{h}^2}{4}}}{-1+e^{\frac{c z_{h}^2}{4}}},\;\;\;\;\mathfrak{c}_{2}=\dfrac{\mu}{-1+e^{\frac{c z_{h}^2}{4}}}.
\end{align}
By inserting solution (\ref{eqqAt}) into the equation $E_{tt}$, one can obtain $V(\phi)$. Then, by inserting $V(\phi)$ into the field equation $E_{xx}$, one can obtain $f_{2}(\phi)$. By inserting $f_{2}(\phi)$ and $V(\phi)$ into $E_{zz}$ one can obtain $\phi^{\prime}$. Finally, from equation $E_{y_{1}y_{1}}$, the differential equation for $g(z)$ obtains as follows
\begin{align}\label{eqqtotal}
&4\nu^{4}z^{4}l(4\alpha+\beta)e^{-\frac{cz^{2}}{4}}g(z)g^{\prime\prime\prime\prime}-4\nu^{3}lz^{3}e^{-\frac{cz^{2}}{4}}[(4\alpha+\beta)(\nu cz^{2}-2\nu+4)g-2z\nu(\alpha-\gamma)g^{\prime}]g^{\prime\prime\prime}\nonumber\\
&-6\nu zle^{-\frac{cz^{2}}{4}}[-4l^{2}z\nu^{3}e^{-\frac{cz^{2}}{2}}-4\nu^{2} z^{2}g^{\prime}(20\gamma-4\alpha+4\beta+(10\alpha+6\beta+14\gamma)\nu+c\nu z^{2}(5\beta+3\alpha+17\gamma))\nonumber\\
&+\nu g(c^{2}\nu^{2}z^{4}(5\beta+24\gamma)+4c\nu z^{2}(16\gamma+2\beta-4\alpha+\nu(4\alpha+\beta))+\nu^{2}(20\beta+32\alpha+48\gamma)+16\nu (\beta+4\gamma)\nonumber\\
&+48\gamma-32\alpha)]g^{\prime\prime}-\nu zle^{-\frac{cz^{2}}{4}}[l^{2}\nu^{2}e^{-\frac{cz^{2}}{2}}(4\nu+6\nu cz^{2}+8)-g((3\beta+12\gamma+2\alpha)c^{3}\nu^{3}z^{6}+c^{2}\nu^{2}z^{4}(80\gamma+\nonumber\\
&16\alpha+20\beta-(19\beta+16\alpha+72\gamma)+4c\nu z^{2}(3(\beta+2\alpha+2\gamma)\nu^{2}-\nu(12\alpha+4\beta+8\gamma)+12\alpha+
10\beta+38\gamma)\nonumber\\
&+4(\nu+2)(\nu^{2}(5\beta+8\alpha+12\gamma)+\nu(16\gamma-8\alpha+2\beta)+4\beta+
12\gamma+8\alpha)))]g^{\prime}-4z^4\nu^3 l(2\gamma+\beta+2\alpha)\nonumber\\
&e^{-\frac{cz^2}{4}}g^{\prime\prime 2}-\nu z^2le^{-\frac{cz^2}{4}}g^{\prime 2}(112\alpha+32\beta+32\gamma+(128\alpha+32\beta+8cz^{2}(18\alpha+5\beta+4\gamma))\nu+(48\alpha+20\beta\nonumber\\
&+32\gamma-16cz^2(3\alpha+
\beta+\gamma)+c^2z^4(40\alpha+8\gamma+11\beta))\nu^2)-c^{2}\mathfrak{c}_{2}^{2}\nu^3 le^{\frac{cz^2}{4}}z^{\frac{4\nu+2}{\nu}}=0.
\end{align}
For generic coupling constant $\alpha$, $\beta$ and $\gamma$ this fourth order differential equation analytically cannot be solved, therefore we consider the case where $\gamma=\alpha,\beta=-4\alpha$. In this case, the theory reduced to Einstien-Gauss-Bonnet gravity (EGB), and the field equation \eqref{eqqtotal}, reduced to second order differential equation for metric function $g(z)$ as follows
\begin{align}
&-4zl((2+\nu cz^{2})^{2}z\nu g(z)\alpha e^{-\frac{cz^{2}}{4}}-z\nu^{3}l^{2}e^{-\frac{3cz^{2}}{4}})g^{\prime\prime}-4\nu z^{2}l\alpha(2+\nu cz^{2})^{2}e^{-\frac{cz^{2}}{4}}g^{\prime 2}-\nonumber\\
&2zl((2+\nu cz^{2})(z^{4}c^{2}\nu^{2}-6cz^{2}\nu^{2}+6\nu c z^{2}+4\nu+8)g(z)\alpha e^{-\frac{cz^{2}}{4}}+2l^{2}\nu^{2}(4+2\nu+3\nu c z^{2})e^{-\frac{3cz^{2}}{4}})g^{\prime}\nonumber\\
&-c^{2}\nu^{3}l\mathfrak{c}_{2}^{2}z^{\frac{4\nu+2}{\nu}}e^{\frac{cz^{2}}{4}}=0.
\end{align}
In the following, we solve the above differential equations in special cases: 
\subsection{The case $c=0$}\label{sext2.1}
In this case the warp function $b(z)=1$ and the field equation for $g(z)$ becomes:
\begin{equation}
z\nu(l^2\nu^2-4\alpha g(z))g^{\prime\prime}-g^{\prime}(4\nu \alpha zg^{\prime}+(2+\nu)(l^2\nu^2-4\alpha g))=0,
\end{equation}
one can exactly solve it and obtain analytic solution for $g(z)$ as 
\begin{align}\label{eqqgz}
g(z)=&\dfrac{1}{4\alpha(1+\nu)}[l^2\nu^2+l^2\nu^3 -\nonumber\\
&\sqrt{l^4\nu^4+2l^4\nu^5+l^4\nu^6-4\nu^2\alpha c_{1}z^{\frac{2(1+\nu)}{\nu}}-16\nu\alpha c_{2}-8\nu^2\alpha c_{2}-4\alpha c_{1}\nu z^{\frac{2(\nu+1)}{\nu}}-8\alpha c_{2}}],
\end{align}
and by taking into account the boundary conditions \eqref{eqqcond}, we get
\begin{equation}
c_{1}=-\dfrac{2(\nu^{3}l^{2}+\nu^{2}l^{2}-2\alpha \nu-2\alpha)}{\nu z_{h}^{\frac{2(\nu+1)}{\nu}}},\;\;\;\;\;c_{2}=\nu^{2}l^{2}-2\alpha.
\end{equation}
For $\alpha\ll 1$, the metric function is given as
{\begin{equation}
g(z)\approx 1-\left(\dfrac{z}{z_{h}}\right)^{\frac{2(\nu+1)}{\nu}}-\dfrac{2\alpha}{\nu^{2}l^{2}}\left(\dfrac{z}{z_{h}}\right)^{\frac{2(\nu+1)}{\nu}}\left(1-\left(\dfrac{z}{z_{h}}\right)^{\frac{2(\nu+1)}{\nu}}\right)+\mathcal{O}(\alpha^{2}).
\end{equation}}
The second term is the correction from the Gauss-Bonnet gravity and in the case of $\alpha\to 0$ the metric function goes to \cite{Arefeva:2018hyo} for Einstein gravity.
\begin{figure}[H]\hspace{0.4cm}
\centering
\subfigure[$\nu=4.5,l=1,z_{h}=2$]{\includegraphics[width=0.3\columnwidth]{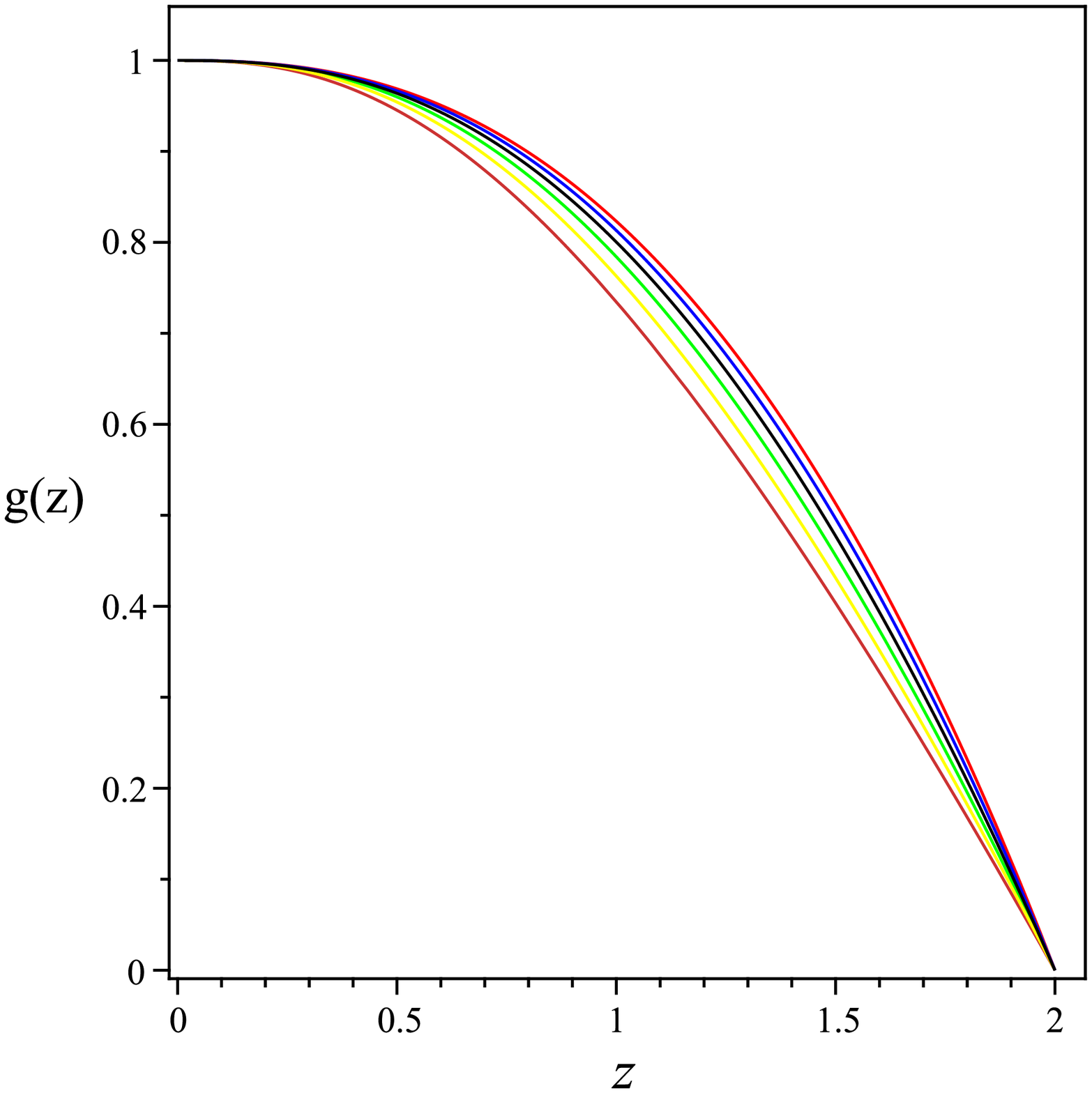}}
\subfigure[$\alpha=0.1,l=1,z_{h}=2$]{\includegraphics[width=0.3\columnwidth]{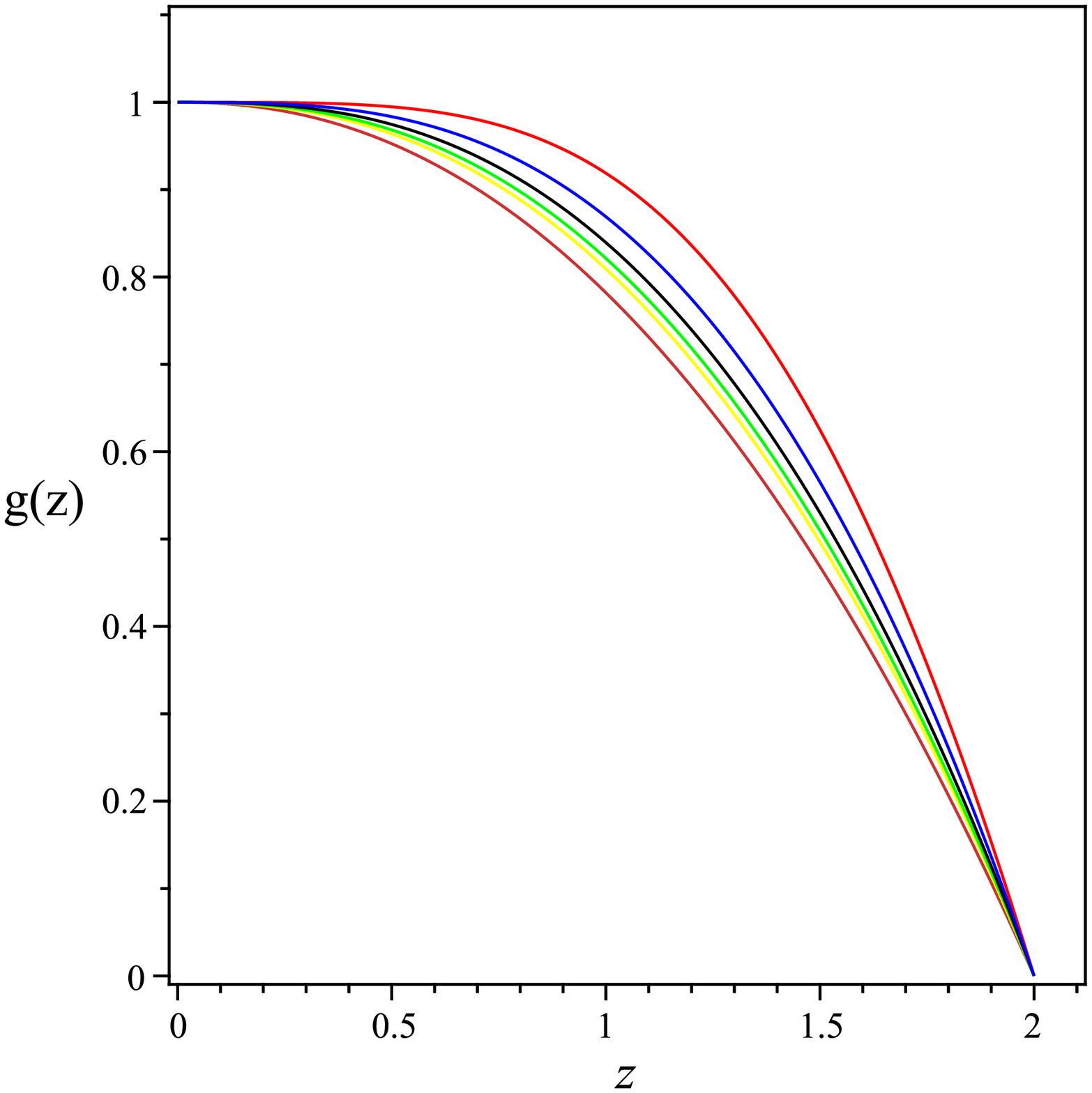}}
\subfigure[$\alpha=0.1,l=1,\nu=4.5$]{\includegraphics[width=0.3\columnwidth]{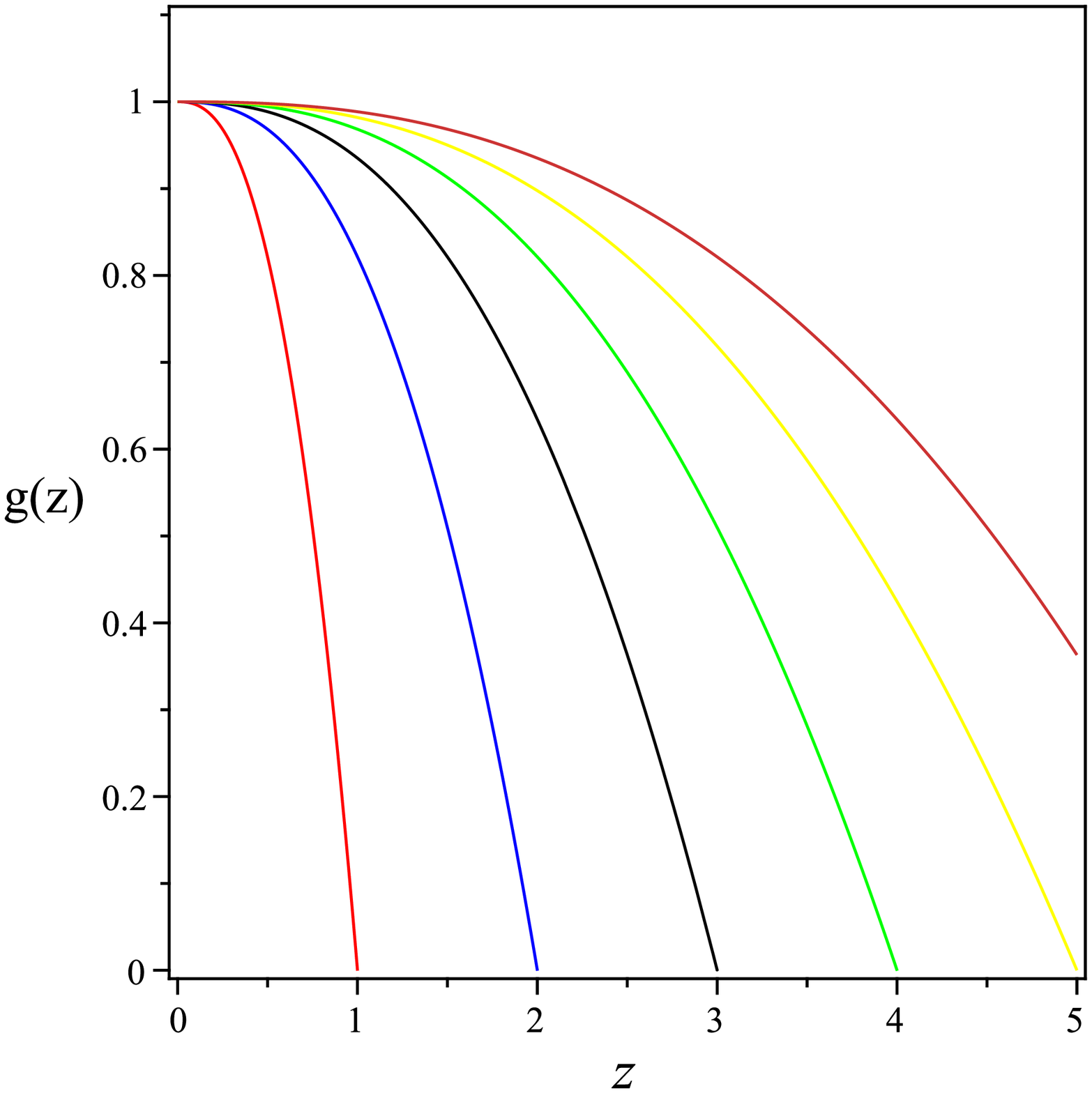}}
\caption{Plot of $g(z)$ in terms of $z$ for $\alpha=\textcolor{red}{0},\textcolor{blue}{0.5},1,\textcolor{green}{1.5}, \textcolor{orange}{2}$ (left), $\nu=\textcolor{red}{1},\textcolor{blue}{2},3,\textcolor{green}{4},\textcolor{orange}{5},6$ (middle) and $z_{h}=\textcolor{red}{1},\textcolor{blue}{2},3,\textcolor{green}{4},\textcolor{orange}{5},6$ (right).} 
\label{gzplote}
\end{figure}
\noindent The behavior of the metric function is depicted in Fig.(\ref{gzplote}). The main feature is that the metric function values
decrease faster for larger $\alpha$ (Fig.\ref{gzplote}a). In the isotropic case ($\nu=1$) the metric function values are larger than in the anisotropic ones ($\nu\neq 1$) (Fig.\ref{gzplote}b). In this panel by increasing $\nu$ the metric function values decrease faster. Changing the values of $\alpha$ and $\nu$ does not influence the horizon position.
In the following we look at the behavior of Ricci and Kretschmann scalar $K=R_{a b c d}R^{a b c d}$ of the black brane. The Ricci scalar is given as follows  
\begin{align}
R&=\dfrac{1}{2\alpha\nu^2 l^2\left((\nu^2l^2-4\alpha)^2z_{h}^{\frac{2\nu+2}{\nu}}+8\alpha (\nu^2l^2-2\alpha)z^{\frac{2\nu+2}{\nu}}\right)^{\frac{3}{2}}}[-l^2\nu^2(4\nu+3(\nu^2+1))\nonumber\\
&\left((\nu^2l^2-4\alpha)^2z_{h}^{\frac{2\nu+2}{\nu}}+8\alpha (\nu^2l^2-2\alpha)z^{\frac{2\nu+2}{\nu}}\right)^{\frac{3}{2}}(3(\nu^2+1)+4\nu)(\nu^2l^2-4\alpha)^{4}z_{h}^{\frac{3\nu+3}{\nu}}\nonumber\\
&+4(\nu^2l^2-2\alpha)((\nu^2l^2-4\alpha)^2(9\nu^2+11\nu+10)(z_{h}z^{2})^{\frac{\nu+1}{\nu}}+8\alpha(\nu^2l^2-2\alpha)(2\nu^2\nonumber\\
&+\nu+3)z_{h}^{\frac{-\nu-1}{\nu}}z^{\frac{4\nu+4}{\nu}})].
\end{align}
The scalars are smooth inside the black hole and start to diverge for $z>z_h$. In larger $\alpha$ it happens earlier for $R$, while for $K$ it happens earlier for smaller $\alpha$.

\begin{figure}[H]\hspace{0.4cm}
\centering
\subfigure[$\nu=4.5,l=1,z_{h}=2$]{\includegraphics[width=0.4\columnwidth]{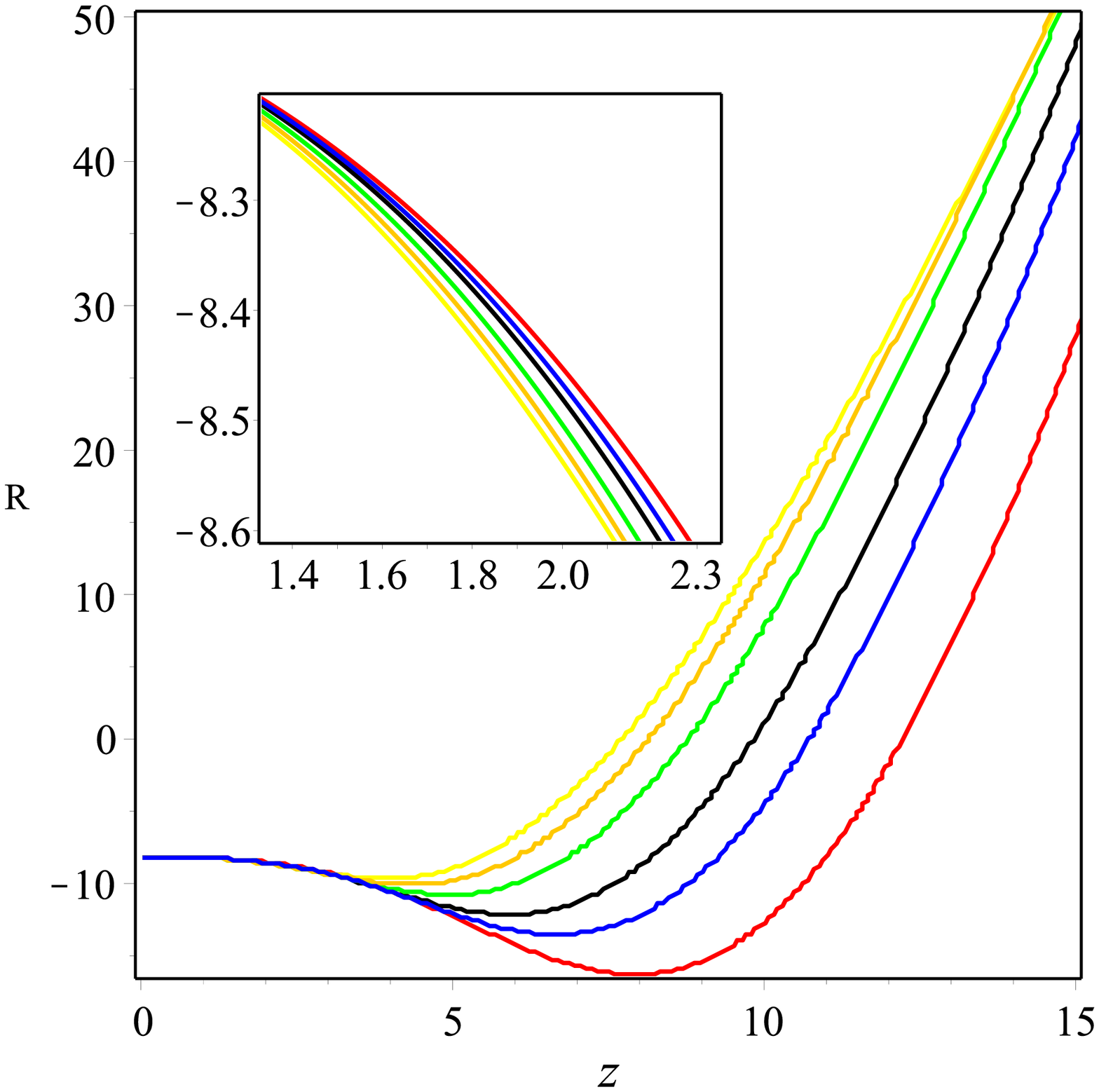}}
\subfigure[$\nu=4.5,l=1,z_{h}=2$]{\includegraphics[width=0.4\columnwidth]{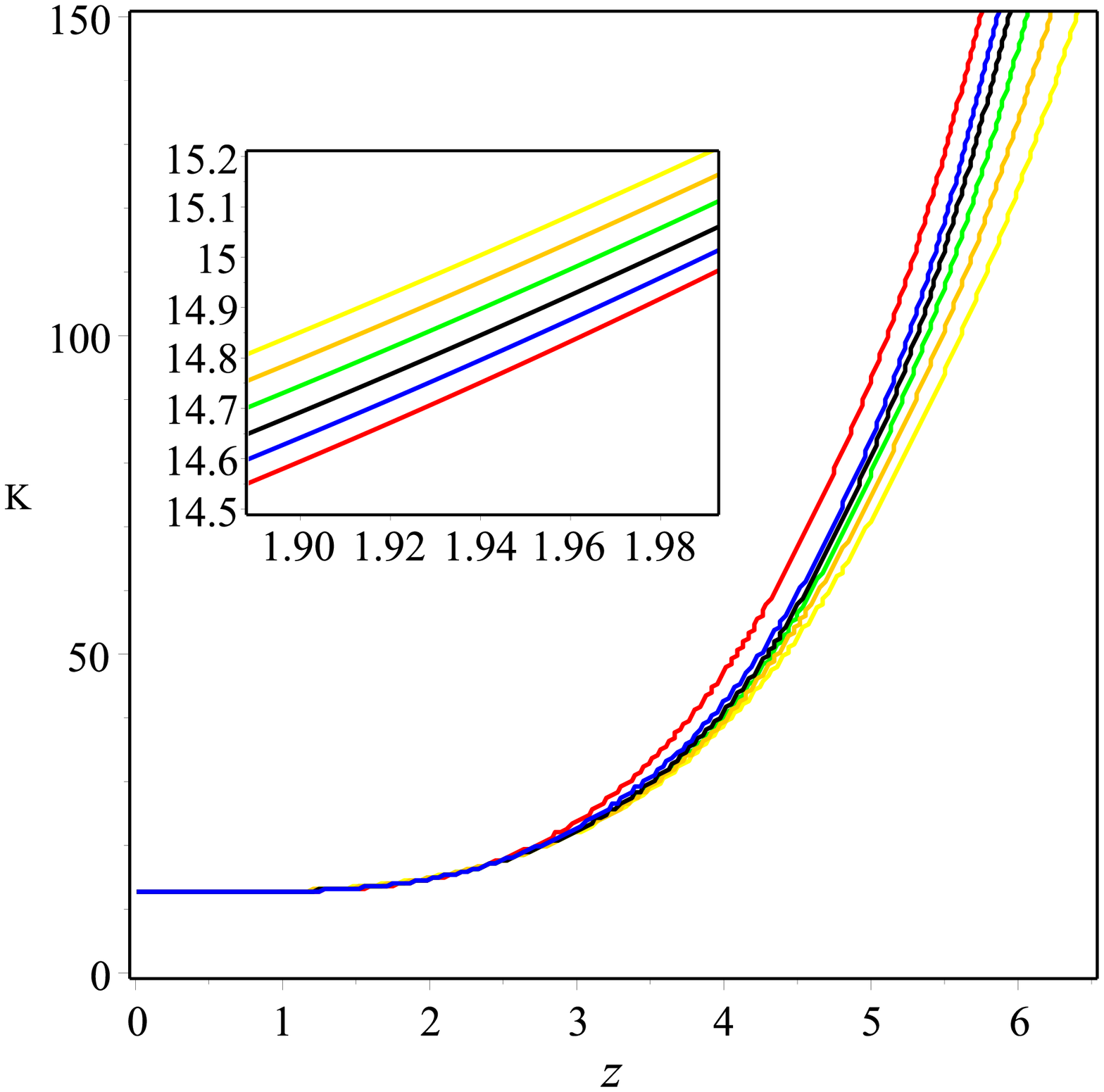}}
\caption{Plot of Ricci scalar and Kretschmann scalar in terms of $z$ for $\alpha=\textcolor{red}{0.1},\textcolor{blue}{0.2},0.3,\textcolor{green}{0.4},\textcolor{orange}{0.5}$.} 
\label{TMSplote}
\end{figure}
\begin{figure}[H]\hspace{0.4cm}
\centering
\subfigure[$\alpha=0.1,l=1,z_{h}=2$]{\includegraphics[width=0.3\columnwidth]{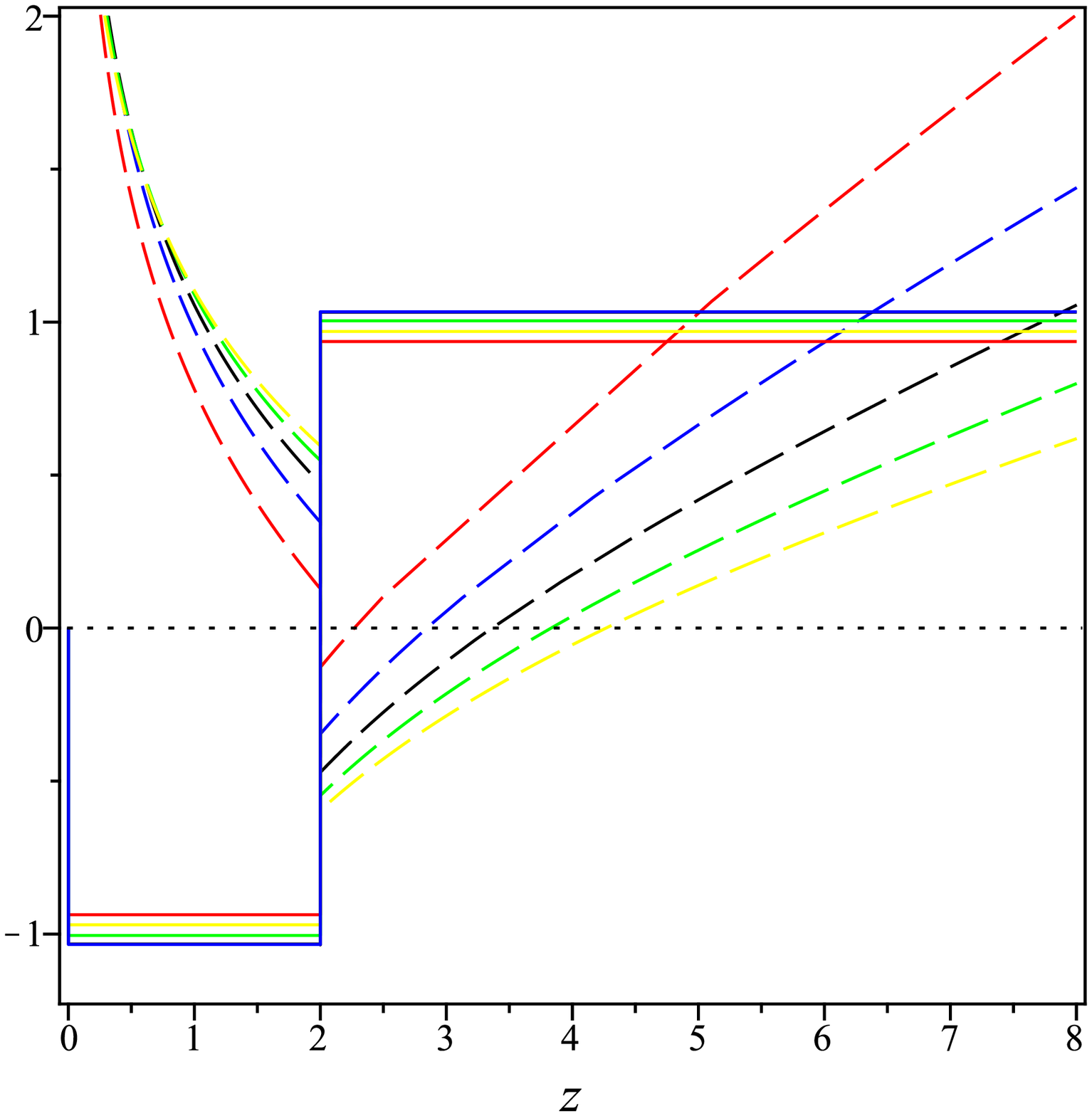}}
\subfigure[$\alpha=0.1,l=1,\nu=4.5$]{\includegraphics[width=0.3\columnwidth]{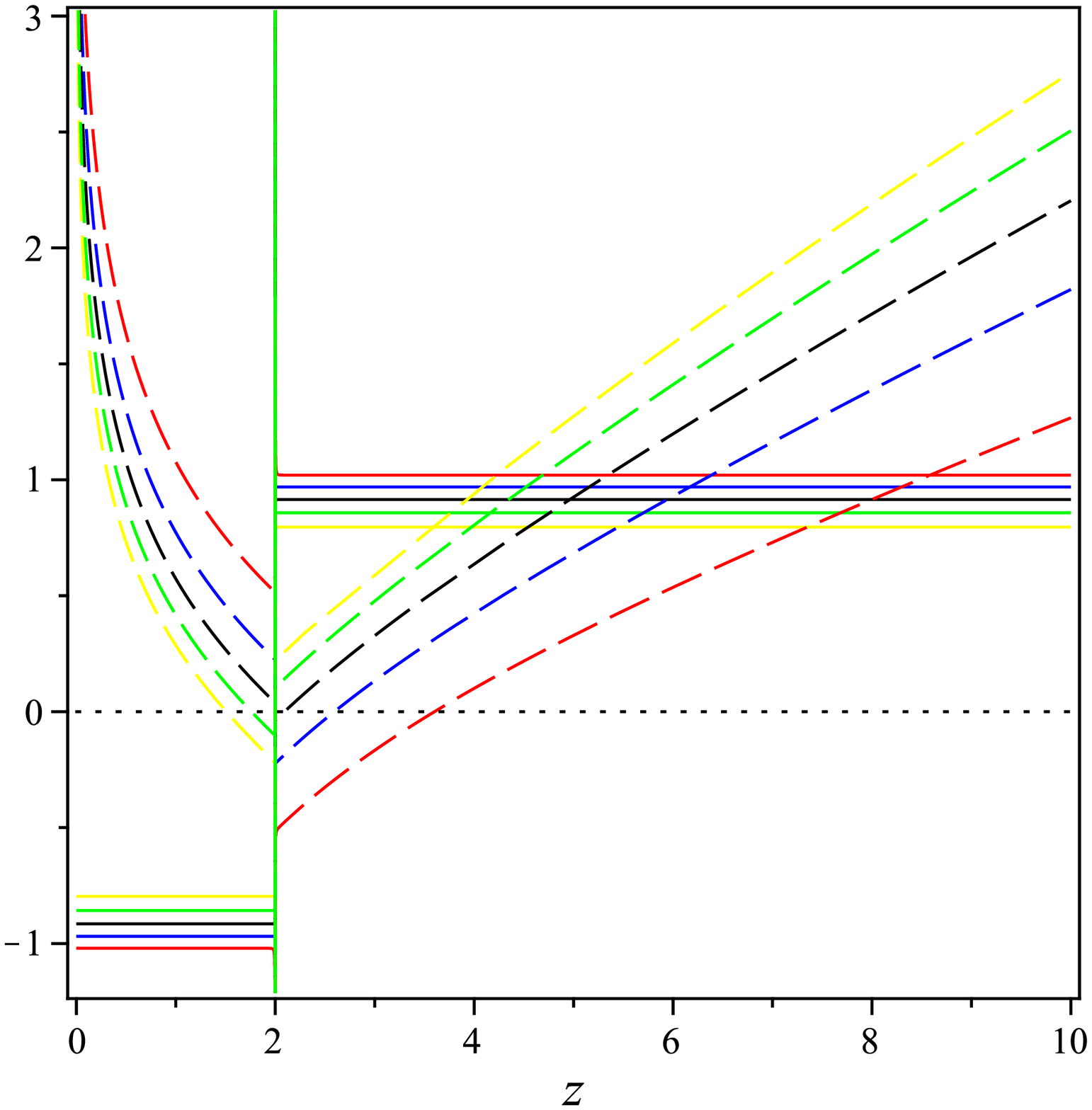}}
\subfigure[$z_{h}=2,\alpha=0.1,l=1$]{\includegraphics[width=0.3\columnwidth]{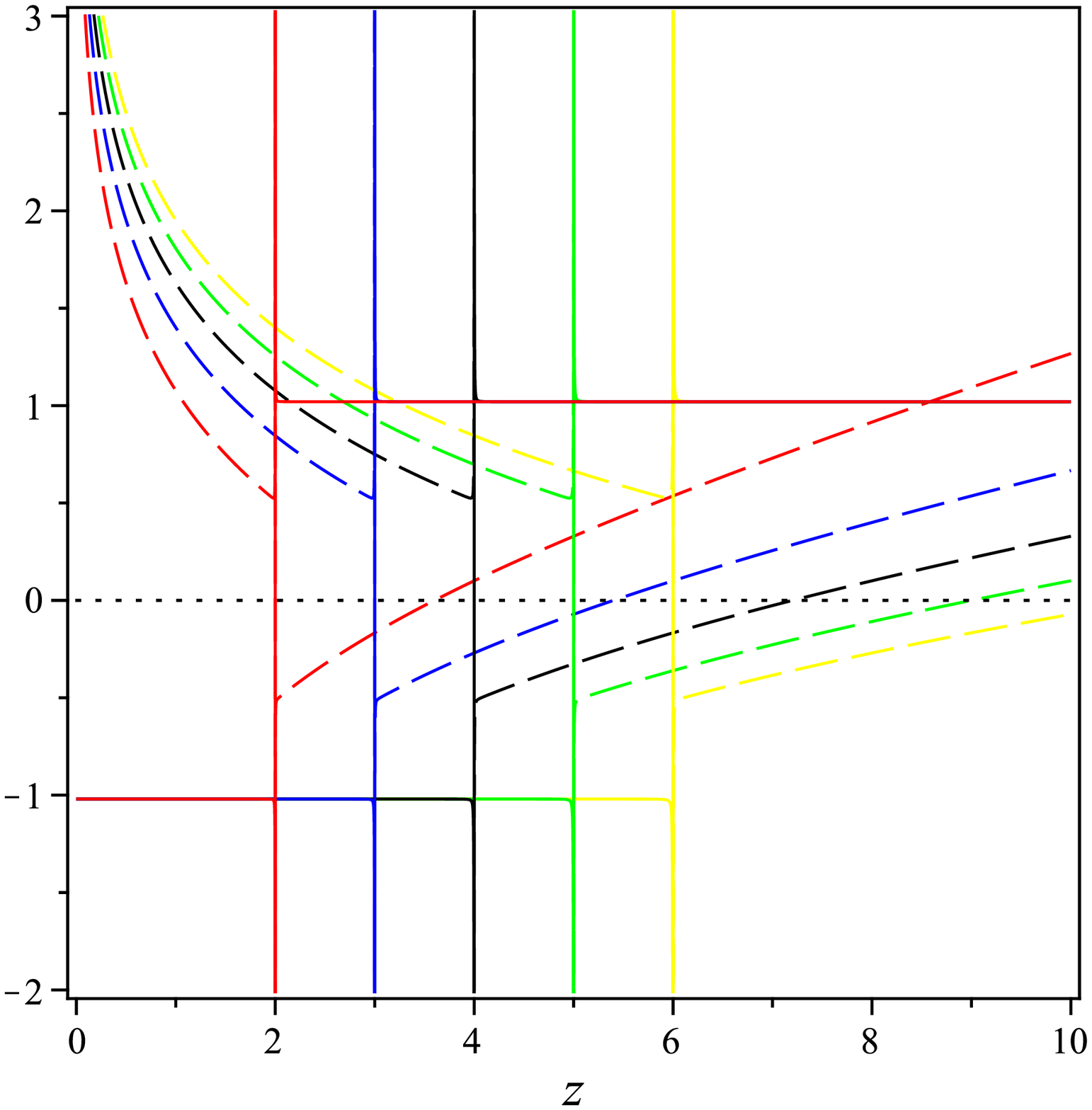}}
\caption{Plots of imagenary (solid lines) and real part (dashed lines) of $\phi$ in terms of $z$ for $\nu=\textcolor{red}{2},\textcolor{blue}{3},4,\textcolor{green}{5},\textcolor{orange}{6}$ (left), for $\alpha=\textcolor{red}{0.1},\textcolor{blue}{0.2},0.3,\textcolor{green}{0.4},\textcolor{orange}{0.5}$ (middle), for $z_{h}=\textcolor{red}{2},\textcolor{blue}{3},4,\textcolor{green}{5},\textcolor{orange}{6}$ (right).} 
\label{phizimreplote}
\end{figure}
\begin{figure}[H]\hspace{0.4cm}
\centering
\subfigure[$\nu=4.5,l=1,z_{h}=2,q=0.5$]{\includegraphics[width=0.3\columnwidth]{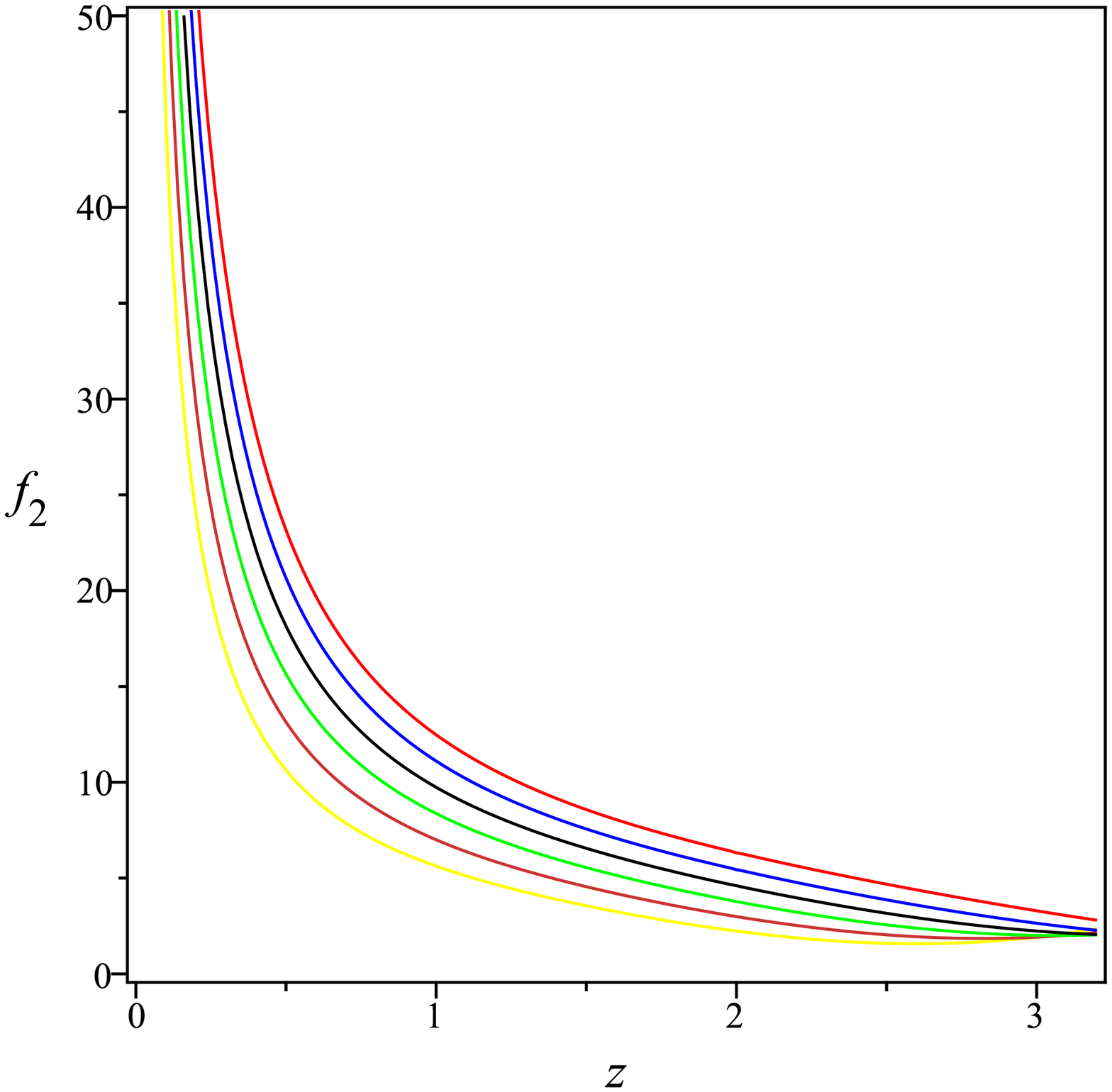}}
\subfigure[$\nu=4.5,l=1,z_{h}=2,\alpha=0.1$]{\includegraphics[width=0.3\columnwidth]{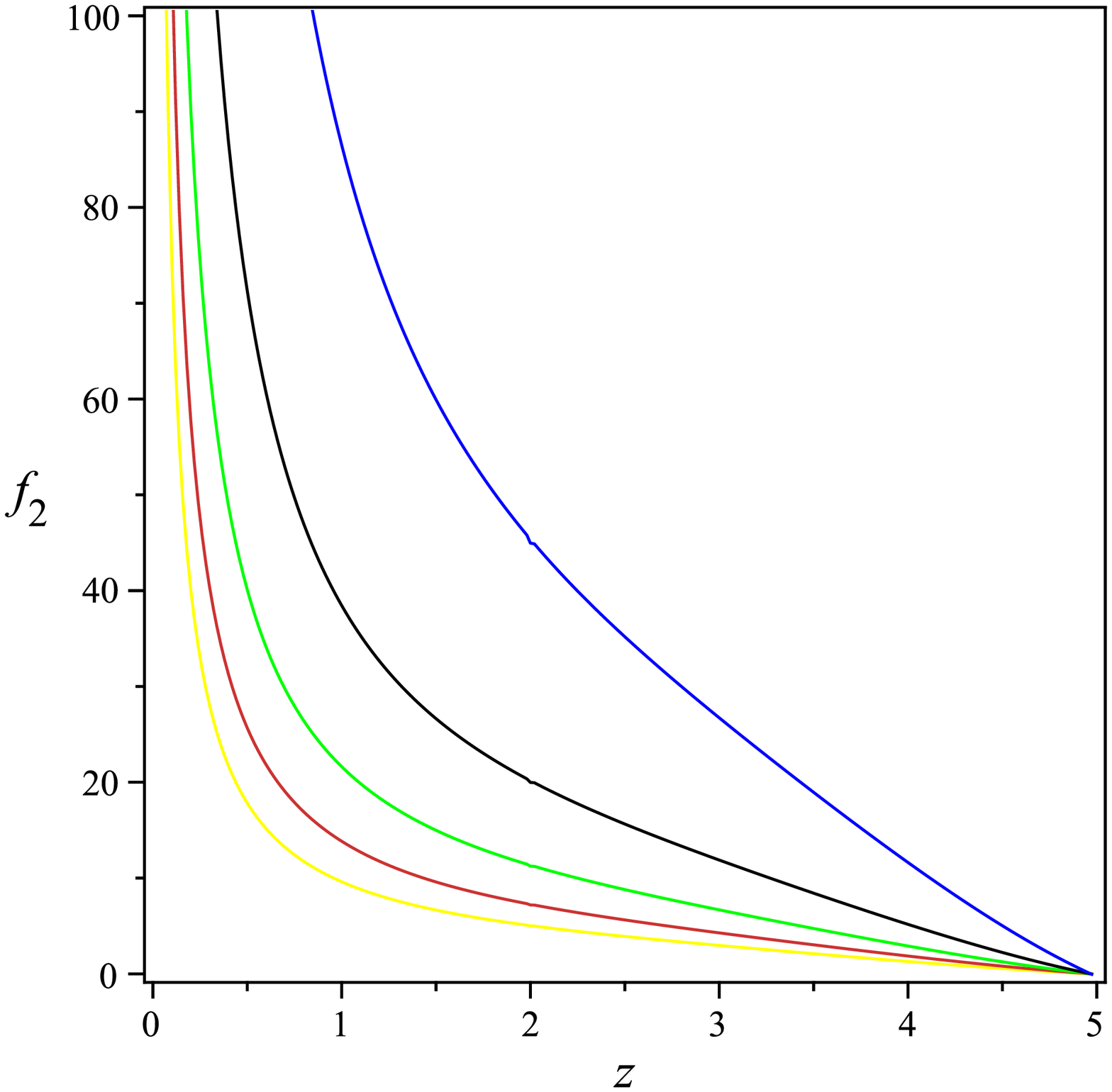}}
\subfigure[$l=1,\alpha=0.1,q=0.5,z_{h}=2$]{\includegraphics[width=0.3\columnwidth]{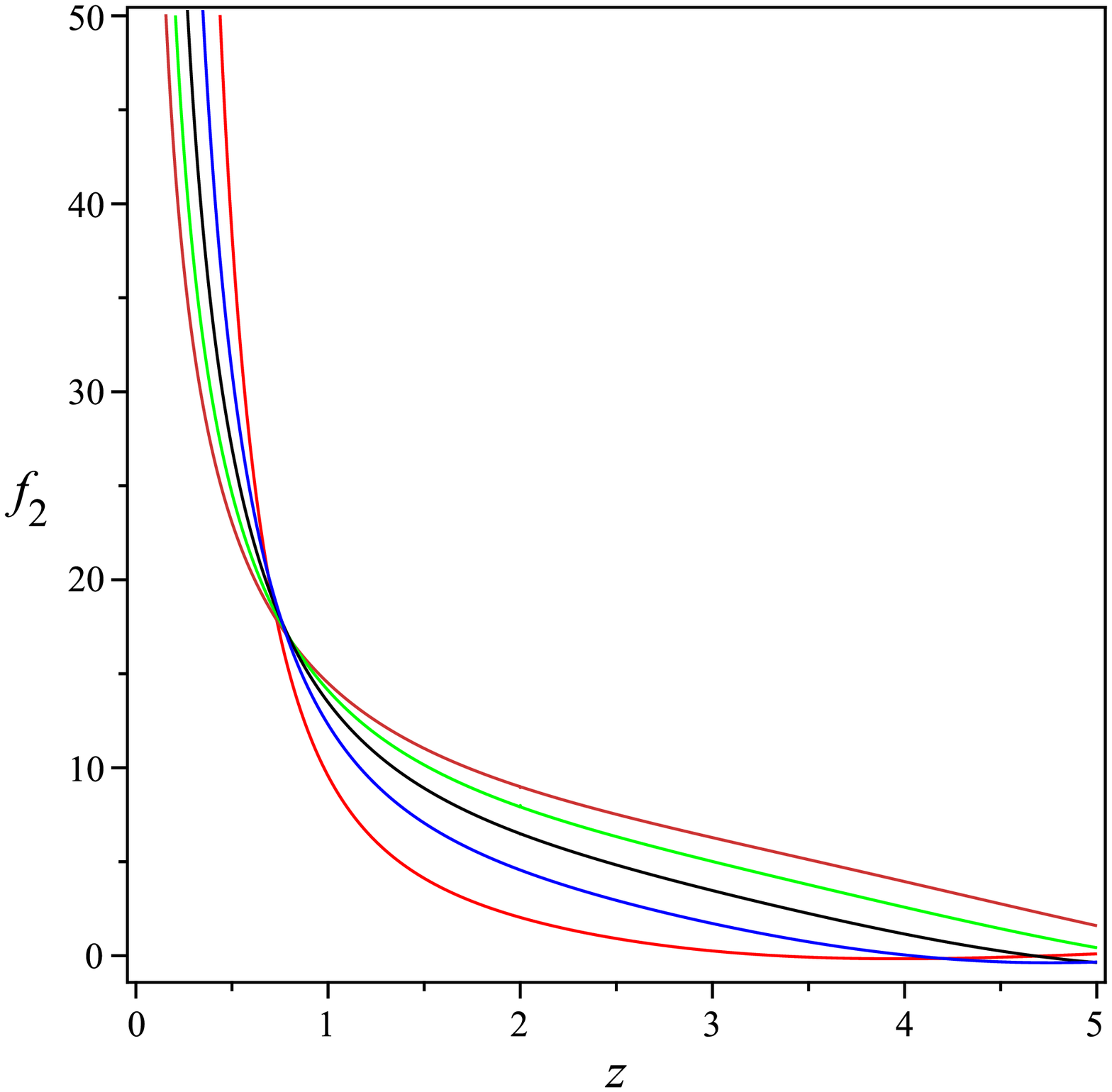}}
\caption{Plots of $f_{2}$ in terms of $z$ for $\alpha=\textcolor{red}{0.1},\textcolor{blue}{0.2},0.3,\textcolor{green}{0.4},\textcolor{orange}{0.5}$ (left), $q=\textcolor{red}{0.1},\textcolor{blue}{0.2},0.3,\textcolor{green}{0.4},\textcolor{orange}{0.5}$ (middle), $\nu=\textcolor{red}{1},\textcolor{blue}{2},3,\textcolor{green}{4},\textcolor{orange}{5}$ (right).} 
\label{f2plote}
\end{figure}
\begin{figure}[H]\hspace{0.4cm}
\centering
\subfigure[$\nu=4.5,z_{h}=2,l=1$]{\includegraphics[width=0.4\columnwidth]{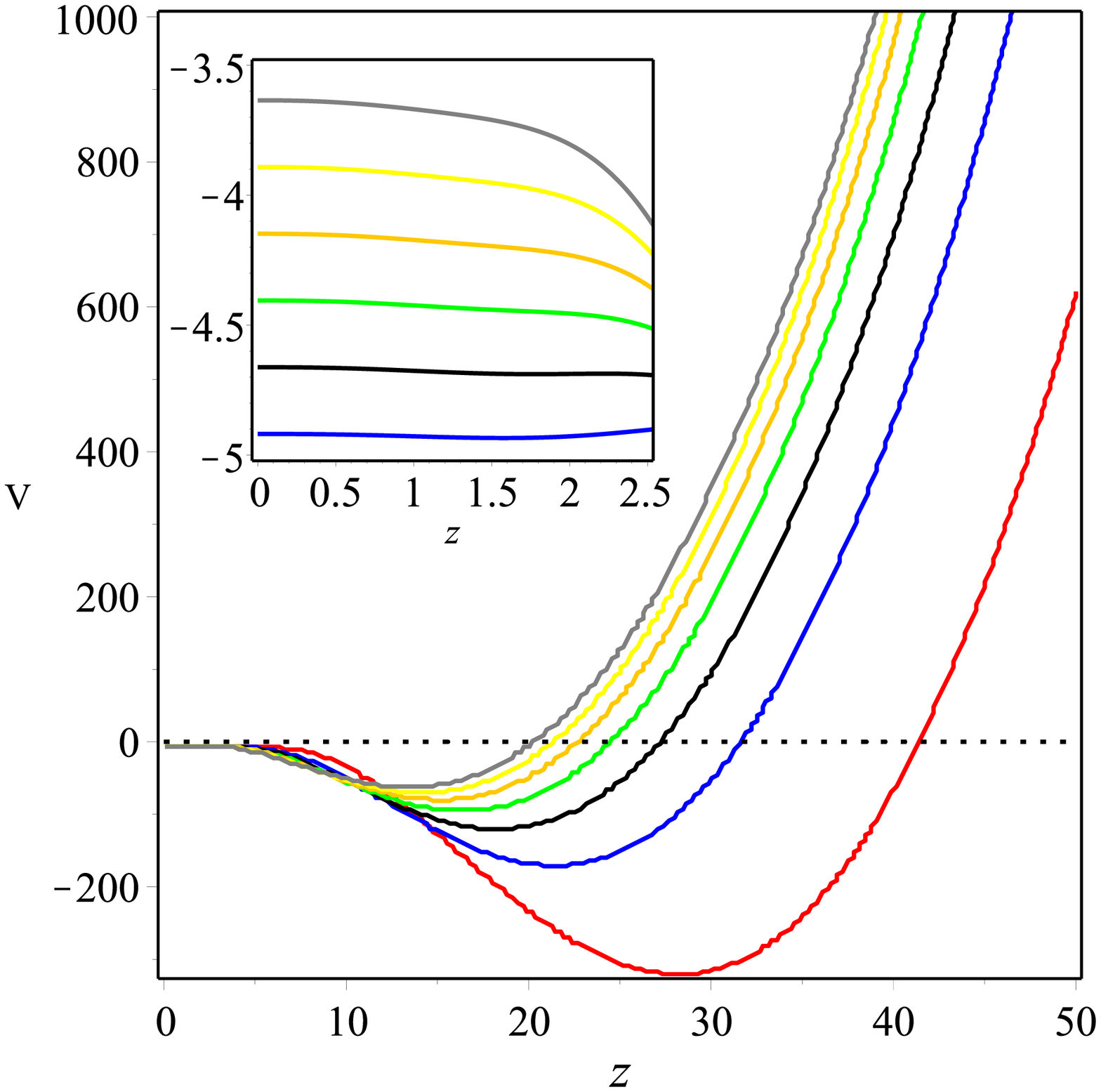}}
\subfigure[$\nu=4.5,l=1,z_{h}=2$]{\includegraphics[width=0.4\columnwidth]{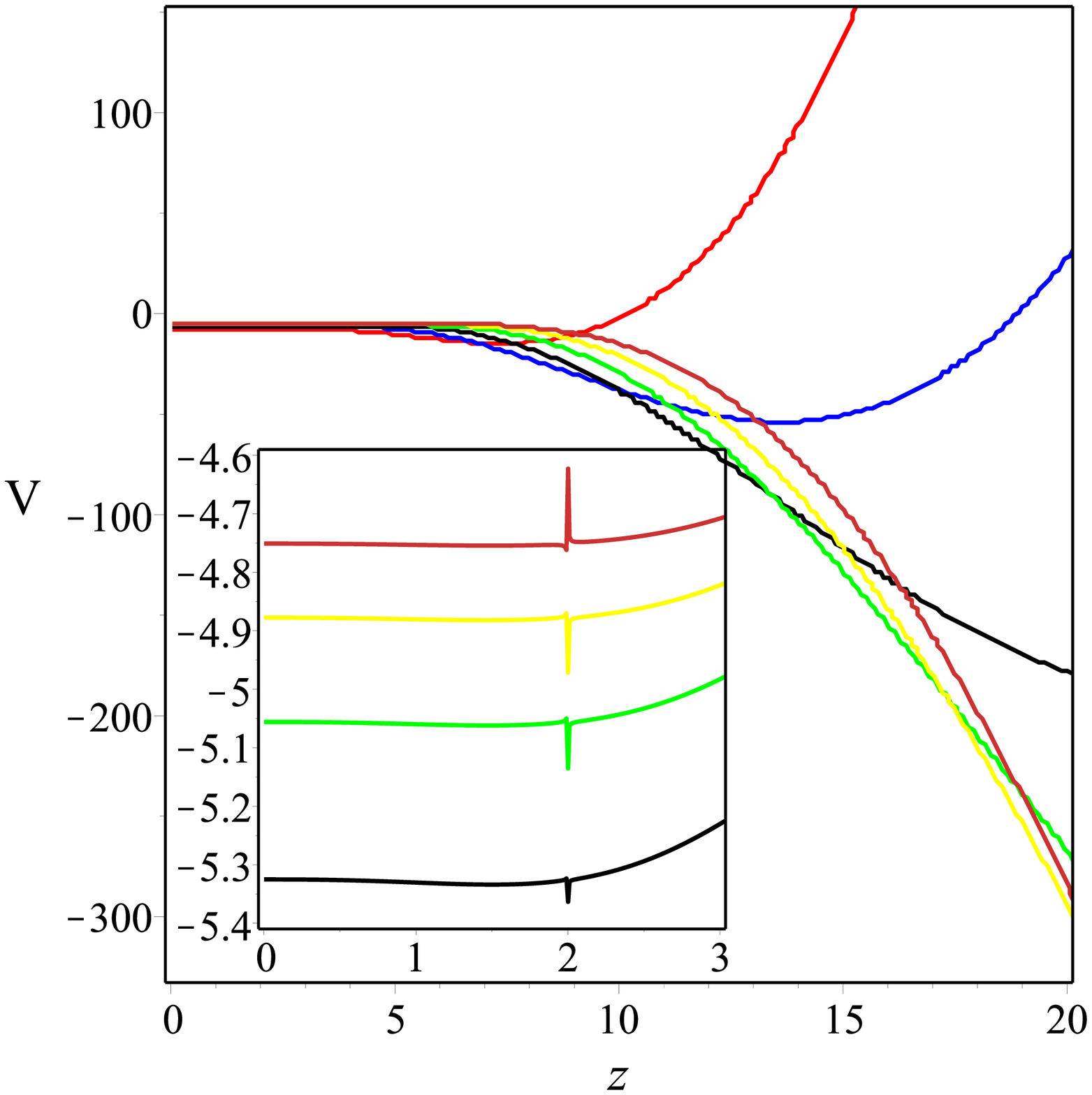}}
\caption{Plots of $V$ in terms of $z$ for $\alpha=\textcolor{red}{0.1},\textcolor{blue}{0.2},0.3,\textcolor{green}{0.4},\textcolor{orange}{0.5}$ (left), $\nu=\textcolor{red}{1},\textcolor{blue}{2},3,\textcolor{green}{4},\textcolor{orange}{5}$ (right).} 
\label{Vzplotc0}
\end{figure}

\noindent By inserting (\ref{eqqgz}) in $E_{tt}$, one can obtain $\phi$ as follows:
\begin{small}
\begin{align}\label{eqphi18}
\phi(z)=\dfrac{-2\sqrt{-2\mathcal{K}(\nu-1)}\left(\sqrt{\mathcal{H}\sqrt{\mathcal{D}}}\arctan\left(\sqrt{\frac{\mathcal{G}}{\mathcal{F}}}\right)-\mathcal{L}\sqrt{\mathcal{F}}\arctan\left(\sqrt{\frac{\mathcal{G}}{\mathcal{D}}}\right)-2\sqrt{\mathcal{DFG}(l^{4}\nu^{4}-8\alpha c_{2})}\right)}{(\nu+1)l\sqrt{\mathcal{GFE}\nu}\sqrt{l^{4}\nu^{4}-8\alpha c_{2}}\sqrt{l^{2}\nu^{3}+l^{2}\nu^{2}-\sqrt{\mathcal{E}}}}+c_{3},
\end{align}
\end{small}
\noindent by imposing the condition $\phi(z_{h})=0$, one can obtain $c_{3}=0$.
The constants $\mathcal{K}, \mathcal{H}, ...$ are provided in \eqref{appeq48}. In the case of $\alpha\ll 1$, one can get
\begin{small}
{\begin{align}\label{eqqphie}
\phi(z) &=\dfrac{2\sqrt{\nu-1}}{\nu}\ln\left(\dfrac{z}{z_{h}}\right)-\dfrac{4\alpha\sqrt{\nu-1}}{l^2\nu^2(\nu+1)}\left[(\nu+1)\ln\left(\dfrac{z}{z_{h}}\right)+\dfrac{\nu}{2}\left(1-\left(\dfrac{z}{z_{h}}\right)^{\frac{2\nu+2}{\nu}}\right)\right]+\mathcal{O}(\alpha^{2}).
\end{align}}
\end{small}
\noindent The first term is the contribution of the Einstien term and the second term is from the Gauss-Bonnet term. In figure (\ref{phizimreplote}), the real and imaginary parts of the scalar field in terms of $z$ for different values of parameters have been shown. As can be seen the imaginary part of scalar field inside and outside the black brane has a non-zero value and is unstable.
\noindent By increasing $\nu$ in $0<z<z_{h}$, the real part and imaginary part of the scalar field increase and decrease respectively and for $z>z_{h}$  vice versa (Fig.(\ref{phizimreplote})a). In panel b, by increasing the coupling of theory in $0<z<z_{h}$ the real and imaginary parts decrease and increase respectively, and for $z>z_{h}$ vice versa. 
\noindent By inserting (\ref{eqqphie}) into $E_{xx}$, one can obtain $f_{2}$ as follows:
\begin{align}\label{eqf221}
f_{2}&=-\dfrac{2}{\nu^{3}q^{2}\alpha (\nu+1)\bar{\mathcal{A}}^{\frac{5}{2}}(\nu^{2}l^{2}(\nu+1)-\sqrt{\bar{\mathcal{A}}})}[-6\nu(\nu-1)(\nu+1)^{5}\alpha c_{1}\bar{\mathcal{C}}(\nu^{4}l^{4}-8\alpha c_{2})(\nu l^{2}(\nu-\frac{1}{2})\nonumber\\
&\sqrt{\bar{\mathcal{A}}}-\bar{\mathcal{K}})z^{\frac{2\nu-2}{\nu}}-8\alpha^{3}c_{1}^{3}\bar{\mathcal{C}}\nu^{2}(2\nu^{2}-5\nu-2+\nu^{3})(\nu+1)^{3}z^{\frac{2+6\nu}{\nu}}+2z^{4}\nu(\nu+1)^{4}\alpha^{2}c_{1}^{2}\bar{\mathcal{C}}(\nu^{2}l^{2}\sqrt{\bar{\mathcal{A}}}\nonumber\\
&(2\nu^{2}+5-5\nu)-\bar{\mathcal{F}})+z^{-\frac{4}{\nu}}[\frac{1}{2}(\nu^{2}l^{2}(\nu+1)^{2}\bar{\mathcal{E}}\bar{\mathcal{A}}^{\frac{3}{2}})+\frac{1}{2}(\nu^{2}l^{2}(\nu^{2}-1)\bar{\mathcal{B}}\bar{\mathcal{A}}^{\frac{5}{2}})+\nu l^{2}(\nu-1)(\nu+1)^{6}\nonumber\\
&(\nu-\frac{1}{2})\bar{\mathcal{C}}\sqrt{\bar{\mathcal{A}}}(\nu^{4}l^{4}-8\alpha c_{2})^{2}-\nu^{15}l^{8}\bar{\mathcal{C}}\bar{\mathcal{G}}-5\nu^{14}l^{8}
\bar{\mathcal{C}}\bar{\mathcal{G}}-9\nu^{13}l^{8}\bar{\mathcal{C}}\bar{\mathcal{G}}-
5\nu^{12}l^{8}\bar{\mathcal{C}}\bar{\mathcal{G}}+5\nu^{11}l^{4}\bar{\mathcal{C}}\bar{\mathcal{G}}(l^{4}+\nonumber\\
&\frac{16}{5}\alpha c_{2})+9\nu^{10}l^{4}\bar{\mathcal{C}}\bar{\mathcal{G}}(l^{4}+\frac{80}{9}\alpha c_{2})+5\nu^{9}l^{4}\bar{\mathcal{C}}\bar{\mathcal{G}}(l^{4}+\frac{144}{5}\alpha c_{2})+l^{4}\nu^{8}\bar{\mathcal{C}}\bar{\mathcal{G}}(l^{4}+80\alpha c_{2})-80\alpha \nu^{7}c_{2}\bar{\mathcal{C}}\bar{\mathcal{G}}\nonumber\\
&(l^{4}+\frac{4}{5}\alpha c_{2})-144\alpha c_{2}\nu^{6}\bar{\mathcal{C}}\bar{\mathcal{G}}(l^{4}+\frac{20}{9}\alpha c_{2})-80\alpha \nu^{5}c_{2}\bar{\mathcal{C}}\bar{\mathcal{G}}(l^{4}+\frac{36}{5}\alpha c_{2})-16\alpha \nu^{4}c_{2}\bar{\mathcal{C}}\bar{\mathcal{G}}(l^{4}+20\alpha c_{2})\nonumber\\
&+320\alpha^{2}\nu^{3}c_{2}^{2}\bar{\mathcal{C}}\bar{\mathcal{G}}+\nu^{2}\bar{\mathcal{L}}
+\nu\bar{\mathcal{J}}+\frac{1}{2}\bar{\mathcal{B}}\bar{\mathcal{A}}^{3}+64\alpha^{2}c_{2}^{2}
\bar{\mathcal{C}}\bar{\mathcal{G}}]],
\end{align}
\noindent where the constants $\bar{\mathcal{A}}$, $\bar{\mathcal{B}}$... are provided in \eqref{appeq49}.
In the case of $\alpha\ll 1$, one can get
\begin{align}
f_{2}&=\dfrac{4l^{2}(\nu^{2}-1)z^{-\frac{4}{\nu}}}{\nu^{2}q^{2}}-\dfrac{8(\nu^{2}-1)z^{4}\alpha}{q^{2}\nu^{4}\left(z^{\frac{2\nu+2}{\nu}}-z_{h}^{\frac{2\nu+2}{\nu}}\right)^{2}}\times\nonumber\\
&\left[-4\nu \left(\dfrac{z_{h}}{z}\right)^{\frac{2\nu+2}{\nu}}+3\left(\dfrac{z}{z_{h}}\right)^{\frac{4\nu+4}{\nu}}-6\left(\dfrac{z}{z_{h}}\right)^{\frac{2\nu+2}{\nu}}+\nu\left(\dfrac{z_{h}}{z}\right)^{\frac{4\nu+4}{\nu}}+(2\nu+3)\right]+\mathcal{O}(\alpha^{2}).
\end{align}
The first terms is the contribution of Einstien gravity and the second term is related to the Gauss-Bonnet gravity. In figure (\ref{f2plote}), the behavior of $f_{2}$ in terms of $z$ are shown. As can be seen, by increasing $\alpha$ and $q$, $f_{2}$ decreases and by increasing $\nu$, $f_{2}$ increases.
\noindent Finally from $E_{tt}$, one can get $V(z)$ (We did not bring it here due to its bulk). In the case of $\alpha\ll 1$, one can get
\begin{align}
V(z)&=-\dfrac{2(\nu+1)(2\nu+1)}{\nu^{2}l^{2}}+\dfrac{\alpha}{\left(z^{\frac{2\nu+2}{\nu}}-z_{h}^{\frac{2\nu+2}{\nu}}\right)^{3}\nu^{4}l^{4}}[-4 z^{\frac{4\nu+4}{\nu}}z_{h}^{\frac{2\nu+2}{\nu}}(-1-7\nu+6\nu^{3}+20\nu^{2})\nonumber\\
&+8\nu (3\nu^{2}+7\nu-1)z^{\frac{2\nu+2}{\nu}}z_{h}^{\frac{4\nu+4}{\nu}}+(8\nu^{3}+64\nu^{2}-36\nu-12)z^{\frac{6\nu+6}{\nu}}+4(\nu-1)(2\nu+1)z^{\frac{10\nu+10}{\nu}}z_{h}^{\frac{-4\nu-4}{\nu}}\nonumber\\
&-4(\nu-1)(8\nu+3)z^{\frac{8\nu+8}{\nu}}z_{h}^{\frac{-2\nu-2}{\nu}}-8\nu^{2}(\nu+2)z_{h}^{\frac{6\nu+6}{\nu}}]+\mathcal{O}(\alpha^{2}).
\end{align}
In Fig. (\ref{Vzplotc0}), the behavior of scalar potential in terms of $z$ for anisotropic case has been shown. {In the left panel, between $0<z<z_{h}$ by increasing $\alpha$, the potential decreases. In the right panel, between $0<z<z_{h}$ by increasing $\nu$ the scalar potential increases. }

\subsubsection{Thermodynamics of the background}
\begin{figure}[H]\hspace{0.4cm}
\centering
\subfigure[$l=1$]{\includegraphics[width=0.3\columnwidth]{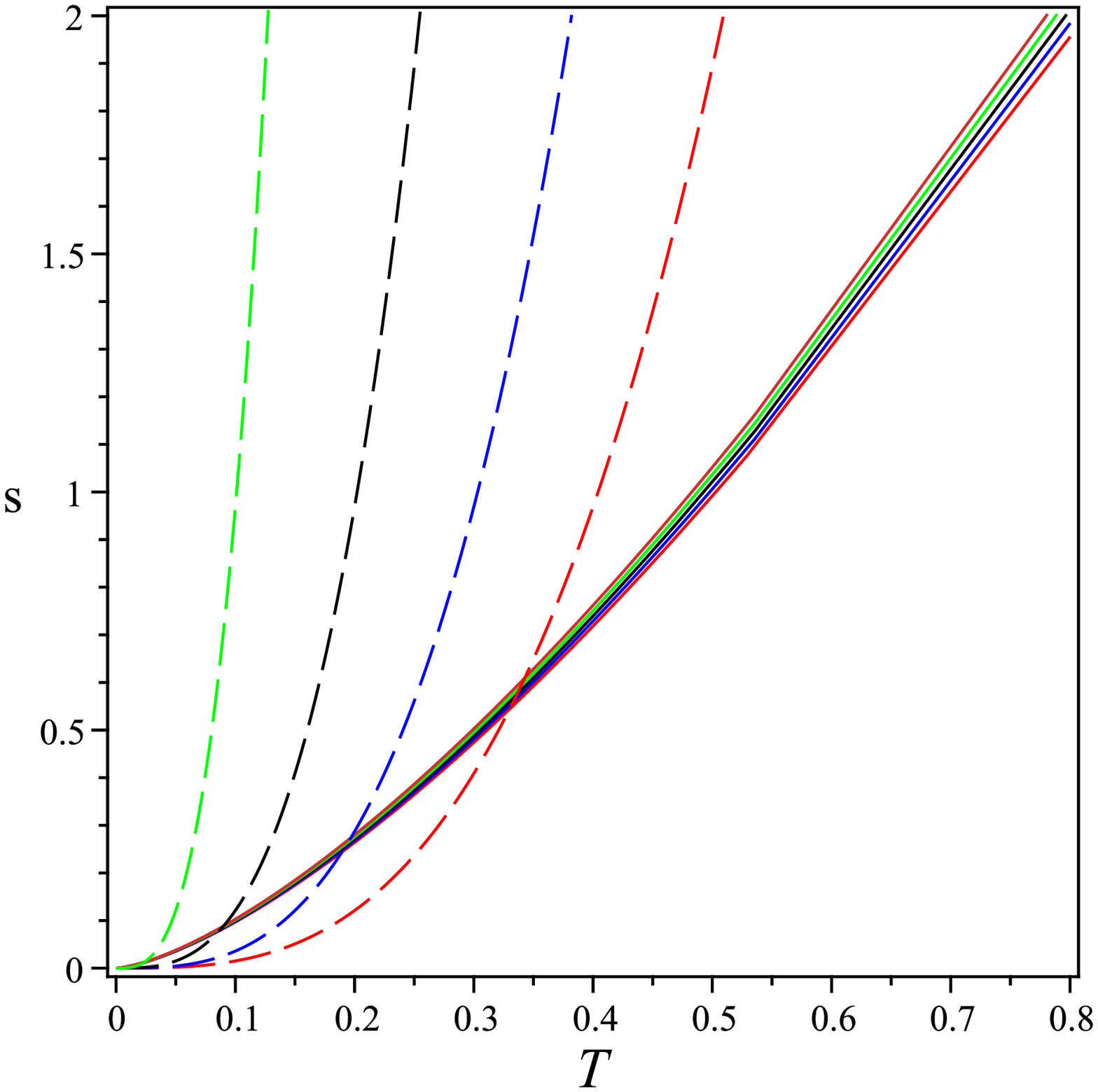}}
\subfigure[$l=1$]{\includegraphics[width=0.3\columnwidth]{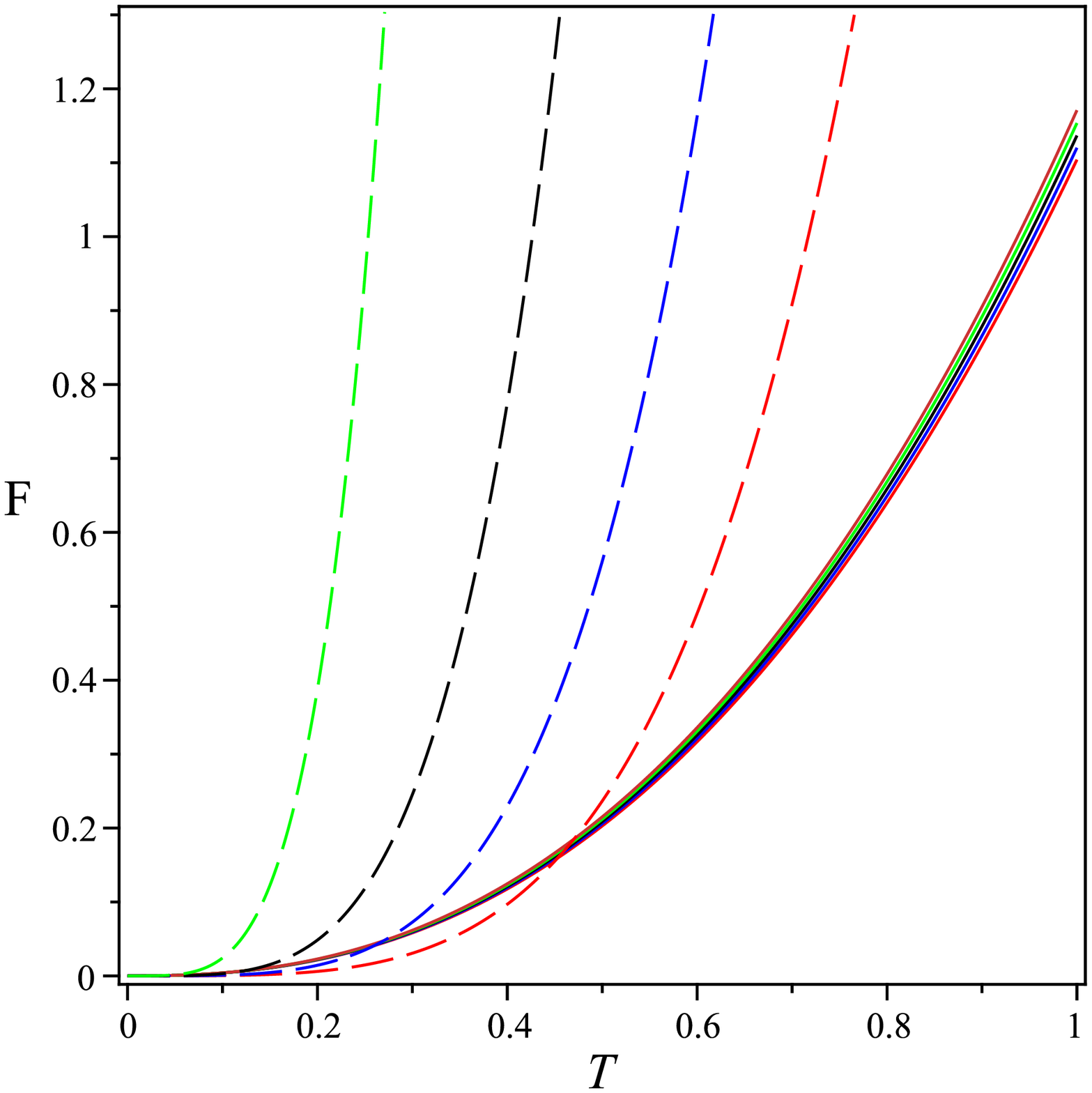}}
\subfigure[$l=1$]{\includegraphics[width=0.3\columnwidth]{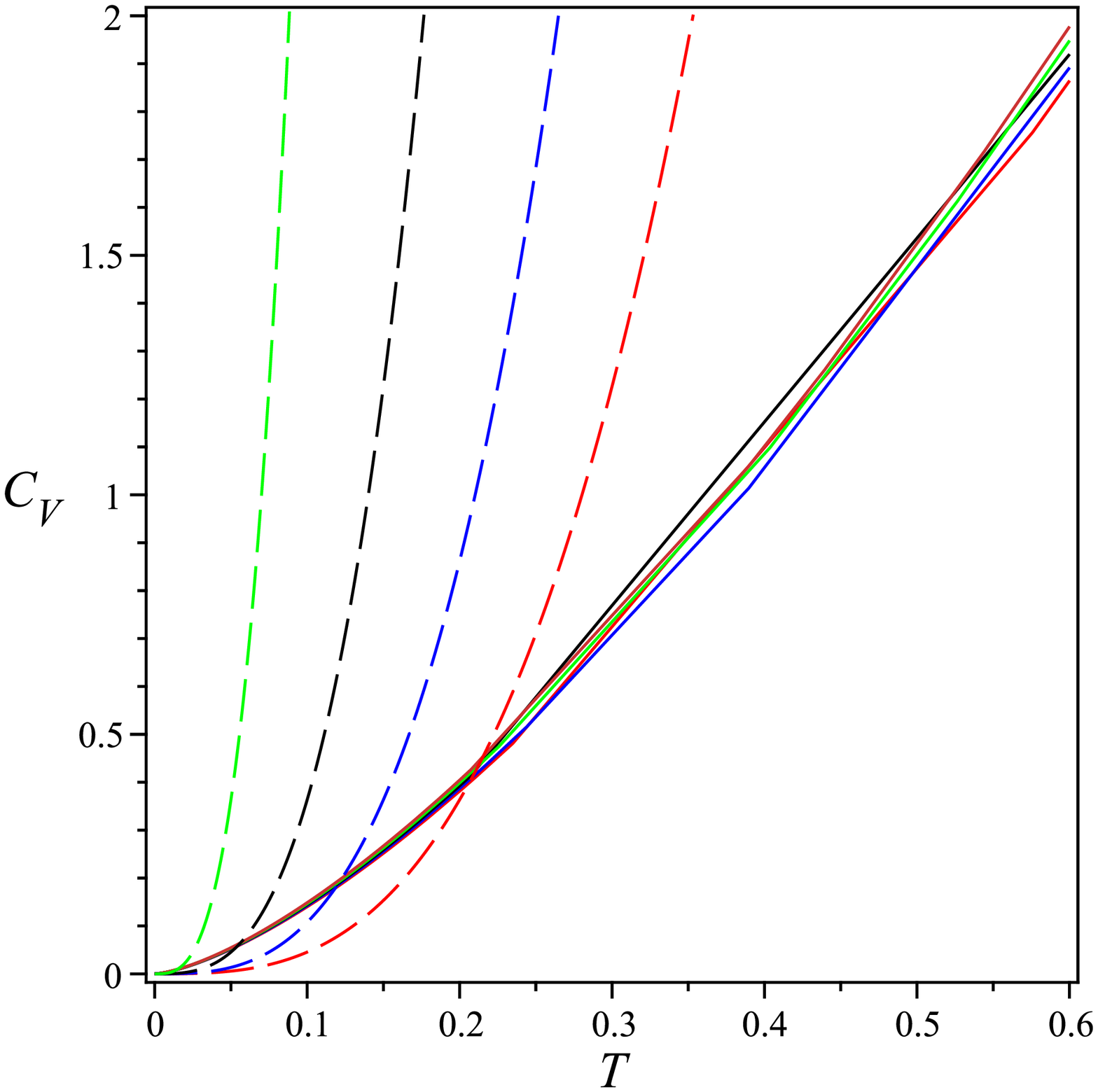}}
\caption{Plots of $s$, $F$ and $C_{V}$ in terms of $T$ for $\alpha=\textcolor{red}{0.1},\textcolor{blue}{0.2},0.3,\textcolor{green}{0.4},\textcolor{orange}{0.5}$ and $\nu=4.5$ (solid line) and $\nu=1$ (dashed lines) .} 
\label{SFCVplote}
\end{figure}

In this subsection, we explore the thermodynamics of the black brane solution (\ref{eqqgz}). In order to investigate the thermodynamic properties of the black brane we need to obtain some relevant thermodynamic quantities. 
The temperature of the black brane is obtained as follows:
\begin{align}
T=\left\vert\dfrac{g^{\prime}}{4\pi}\right\vert=\dfrac{\nu+1}{2\pi z_{h}\nu}\left(1-\dfrac{2\alpha}{\nu^{2}l^{2}}\right).
\end{align}
It is noticed that, the temperature monotonically decreases with the increase of the horizon.
By increasing $\alpha$, the  temperature is decreased, and for $\alpha=0$, one can get the result of \cite{Arefeva:2018hyo}. The entropy is given as follows \cite{Wald1},\cite{Wald2}
\begin{align}
S=-\dfrac{1}{8}\int_{\Sigma}d^{n-2}x\sqrt{\eta}\dfrac{\delta L}{\delta R_{\mu \alpha \beta \nu}}\epsilon_{\mu \alpha}\epsilon_{\beta \nu},
\end{align}
where
\begin{align}
\dfrac{\delta L}{\delta R_{\mu \alpha \beta \nu}}=&\left(\dfrac{1}{2}+\alpha R\right)\left(g^
{\mu \beta}g^{\alpha \nu}-g^{\mu \nu}g^{\alpha \beta}\right)+\dfrac{1}{2}\beta\left(R^{\mu \beta}g^{\alpha \nu}-R^{\alpha \beta}g^{\mu \nu}-R^{\mu \nu}g^{\alpha \beta}+R^{\alpha \nu}g^{\mu \beta}\right)\nonumber\\
&+2\gamma R^{\mu \alpha \beta \nu},
\end{align}
and $\epsilon_{\mu \nu}=-2\sqrt{-h}\delta^{t}_{[\mu}\delta^{z}_{\nu]}$.
For $c=0$ we have
\begin{equation}\label{eqqen28}
s=\dfrac{S}{\mathcal{V}}=\dfrac{l^3P(z_{h}) b(z_{h})^{\frac{3}{2}}}{4z_{h}^{3}}=\dfrac{l^3}{4 z_{h}^{\frac{\nu+2}{\nu}}},
\end{equation}
which is independent of parameter of the EGB gravity.
In terms of the temperature, the entropy is given as
\begin{equation}
s=\dfrac{l^3}{4}\left(\dfrac{2\pi T\nu^{3}l^{2}}{(\nu+1)(\nu^{2}l^{2}-2\alpha)}\right)^{\frac{\nu+2}{\nu}}.
\end{equation}
For isotropic case $s\approx T^{3}$ and for anisotropic case $s\approx T^{\frac{\nu+1}{\nu}}$.
The free energy density $F(T)$ can be calculated from the entropy density $s(T)$ by integrating as follows
\begin{equation}
F=\int s dT=\dfrac{\nu}{2(\nu+1)}T s,
\end{equation}
which is related to temperature as $F\approx T^{\frac{2\nu+1}{\nu}}$. 
The sound velocity $c_{s}^{2}$ which can directly measure the conformality of the system, can be obtained from the temperature and entropy:
\begin{equation}
c_{s}^{2}=\dfrac{d\log T}{d\log s}=\dfrac{\nu}{\nu+2}.
\end{equation}
For isotropic case ($\nu=1$), $c_{s}^{2}= 1/3$, the system is conformal, for anisotropic ($\nu\neq 1$), $c_{s}^{2}\neq 1/3$ the system is non-conformal.
The heat capacity is given as 
\begin{equation}
C_{V}=T\dfrac{ds}{dT}=\dfrac{(\nu+2)}{\nu}s=\dfrac{s}{c_{s}^{2}}.
\end{equation}
In terms of temperature, 
\begin{equation}\label{eqqcvt}
\dfrac{C_{V}}{T^{3}}=\dfrac{l^3(\nu+2)}{4\nu}\left(\dfrac{2\pi\nu^3 l^2}{(\nu+1)(\nu^2l^2-2\alpha)}\right)^{\frac{\nu+2}{\nu}}T^{\frac{2-2\nu}{\nu}}.
\end{equation}
For isotropic case, the right hand side of \eqref{eqqcvt} has a constant value and for $\nu> 1$ depends to the temperature and at high temperature goes to zero. Since entropy is positive therefore $C_{V}$ is positive and the black hole is stable. In figure \ref{SFCVplote}, the behavior of $s$, $F$ and $C_{V}$ in terms of $T$ for isotropic (dashed lines) and anisotropic (solid lines) and different values of $\alpha$ have been shown. As can be seen, by increasing $\alpha$, the thermodynamical quantities $s$, $F$, and $C_{V}$ increase. 
For $T<T^{i}_{cross}$, the entropy, free energy, and heat capacity of anisotropic case is larger than isotropic, and for $T>T^{i}_{cross}$ vice versa. Where $i=s, F, C_{V}$ and $T^{i}_{cross}$ are given as follows 
\begin{align}
T^{s}_{cross}=&\left[\dfrac{l^{4}((l^2-2\alpha)^{3}\nu^{\frac{3(\nu+2)}{\nu}}\pi^{\frac{2(1-\nu)}{\nu}}2^{\frac{\nu+2}{\nu}})^{\nu}}{l^{4\nu}((\nu+1)(l^2\nu^2-2\alpha))^{2}(\nu+1)(l^2\nu^2-2\alpha)^{\nu}}\right]^{\frac{1}{2\nu-2}},\\
T_{cross}^{F}=&\dfrac{\left(\dfrac{(l^2-2\alpha)^{2}4^{\frac{\nu+1}{\nu}}\pi^{\frac{2-2\nu}{\nu}}}{\nu^{\frac{-3\nu-6}{\nu}}+\nu^{\frac{-4\nu-6}{\nu}}}\right)^{\frac{\nu}{2(\nu-1)}}}{[(\nu+1)(l^2\nu^{2}-2\alpha)^{\nu+2}l^{4(\nu-1)}]^{\frac{1}{2\nu-2}}},\\
T^{C_V}_{cross}=&\left[\dfrac{l^{4(\nu+1)}((l^2\nu^2-2\alpha)^{3}(\nu+2)\nu^{\frac{2(\nu+1)}{\nu}}\pi^{\frac{2(1-\nu)}{\nu}}2^{\frac{\nu+2}{\nu}})^{\nu}}{3^{\nu}(\nu+1)(l^2\nu^2-2\alpha)^{\nu+2}}\right]^{\frac{1}{2\nu-2}}.
\end{align}

\subsection{The case $c\neq0, \mu=0$}\label{sect2.2}
In this case $A_{t}=0$ and the differential equation \eqref{eqqtotal}, becomes
\begin{align}\label{eqq34}
&4\nu lz^{2}e^{-\frac{cz^{2}}{4}}(l^{2}\nu^{2}e^{-\frac{cz^{2}}{2}}-\alpha(2+\nu cz^{2})^{2}g(z))g^{\prime\prime}-2zle^{-\frac{cz^{2}}{4}}(\nu^{2}l^{2}(4+(2+3z^{2})\nu)-\nonumber\\
&\alpha(2+\nu c z^{2})(8+4\nu+6\nu c z^{2}-6c\nu^{2}z^{2}+c^{2}\nu^{2}z^{4})g)g^{\prime}-4\alpha \nu z^{2}le^{-\frac{cz^{2}}{4}}(2+\nu cz^{2})^2g^{\prime 2}=0.
\end{align}
In order to solve equation \eqref{eqq34}, we assume $g(z)$ as follows
\begin{equation}\label{eqqmetr35}
g(z)=1+\epsilon g_{1}(z)+\mathcal{O}(\epsilon^{2}),
\end{equation}
by inserting it into the \eqref{eqq34}, one can achieve a homogeneous differential equation for $g_{1}(z)$ as
\begin{equation}\label{eqq360}
g_{1}^{\prime\prime}+\dfrac{[-l^{2}\nu^{3}(2+3cz^{2})-4\nu^{2}(l^{2}+4\alpha cz^{2})+4\alpha \nu (2+cz^{2})+16\alpha]g_{1}^{\prime}}{\nu z(l^{2}\nu^{2}-4\alpha)}=0.
\end{equation}
 Solving \eqref{eqq360} give $g_{1}(z)$ as
\begin{align}\label{eqqg1z}
g_{1}(z)=c_{1}+c_{2}\text{erf}\left(\dfrac{1}{2}\sqrt{\dfrac{c(-6l^{2}\nu^{2}-32\nu\alpha+8\alpha)}{(l^{2}\nu^{2}-4\alpha)}}z\right),
\end{align}
where $c_{1}$ and $c_{2}$ are constants of integration. Using \eqref{eqqg1z}, the metric \eqref{eqqmetr35} becomes
\begin{equation}
g(z)=1+c_{1}+c_{2}\text{erf}\left(\dfrac{1}{2}\sqrt{\dfrac{c(-6l^{2}\nu^{2}-32\nu\alpha+8\alpha)}{(l^{2}\nu^{2}-4\alpha)}}z\right).
\end{equation}
The conditions \eqref{eqqcond} give us $c_{1}$ and $c_{2}$ as 
\begin{equation}\label{eqqconst}
c_{1}=0,\;\;\;\;c_{2}=-\dfrac{1}{\text{erf}\left(\dfrac{1}{2}\sqrt{\dfrac{c(-6l^{2}\nu^{2}-32\nu\alpha+8\alpha)}{(l^{2}\nu^{2}-4\alpha)}}z_{h}\right)}.
\end{equation}
Finally, using \eqref{eqqconst} the metric function becomes
\begin{equation}\label{eqmetric40}
g(z)=1-\dfrac{\text{erf}\left(\dfrac{1}{2}\sqrt{\dfrac{c(-6l^{2}\nu^{2}-32\nu\alpha+8\alpha)}{(l^{2}\nu^{2}-4\alpha)}}z\right)}{\text{erf}\left(\dfrac{1}{2}\sqrt{\dfrac{c(-6l^{2}\nu^{2}-32\nu\alpha+8\alpha)}{(l^{2}\nu^{2}-4\alpha)}}z_{h}\right)}.
\end{equation}
In the case of $\alpha\ll 1$ and $c<0$, the blakening function become
\begin{small}
\begin{equation}
g(z)\approx 1-\dfrac{\text{erf}\left(\dfrac{\sqrt{-6c}z}{2}\right)}{\text{erf}\left(\dfrac{\sqrt{-6c}z_{h}}{2}\right)}-\dfrac{4(2\nu+1)\sqrt{-6c}\alpha\left(ze^{\frac{3cz^{2}}{2}}\text{erf}\left(\dfrac{\sqrt{-6c}z_{h}}{2}\right)-z_{h}\text{erf}\left(\dfrac{\sqrt{-6c}z}{2}\right)e^{\frac{3cz_{h}^{2}}{2}}\right)}{3\sqrt{\pi}l^{2}\nu^{2}\text{erf}\left(\dfrac{\sqrt{-6c}z_{h}}{2}\right)^2}+\mathcal{O}\left(\alpha^{2}\right),
\end{equation}
\end{small}and for $\alpha\ll 1$ and $c>0$, we have
 \begin{equation}
 g(z)\approx 1-\dfrac{z}{z_{h}}e^{\frac{3c(z^{2}-z_{h}^2)}{2}}-\dfrac{2\alpha c(2\nu+1)}{\nu^2 l^2}\dfrac{z(z^2-z_{h}^2)}{z_{h}}e^{\frac{3c(z^{2}-z_{h}^2)}{2}}+\mathcal{O}(\alpha^2).
 \end{equation}
 In figure \eqref{gzplote11}, the behavior of $g(z)$ for positive and negative values of warp function is depicted. As $c$ increases, the metric slope becomes more decreasing. Also, changing the value of $c$ has no effect on the value of the horizon. By substituting the obtained metric \eqref{eqmetric40}, we arrive at the differential equation for the scalar field as:
\begin{figure}[H]\hspace{0.4cm}
\centering
\subfigure[$\nu=4.5,\alpha=0.1,l=1,z_{h}=2$]{\includegraphics[width=0.4\columnwidth]{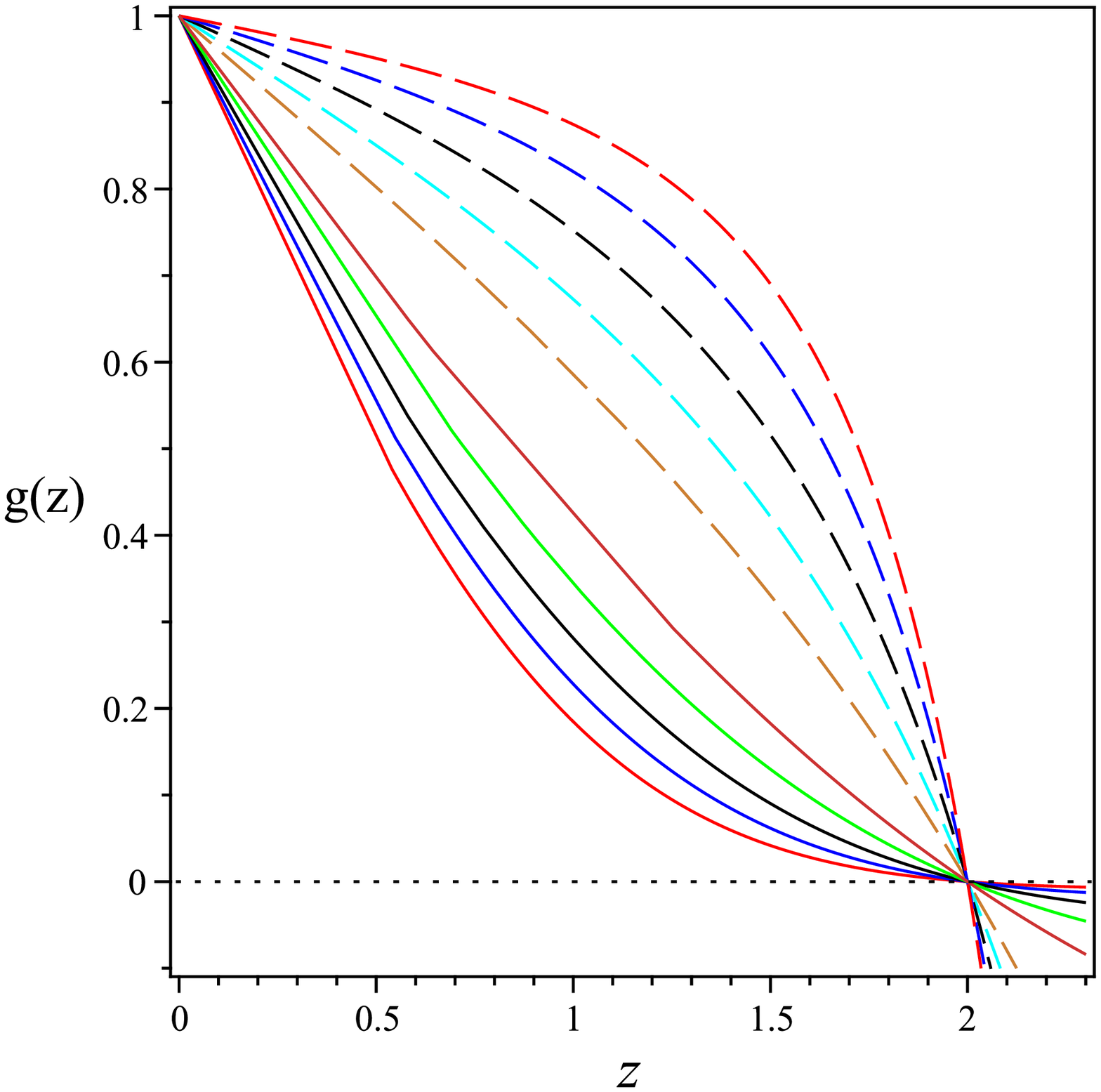}}
\subfigure[$\alpha=0.1,l=1,z_{h}=2$]{\includegraphics[width=0.4\columnwidth]{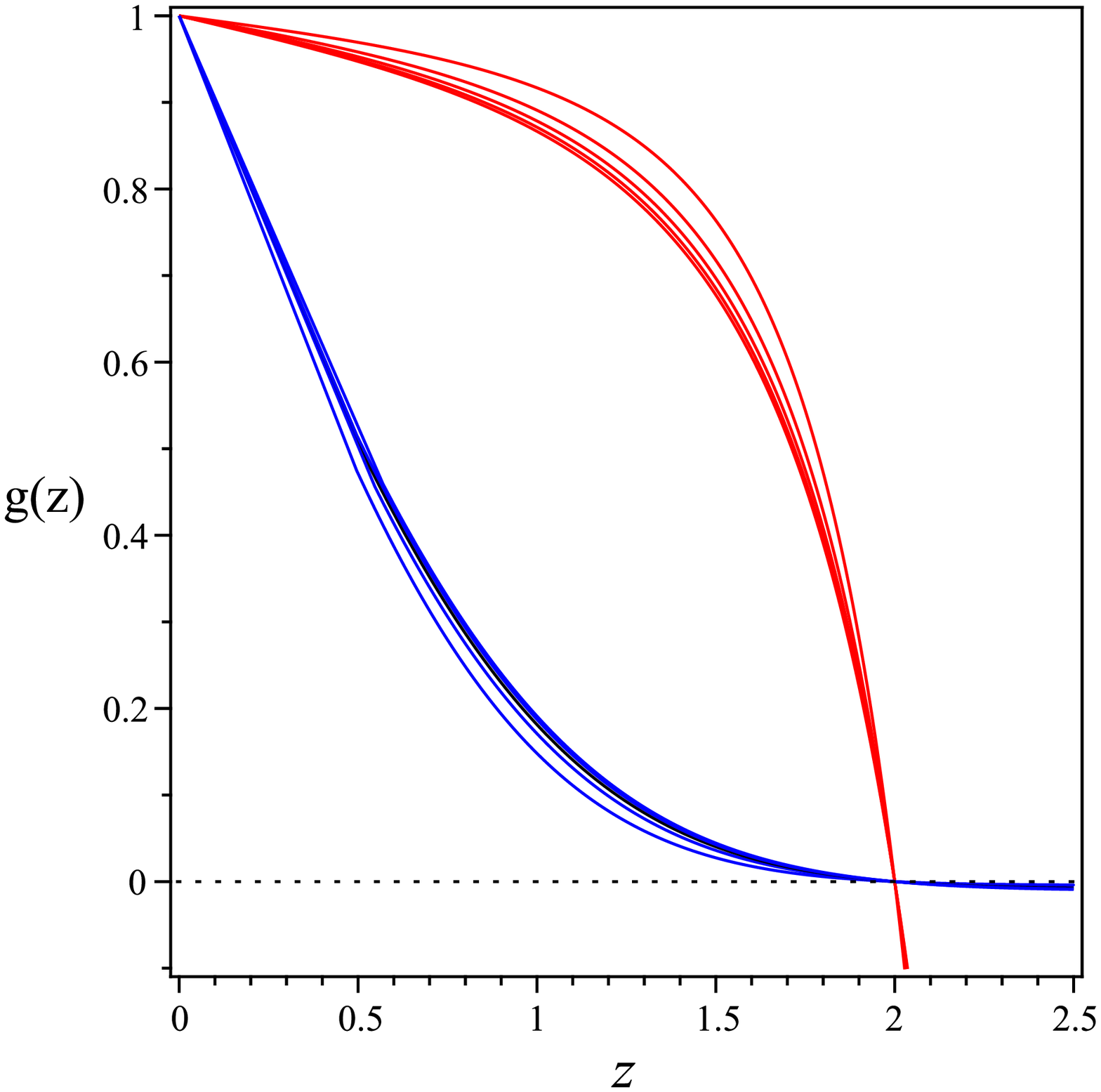}}
\caption{Plots of $g(z)$ in terms of $z$ for $c=\textcolor{red}{0.5},\textcolor{blue}{0.4},0.3,\textcolor{cyan}{0.2},\textcolor{orange}{0.1}$ (dashed lines) and $\textcolor{red}{-0.5},\textcolor{blue}{-0.4},-0.3,\textcolor{green}{-0.2},\textcolor{orange}{-0.1}$ (solid lines) (left), $\nu=1,2,3,4,5,6$, $\textcolor{red}{c=0.5}$  and $\textcolor{blue}{c=-0.5}$ (right).} 
\label{gzplote11}
\end{figure}

\begin{figure}[H]\hspace{0.4cm}
\centering
\subfigure[$\nu=4.5,\alpha=0.1,l=1,z_{h}=2$]{\includegraphics[width=0.4\columnwidth]{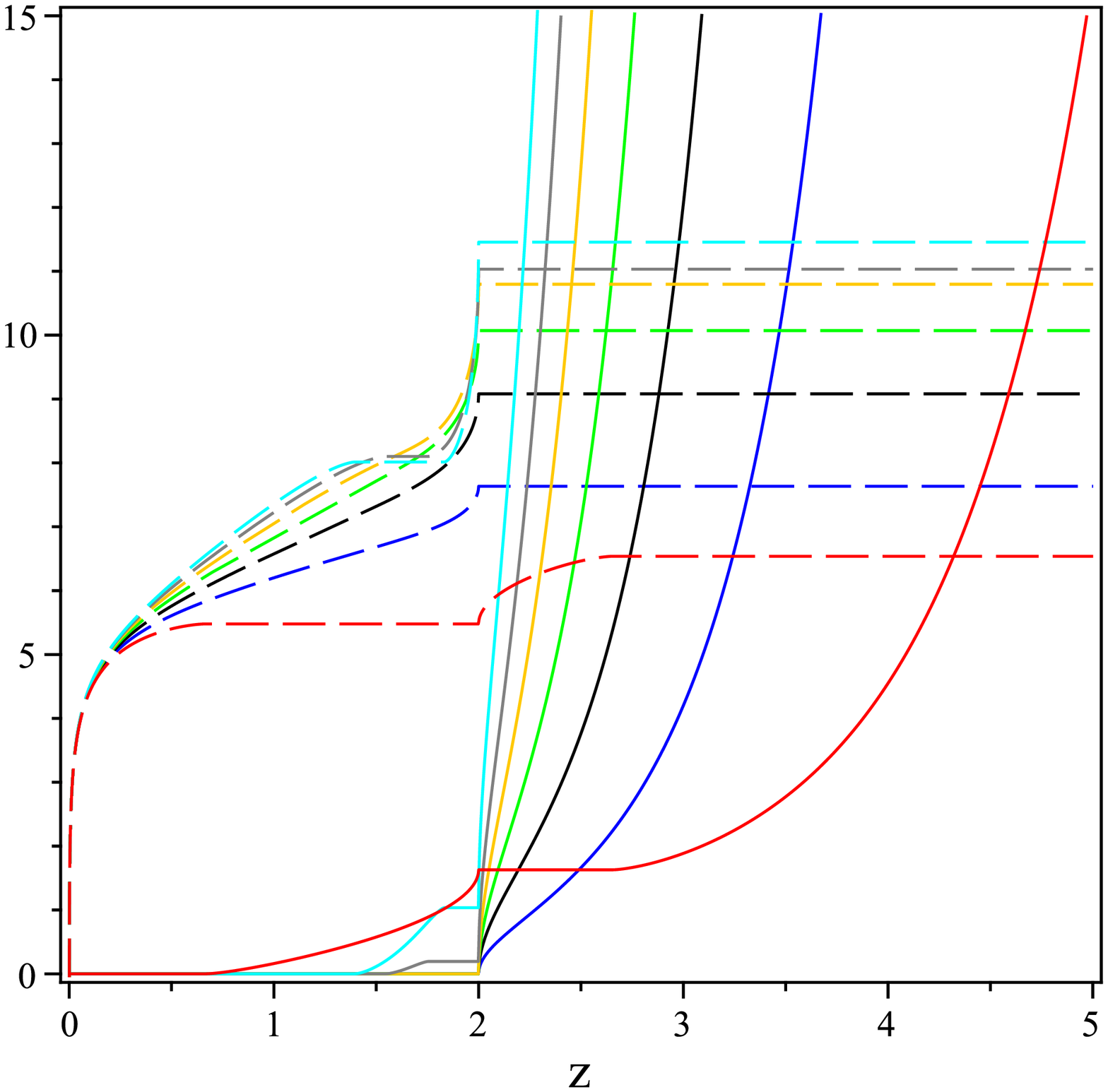}}
\subfigure[$\nu=4.5,l=1,z_{h}=2,c=-0.5$]{\includegraphics[width=0.4\columnwidth]{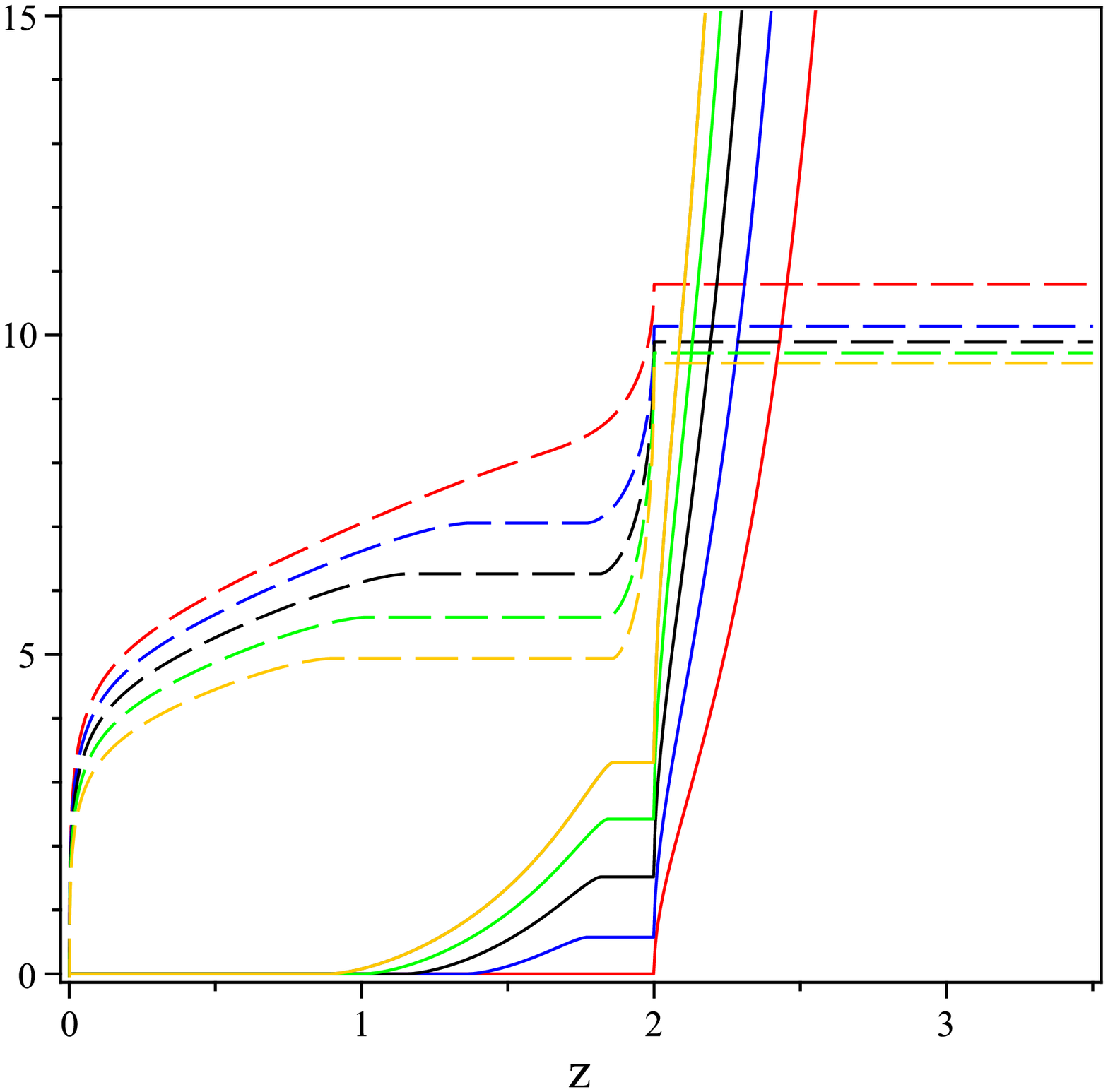}}
\caption{Plots of real (dashed lines) and imagenary (solid lines) part of $\phi$ in terms of $z$ for $c=\textcolor{red}{-0.1},\textcolor{blue}{-0.2},-0.3,\textcolor{green}{-0.4},\textcolor{yellow}{-0.5}$ (left), $\alpha=\textcolor{blue}{0.1},0.2,\textcolor{green}{0.3},\textcolor{yellow}{0.4}$,0.5 (right).} 
\label{phizplote11}
\end{figure}

\begin{figure}[H]\hspace{0.4cm}
\centering
\subfigure[$\nu=4.5,\alpha=0.1,l=1,z_{h}=2,q=0.5$]{\includegraphics[width=0.3\columnwidth]{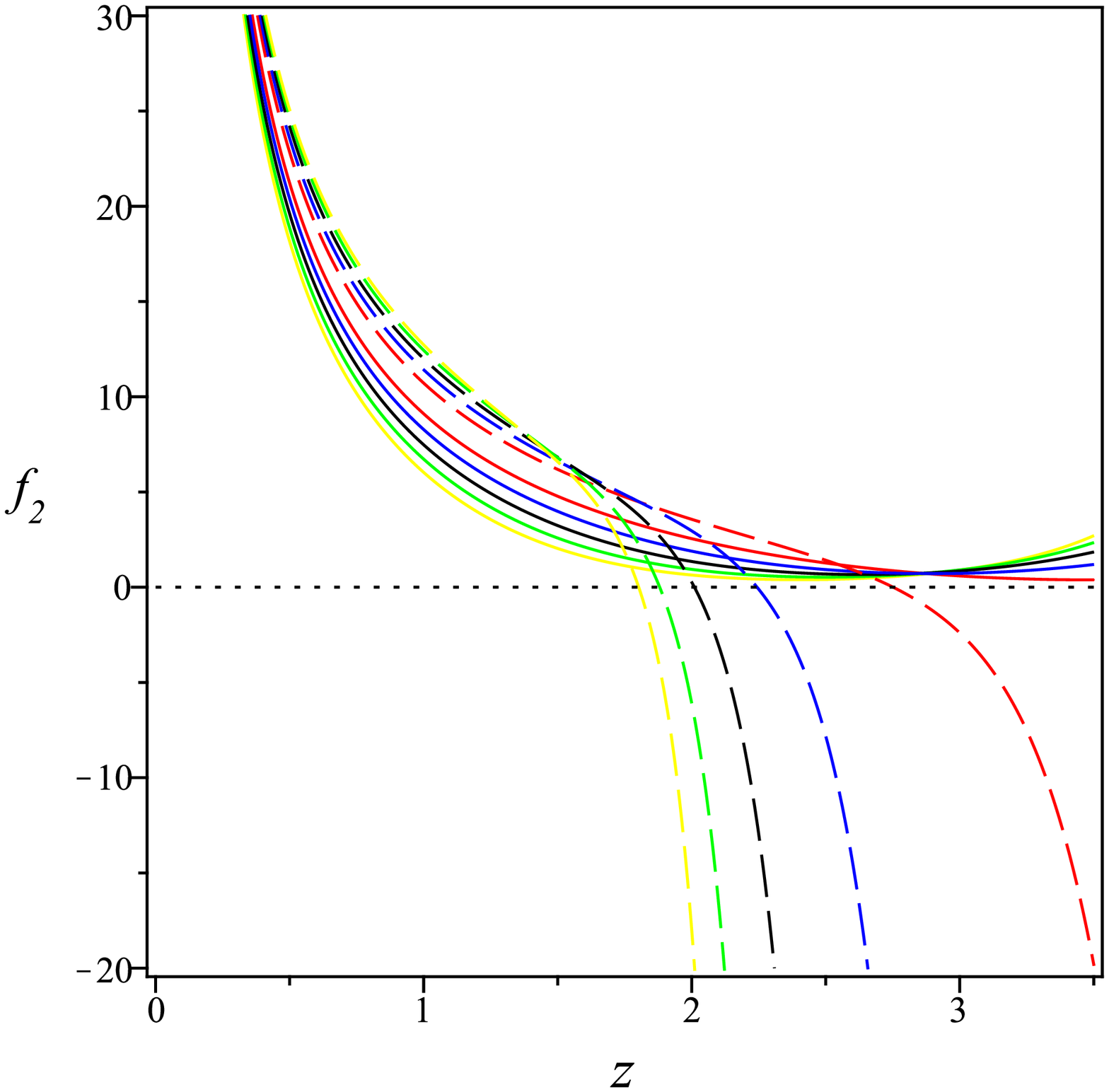}}
\subfigure[$\alpha=0.1,l=1,z_{h}=2,q=0.5$]{\includegraphics[width=0.3\columnwidth]{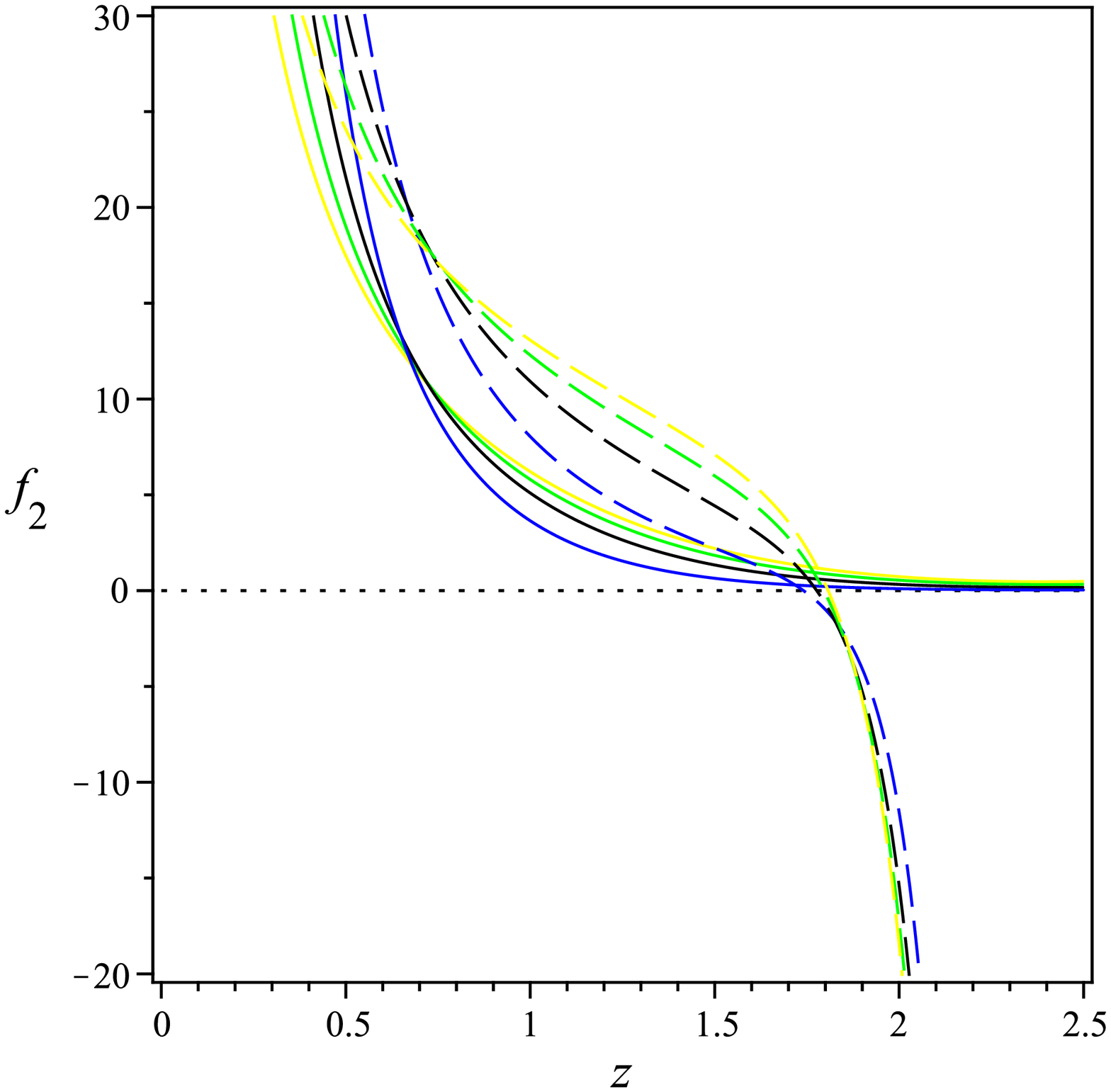}}
\subfigure[$l=1,z_{h}=2,q=0.5,\nu=4.5$]{\includegraphics[width=0.3\columnwidth]{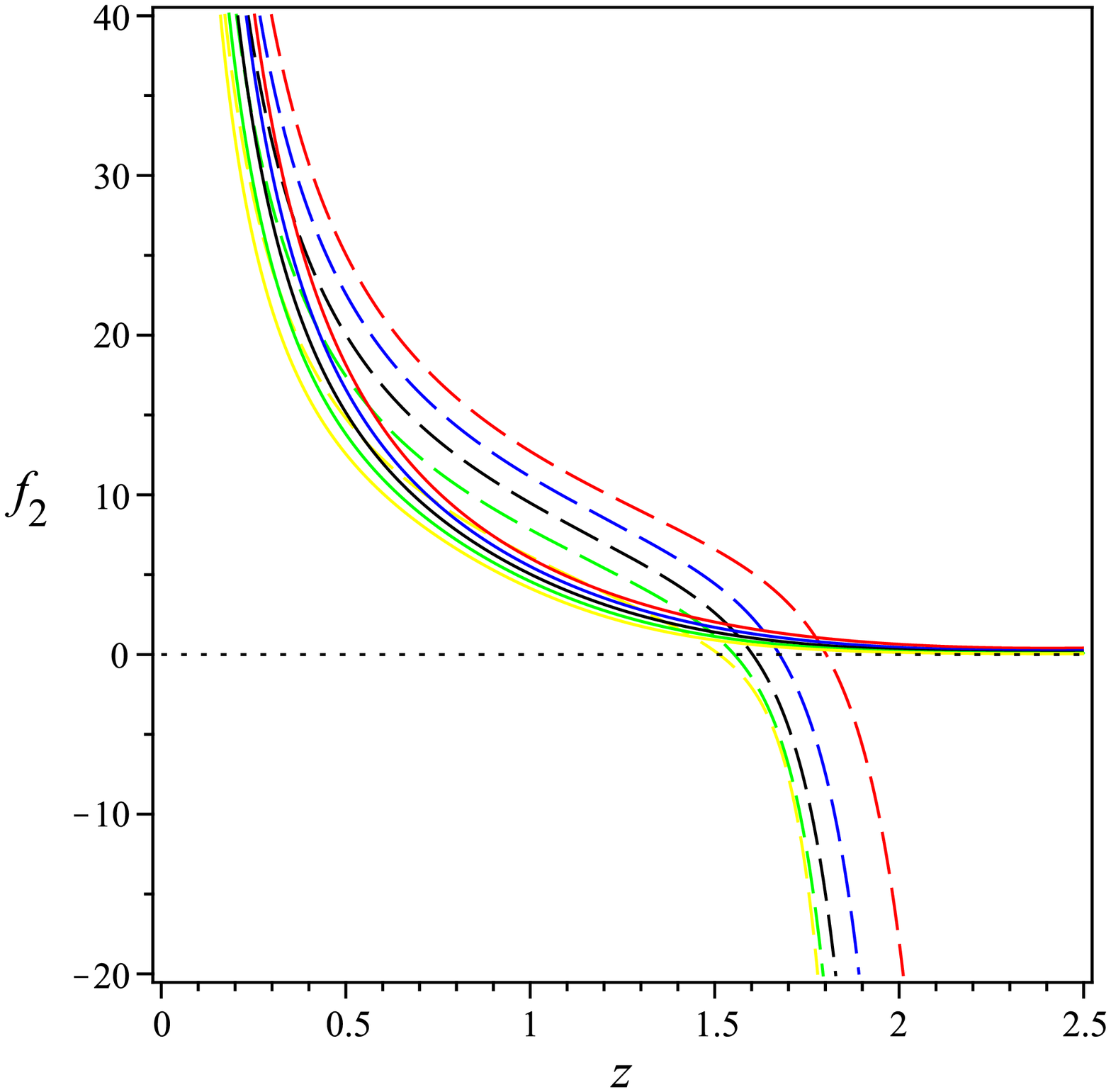}}
\caption{Plots of $f_{2}(z)$ in terms of $z$ for $c=\textcolor{red}{0.1},\textcolor{blue}{0.2},0.3,\textcolor{green}{0.4},\textcolor{yellow}{0.5}$ (solid lines) and $c=\textcolor{red}{-0.1},\textcolor{blue}{-0.2}, -0.3,\textcolor{green}{-0.4},\textcolor{yellow}{-0.5}$ (dashed lines) (left), $\nu=\textcolor{blue}{2},3,\textcolor{green}{4},\textcolor{yellow}{5}$, $c=0.5$ (solid lines) and $c=-0.5$ (dahsed lines) (middle), $\alpha=\textcolor{blue}{0.1},0.2,\textcolor{green}{0.3},\textcolor{yellow}{0.4}$, 0.5, $c=0.5$ (solid lines) and $c=-0.5$ (dashed lines) (right).} 
\label{f2zplote}
\end{figure}

\begin{small}
\begin{align}
\phi^{\prime 2}&=\dfrac{e^{\frac{cz^{2}}{2}}}{z^2 l^2\nu^{3}{2\pi (l^{2}\nu^{2}-4\alpha)\text{erf}\left(\frac{\sqrt{-2c\mathbb{A}}z_{h}}{2}\right)\left[\text{erf}\left(\frac{\sqrt{-2c\mathbb{A}}z_{h}}{2}\right)-\text{erf}\left(\frac{\sqrt{-2c\mathbb{A}}z}{2}\right)\right]}}[2\alpha\sqrt{-2c\pi \mathbb{A}}z(2+\nu cz^{2})\nonumber\\
&\left[-\text{erf}\left(\frac{\sqrt{-2c\mathbb{A}}z_{h}}{2}\right)+\text{erf}\left(\frac{\sqrt{-2c\mathbb{A}}z}{2}\right)\right]e^{\frac{\mathbb{A}cz^{2}}{2}}(\nu^2 c^{2}z^{4}(5l^2 \nu^{2}+32\alpha \nu-4\alpha)-4(\nu+2)(l^2\nu^2-4\alpha)\nonumber\\
&+2\nu cz^{2}(3\nu^3 l^2+20\alpha \nu+3l^2\nu^2+4\alpha))-8c\alpha \nu z^{2}(2+\nu cz^{2})^{2}(3l^2\nu^2+16\alpha \nu-4\alpha)e^{\mathbb{A}cz^{2}}\nonumber\\
&+2l^2\nu^2 z\sqrt{-2c\pi \mathbb{A}}\text{erf}\left(\frac{\sqrt{-2c\mathbb{A}}z_{h}}{2}\right)e^{\frac{cz^{2}\nu(\nu l^2+8\alpha)}{\nu^2l^2-4\alpha}}(\nu c z^{2}(3l^2\nu^2+32\alpha \nu+4\alpha)-2(\nu+2)(l^2\nu^2-4\alpha))\nonumber\\
&-\pi(l^2\nu^2-4\alpha)\left(\text{erf}(\frac{\sqrt{-2c\mathbb{A}}z}{2})-\text{erf}(\frac{\sqrt{-2c\mathbb{A}}z_{h}}{2})\right)(l^2\nu e^{-\frac{cz^{2}}{2}}\text{erf}(\frac{\sqrt{-2c\mathbb{A}z_{h}}}{2})(-8+8\nu+3\nu^2c^2z^{4}\nonumber\\
&+18\nu^2 c z^{2})+\alpha(2+\nu cz^{2})(\text{erf}(\frac{\sqrt{-2c\mathbb{A}}z}{2})-\text{erf}(\frac{\sqrt{-2c\mathbb{A}}z_{h}}{2}))(-16+16\nu+3\nu^2 z^{6}c^{3}+2\nu c^{2}z^{4}(1+\nonumber\\
&11\nu)+4cz^{2}(6\nu^2-2+5\nu)))]
\end{align}
\end{small}where 
\begin{equation}\label{eqA44}
\mathbb{A}=\dfrac{3l^2\nu^2+16\nu\alpha-4\alpha}{l^2\nu^2-4\alpha}.
\end{equation}
In figure \ref{phizplote11}, the behavior of imaginary and real part of $\phi(z)$ in terms of $z$ for different values of parameters has been shown. As can be seen, the imaginary part of scalar field inside the black hole is zero and outside the black hole the scalar field is unstable, and by increasing $\alpha$ instability increase. 
\noindent In equation (\ref{eqqqf211}) the exact coupling function $f_{2}$ and approximatly to first order in $\alpha$ in equation \eqref{eqqqf212} has been obtained. In figure \eqref{f2zplote}, $f_{2}$ for positive/negative $c$ and for different values of parameters has been plotted. The important feature of the figures is that for the negative $c$, $f_{2}$ goes to the negative values by increasing $z$, while it does not become negative anywhere for the positive $c$.
\begin{small}
\begin{align}\label{eqqqf211}
f_{2}&=\dfrac{(1-\nu)z^{-\frac{4}{\nu}}}{\pi^{\frac{3}{2}}\nu^{3}q^{2}(l^2\nu^2-4\alpha)\text{erf}\left(\frac{\sqrt{-2c\mathbb{A}}z_{h}}{2}\right)^{2}}[\pi z\alpha\sqrt{\mathbb{A}}e^{\frac{c\mathbb{A}z^{2}}{2}}\left[\text{erf}\left(\frac{\sqrt{-2c\mathbb{A}}z}{2}\right)-\text{erf}\left(\frac{\sqrt{-2c\mathbb{A}}z_{h}}{2}\right)\right](cz^{2}l^{2}\nu^{4}(cz^{2}-6)\nonumber\\
&+\nu^{3}(32c^{2}z^{4}\alpha-2l^2(cz^{2}+10))+\nu^{2}(-8l^{2}+\alpha(12c^{2}z^{4}+88cz^{2}))+40\alpha\nu(cz^{2}+2)+32\alpha)-4\sqrt{\pi}\alpha cz^{2}\nu(2+\nonumber\\
&\nu cz^{2})(3l^2\nu^2+16\alpha \nu-4\alpha)e^{c\mathbb{A}z^{2}}+(l^2\nu^2-4\alpha)(-2l^2\nu^2\pi z\sqrt{-2c\mathbb{A}}e^{\frac{cz^{2}\nu(l^2\nu+8\alpha)}{l^2\nu^2-4\alpha}}\text{erf}\left(\frac{\sqrt{-2c\mathbb{A}}z_{h}}{2}\right)+\pi^{\frac{3}{2}}\nonumber\\
&\left[\text{erf}\left(\frac{\sqrt{-2c\mathbb{A}}z}{2}\right)-\text{erf}\left(\frac{\sqrt{-2c\mathbb{A}}z_{h}}{2}\right)\right](l^2\nu\text{erf}\left(\frac{\sqrt{-2c\mathbb{A}}z_{h}}{2}\right)e^{-\frac{cz^{2}}{2}}(3\nu cz^{2}+4\nu+4)+[\text{erf}\left(\frac{\sqrt{-2c\mathbb{A}}z}{2}\right)-\nonumber\\
&\text{erf}\left(\frac{\sqrt{-2c\mathbb{A}}z_{h}}{2}\right)]\alpha(16+8cz^{2}+\nu^{2}(z^{6}c^{3}-2c^{2}z^{4})+\nu(16+6c^{2}z^{4}+12cz^{2}))],
\end{align}
\end{small}In the case of $\alpha\ll 1$ and $c<0$ one can get
\begin{small}
\begin{align}\label{eqqqf212}
&f_{2}\approx \dfrac{l^2(\nu-1)e^{-\frac{cz^{2}}{2}}\left[2\sqrt{-6c}z\nu e^{\frac{3cz^{2}}{2}}-\pi^{\frac{1}{2}}\left(4+4\nu+3\nu cz^{2}\right)\left(\text{erf}\left(\frac{\sqrt{-6c}z}{2}\right)-\text{erf}\left(\frac{\sqrt{-6c}z_{h}}{2}\right)\right)\right]}{\nu^2 q^{2}z^{\frac{4}{\nu}}q^2\pi^{\frac{1}{2}}\text{erf}\left(\dfrac{\sqrt{-6c}z_{h}}{2}\right)}\nonumber\\
&-\dfrac{(\nu-1)\alpha}{\nu^{4}q^{2}\pi^{3}z^{\frac{4}{\nu}}\text{erf}\left(\frac{\sqrt{-6c}z_{h}}{2}\right)^{2}}[-12z^{2}\pi^{2}c\nu^{2}(2+\nu cz^{2})e^{3cz^{2}}-4(2\nu+1)ze^{cz^{2}}(4\pi^2 c\nu z_{h}e^{\frac{3cz_{h}}{2}}+\pi^{\frac{5}{2}}\nonumber\\
&\sqrt{-6c}\text{erf}\left(\frac{\sqrt{-6c}z_{h}}{2}\right)\left(\nu c z^{2}-\frac{2}{3}\nu-\frac{4}{3}\right)+\pi^{\frac{5}{2}}\sqrt{-6c}\nu z)e^{\frac{3cz^{2}}{2}}(c^{2}z^{4}\nu^{2}-6c\nu^2 z^{2}-2\nu cz^{2}-20\nu-8)\nonumber\\
&\left(\text{erf}\left(\frac{\sqrt{-6c}z}{2}\right)-\text{erf}\left(\frac{\sqrt{-6c}z_{h}}{2}\right)\right)-4(2\nu+1)\sqrt{-6c}\pi^{\frac{5}{2}}z_{h}(\frac{4}{3}+\frac{4}{3}\nu+\nu cz^{2})\text{erf}\left(\frac{\sqrt{-6c}z}{2}\right)e^{-\frac{cz^{2}}{2}}e^{\frac{3cz^{2}}{2}}\nonumber\\
&+\nu \pi^{3}\left(\text{erf}\left(\frac{\sqrt{-6c}z}{2}\right)-\text{erf}\left(\frac{\sqrt{-6c}z_{h}}{2}\right)\right)^{2}(16+8cz^{2}+\nu(16+16c^{2}z^{4}+12cz^{2})+\nu^{2}(z^{6}c^{3}-2c^{2}z^{4}))]\nonumber\\
&+\mathcal{O}(\alpha^{2}).
\end{align}
\end{small}
\noindent From equation  $E_{tt}$ we get the expression for the scalar potential $V$ as a function of $z$ as follows:
\begin{small}
\begin{align}
V&=\dfrac{e^{cz^{2}}}{4\nu^3l^{4}\pi^{\frac{3}{2}}(l^2\nu^2-4\alpha)\text{erf}\left(\frac{\sqrt{-2c\mathbb{A}}z_{h}}{2}\right)^{2}}[-2\alpha \pi ze^{\frac{c\mathbb{A}z^{2}}{2}}\sqrt{-2c\mathbb{A}}\left[\text{erf}\left(\frac{\sqrt{-2c\mathbb{A}}z}{2}\right)-\text{erf}\left(\frac{\sqrt{-2c\mathbb{A}}z_{h}}{2}\right)\right]\nonumber\\
&[l^2z^{2}c\nu^{5}(5cz^{2}+3c^{2}z^{4}+6)+\nu^{4}(-32c^{2}z^{4}\alpha+l^{2}(20cz^{2}+20+c^{2}z^{4}))+\nu^{3}(4l^2+10l^2cz^{2}-\alpha(88cz^2\nonumber\\
&+4c^2z^4+12c^3z^6))-\nu^2(8l^2+\alpha(80+36c^{2}z^{4}+48cz^{2}))-\nu\alpha
(8cz^{2}+48)+32\alpha]+\nu[8c\alpha z^{2}\sqrt{\pi}(1\nonumber\\
&-\nu)(2+\nu c z^{2})(3l^2\nu^2+16\alpha \nu-4\alpha)e^{c\mathbb{A}z^{2}}+(l^2\nu^2-4\alpha)[-2\pi \nu zl^2\sqrt{-2c\mathbb{A}}(3\nu cz^{2}+4\nu+2)\text{erf}\left(\frac{\sqrt{-2c\mathbb{A}}z_{h}}{2}\right)\nonumber\\
&e^{\frac{c\nu z^{2}(l^2\nu+8\alpha)}{l^2\nu^2-4\alpha}}+\left[\text{erf}\left(\frac{\sqrt{-2c\mathbb{A}}z}{2}\right)-\text{erf}\left(\frac{\sqrt{-2c\mathbb{A}}z_{h}}{2}\right)\right][l^2\text{erf}\left(\frac{\sqrt{-2c\mathbb{A}}z_{h}}{2}\right)e^{-\frac{cz^{2}}{2}}(8+\nu^2(9c^2z^4+12cz^{2}+16)\nonumber\\
&+\nu(6cz^{2}+8))+(2+cz^{2})\alpha\left[\text{erf}\left(\frac{\sqrt{-2c\mathbb{A}}z}{2}\right)-\text{erf}\left(\frac{\sqrt{-2c\mathbb{A}}z_{h}}{2}\right)\right](\nu^2(3z^{6}c^{3}-2c^{2}z^{4})+\nu(20cz^{2}+16+\nonumber\\
&14c^{2}z^{4}+32+16cz^{2}))]]]].
\end{align}
\end{small}
\begin{figure}[H]\hspace{0.4cm}
\centering
\subfigure[$\nu=4.5,\alpha=0.1,l=1,z_{h}=2$]{\includegraphics[width=0.4\columnwidth]{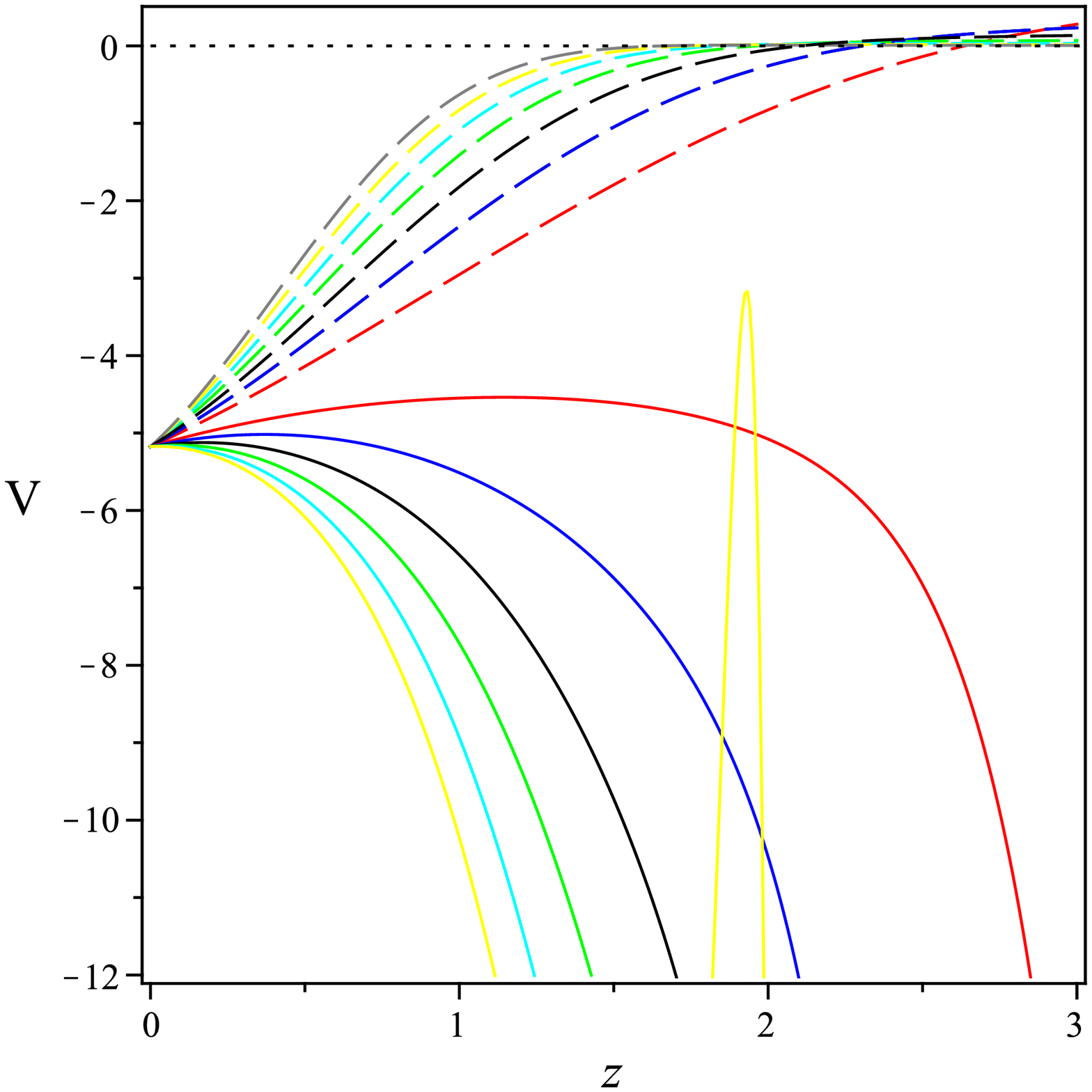}}
\subfigure[$\alpha=0.1,l=1,z_{h}=2$]{\includegraphics[width=0.4\columnwidth]{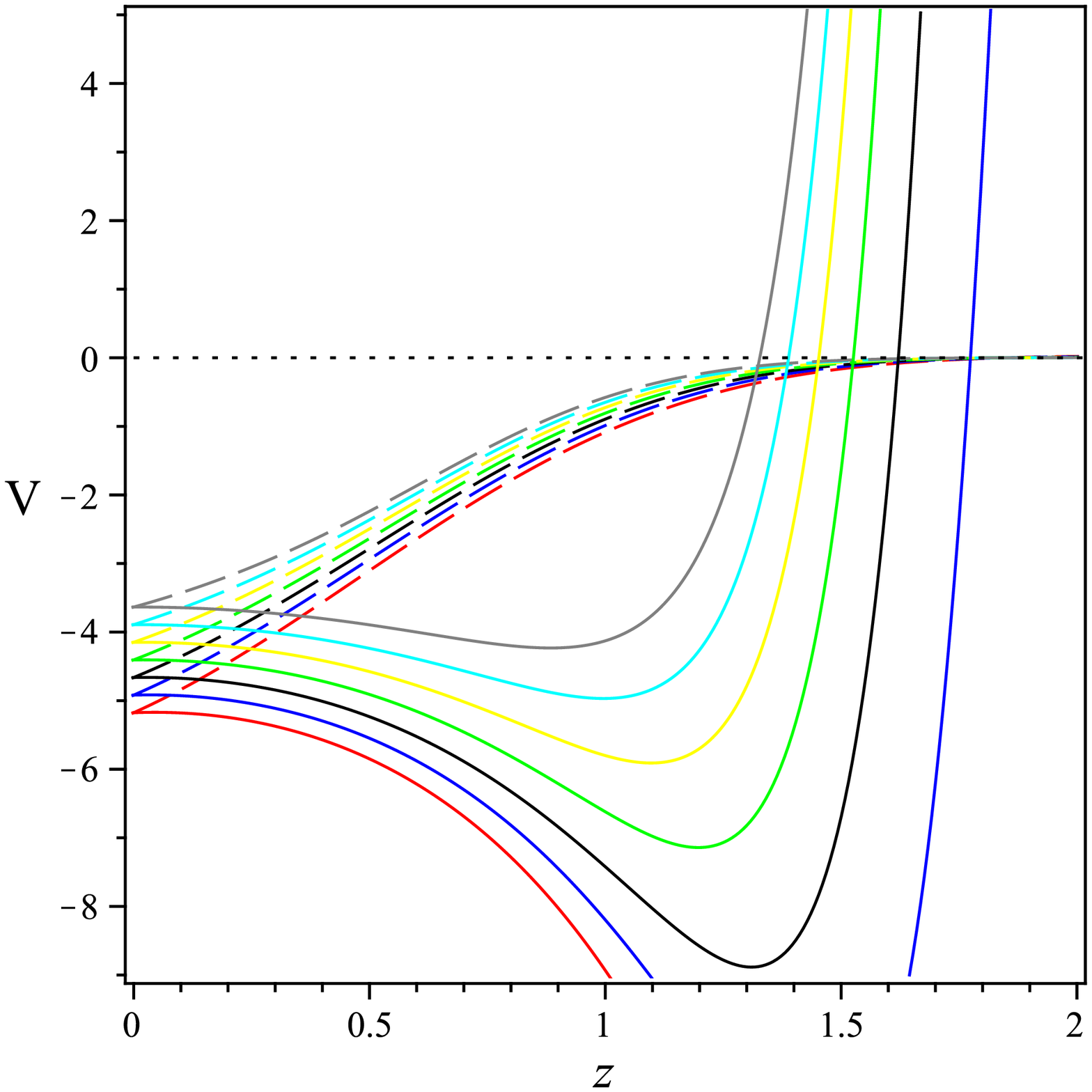}}
\caption{Plots of $V(z)$ in terms of $z$ for $c=\textcolor{red}{0.1},\textcolor{blue}{0.2},0.3,\textcolor{green}{0.4},\textcolor{yellow}{0.5}$ (dashed lines) and $c=\textcolor{red}{-0.1},\textcolor{blue}{-0.2}, -0.3,\textcolor{green}{-0.4},\textcolor{yellow}{-0.5}$ (solid lines) (left), $\alpha=\textcolor{red}{0.1},\textcolor{blue}{0.2},0.3,\textcolor{green}{0.4},\textcolor{yellow}{0.5}$, $c=0.5$ (dashed lines) and $c=-0.5$ (solid lines) (right).} 
\label{Vzplote}
\end{figure}
\noindent{Regardless of the sign of $c$, $V(z)$ is negative under the horizon. In panel a, with the increase of $\vert c\vert$, $\vert V\vert$ decreases. In panel b, as $\alpha$ increases, $\vert V\vert$ increases.}

\subsubsection{Thermodynamics of background}
\begin{figure}[H]\hspace{0.4cm}
\centering
\subfigure[$\nu=4.5,l=1,c=-0.5$]{\includegraphics[width=0.4\columnwidth]{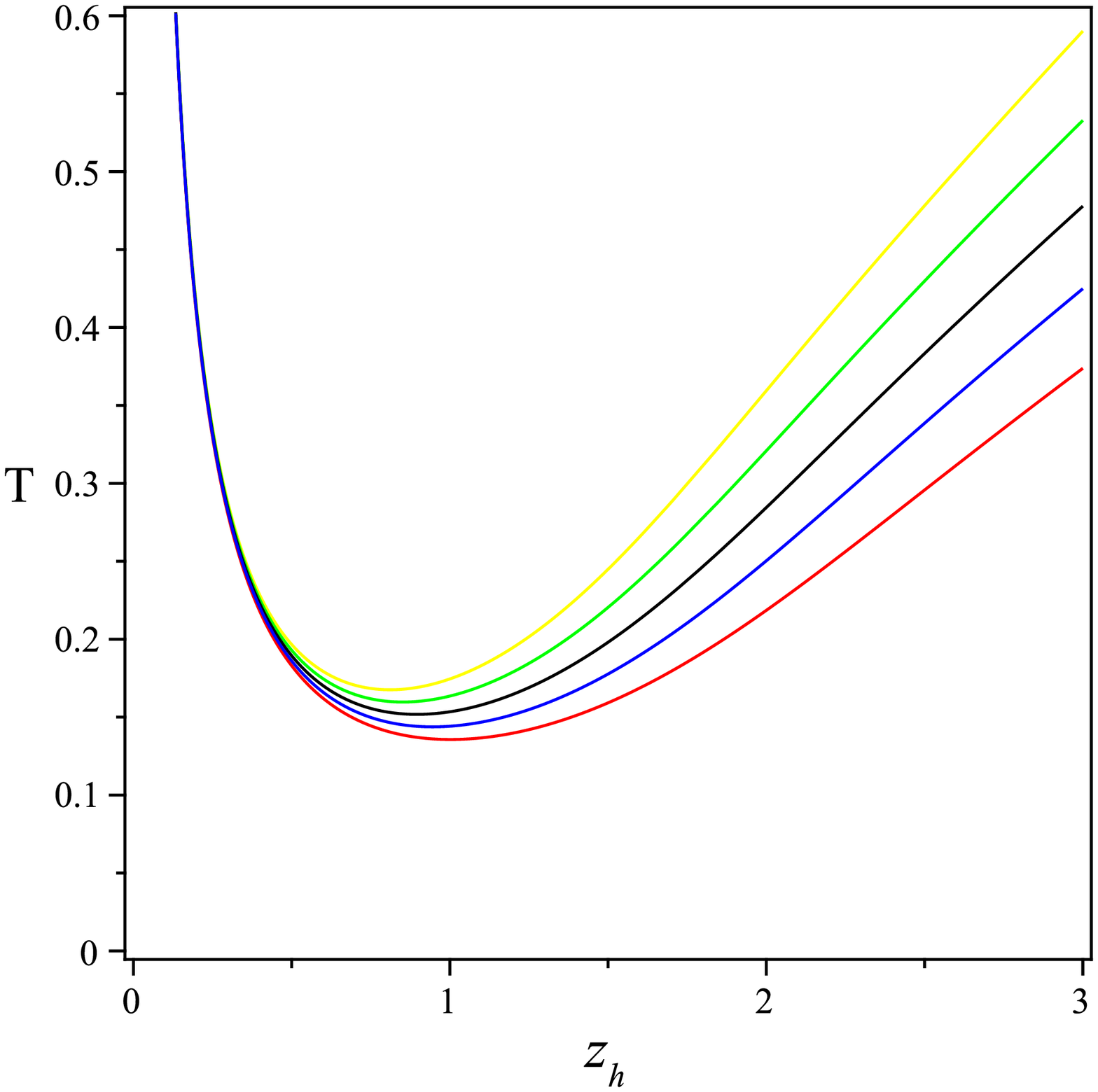}}
\subfigure[$\nu=4.5,l=1,c=-0.5$]{\includegraphics[width=0.4\columnwidth]{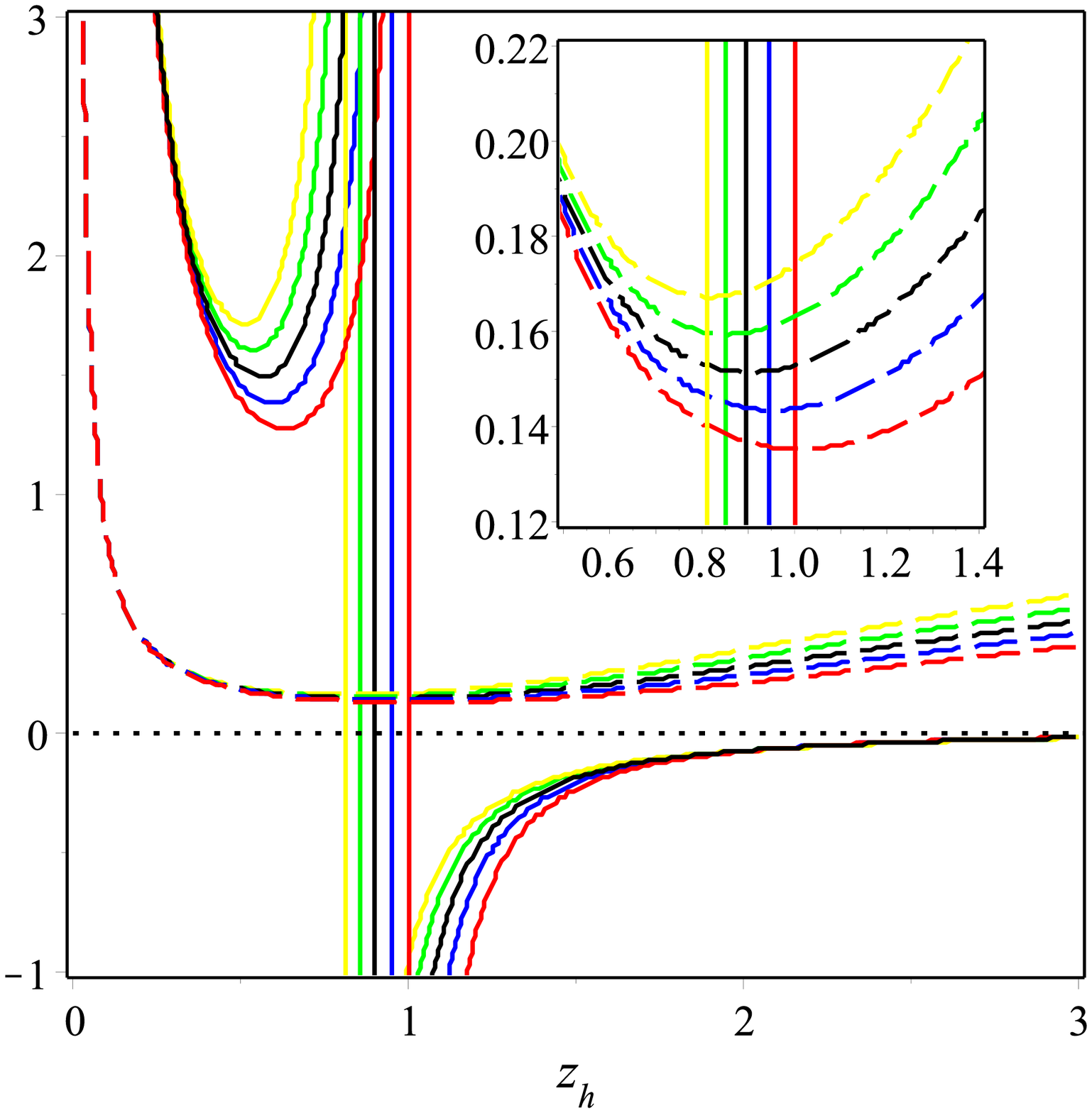}}
\caption{Plots of $T$ in terms of $z_{h}$ for $\alpha=\textcolor{red}{0.1},\textcolor{blue}{0.2},0.3,\textcolor{green}{0.4},\textcolor{orange}{0.5}$ and $\nu=4.5$ (left). Plots of $C_{V}$ and $T$ in terms of $z_{h}$ for $\alpha=\textcolor{red}{0.1},\textcolor{blue}{0.2},0.3,\textcolor{green}{0.4},\textcolor{orange}{0.5}$ (right)} 
\label{Tplotc}
\end{figure}

\begin{figure}[H]\hspace{0.4cm}
\centering
\subfigure[$l=1,c=-0.5,\alpha=0.1$]{\includegraphics[width=0.3\columnwidth]{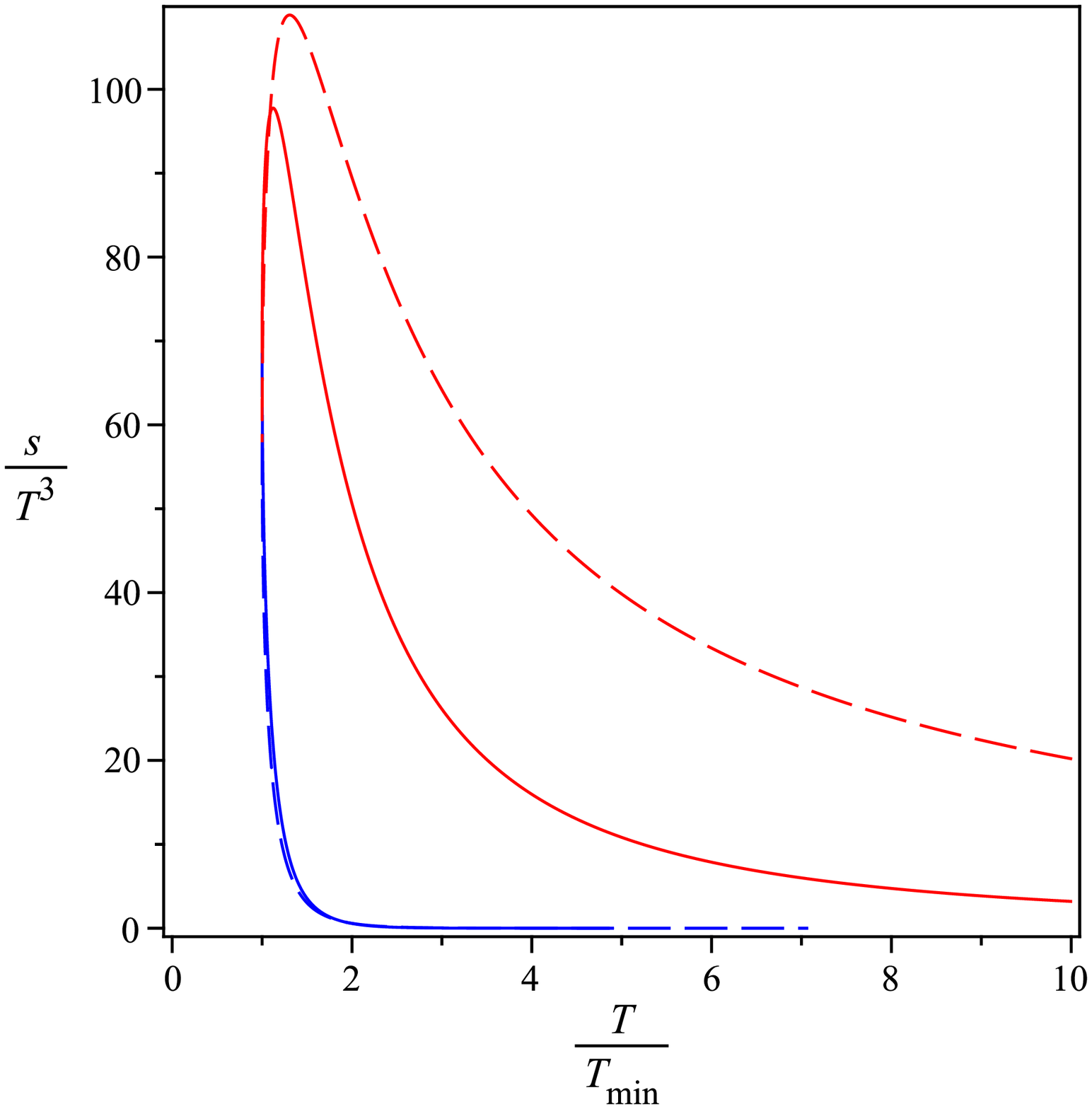}}
\subfigure[$l=1,c=-0.5,\alpha=0.1$]{\includegraphics[width=0.3\columnwidth]{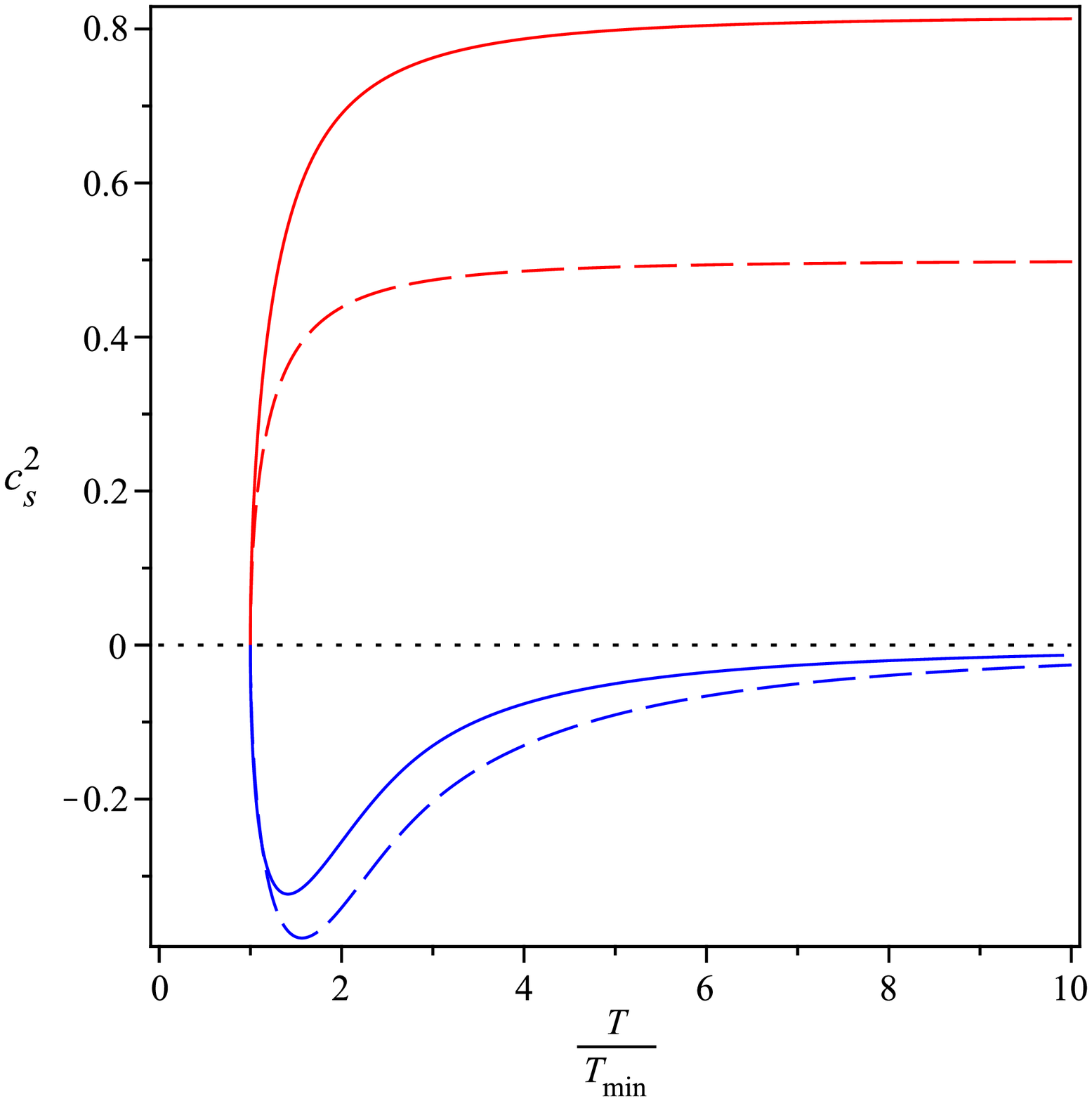}}
\subfigure[$l=1,c=-0.5,\alpha=0.1$]{\includegraphics[width=0.3\columnwidth]{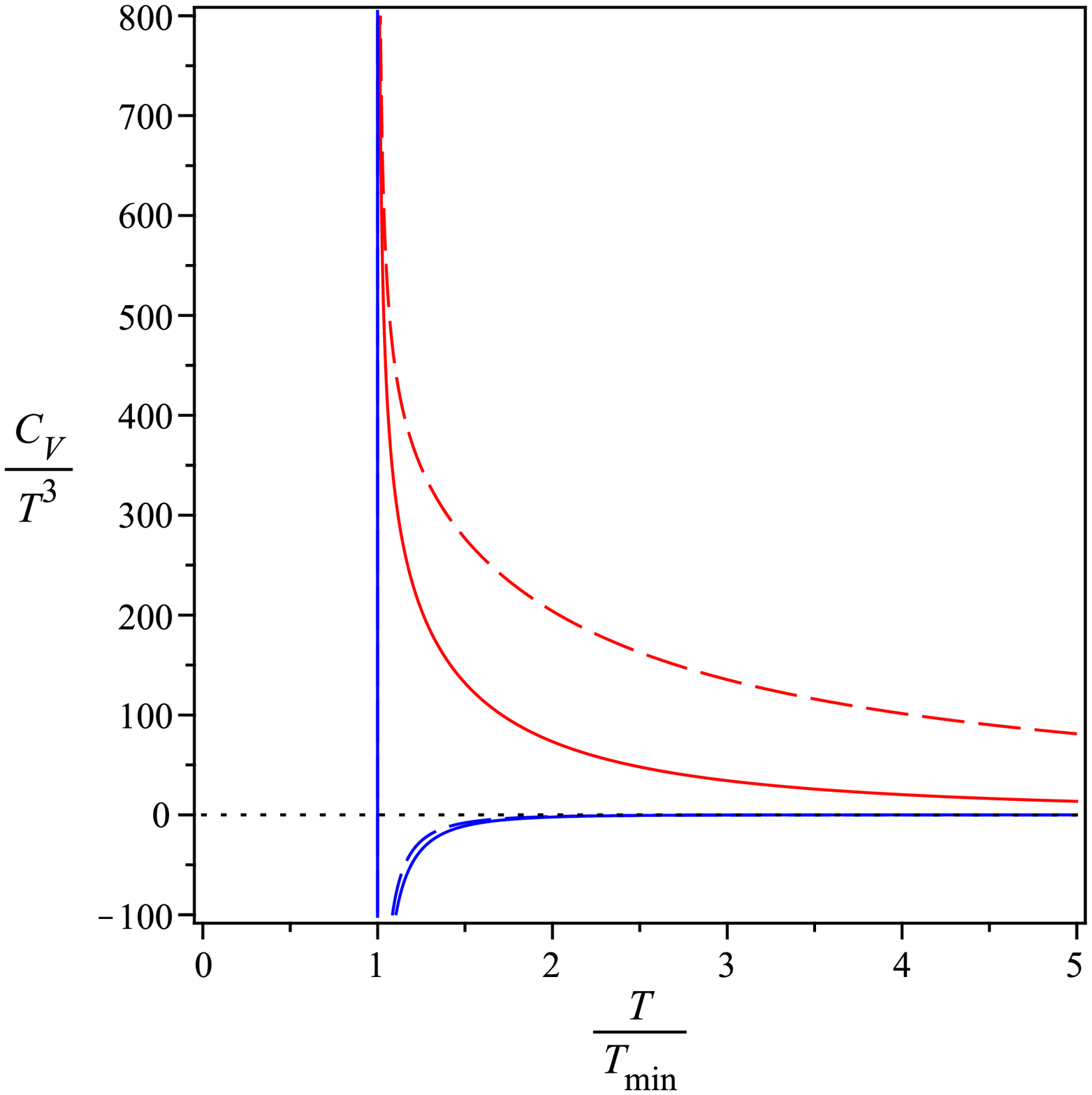}}
\caption{Plots of  scaled entropy density $s/T^3\;(\text{left}), c_{s}^{2}\;(\text{middle})$ and $C_{V}/T^3\;(\text{right})$ in terms of scaled temperature $T/T_{min}$ for $\nu=1,T_{min}=0.19457\;(\text{dashed lines}),\nu=4.5,T_{min}=0.1357\;(\text{solid lines})$. In each panel red lines correspond to the large stable solution and blue lines correspond to unstable small solution.} 
\label{st3plot}
\end{figure}

\begin{figure}[H]\hspace{0.4cm}
\centering
\subfigure[$l=1,c=-0.5,\alpha=0.1$]{\includegraphics[width=0.45\columnwidth]{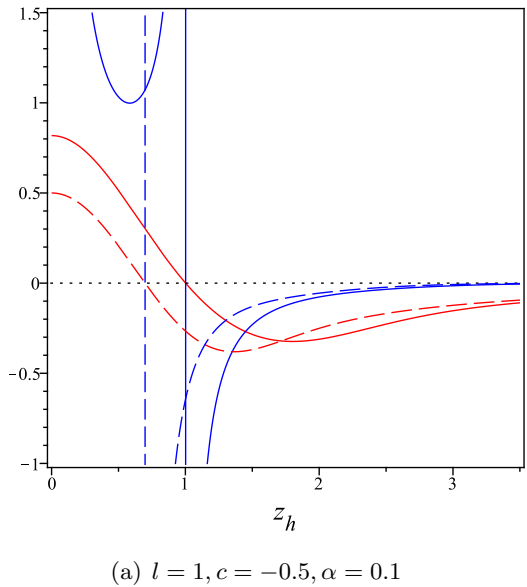}}
\caption{Plots of $c_{s}^{2}$ (\textcolor{red}{red lines}) and $C_{V}$ (\textcolor{blue}{blue lines}) in terms of $z_{h}$ for $\nu=1\;(\text{dashed lines}),4.5\;(\text{solid lines})$.} 
\label{cscvplot}
\end{figure}

The Hawking temperature can be obtained by using the surface gravity interpretation
\begin{equation}
T=\dfrac{\kappa}{2\pi}=\left\vert\dfrac{g^{\prime}}{4\pi}\right\vert=\dfrac{e^{\frac{c\mathbb{A}z_{h}^{2}}{2}}\sqrt{-2c\mathbb{A}}}{4\pi^{\frac{3}{2}}\text{erf}\left(\dfrac{z_{h}}{2}\sqrt{-2c\mathbb{A}}\right)},
\end{equation}
where $\mathbb{A}$ provided in (\ref{eqA44}).
In the case of small $\alpha$ and $c<0$, one can get
\begin{small}
\begin{align}
T\approx \dfrac{\sqrt{-6c}e^{\frac{3c z_{h}^{2}}{2}}}{4\pi^{\frac{3}{2}}\text{erf}\left(\frac{\sqrt{-6c}z_{h}}{2}\right)}+\dfrac{(2\nu+1)e^{\frac{3cz_{h}^{2}}{2}}\alpha[\pi\sqrt{-6c}(cz_{h}^2+\frac{1}{3})\text{erf}\left(\frac{\sqrt{-6c}z_{h}}{2}\right)-2\sqrt{\pi}cz_{h}e^{\frac{3cz_{h}^2}{2}}]}{\pi^{\frac{5}{2}}l^2\nu^2\text{erf}\left(\frac{\sqrt{-6c}z_{h}}{2}\right)^2}+\mathcal{O}\left(\alpha^2\right).
\end{align}
\end{small}The behavior of temperature is shown in Figs. (\ref{Tplotc}). In Fig. (\ref{Tplotc})a, the variation of temperature with respect to the horizon radius $z_{h}$ for different values of the coupling of theory $\alpha$ is shown. As can be seen there exists a minimum temperature $T_{min}$ below which no black hole solution exist (thermal gas). However, for $T>T_{min}$, there are two black hole solutions, a large and a small one (deconfined quark gluon plasma phase). The small black hole phase for which $T$ increases with $z_{h}$ whereas the large black hole phase for which $T$ decreases with $z_{h}$. In Fig. (\ref{Tplotc})b, to study the stability of the solutions, we have shown the behavior of heat capacity $C_{V}$ and temperature. As can be seen the large black hole has positive heat capacity and therefore is stable and small black hole is unstable and thus not physical. 
\noindent Following the standard Bekenstein-Hawking formula \eqref{eqqen28}, one can easily read the black hole entropy density $s$, which is defined as
\begin{equation}
s=\dfrac{l^3e^{-\frac{3cz_{h}^{2}}{4}}}{4z_{h}^{\frac{\nu+1}{\nu}}}.
\end{equation}
The scaled entropy density $s/T^{3}$ as a function of scaled temperature $T/T_{min}$ is shown in Fig.\ref{st3plot}a. The red lines correspond to the large stable solution and blue lines are for the small unstable solution. The numerical result of the square of the sound velocity is shown in \ref{st3plot}b. At $T_{min}$, the sound velocity square is around $0$ which is in agreement with lattice data $0.05$. At high temperature, the sound velocity square goes to $0.45$ for $\nu=1$, which means that the system is approximatly asymptotically conformal, while for $\nu=4.5$, the sound velocity square goes to $0.8$. The numerical result of the specific heat is shown in Fig. \ref{st3plot}c. It can be clearly seen that the specific heat $C_{V}$ diverges at $T_{min}$. At $T\to \infty$, the scaled specific
heat $C_{V}/T^{3}$ approaches to the zero for this approximate solution \cite{Li:2011hp}.
\noindent In figure \ref{cscvplot}, the behavior of $c_{s}^{2}$ and $C_{V}$ have been shown. As can be seen from the figure and obtained the result in the previous section, it can be concluded that the heat capacity has an inverse relationship with the speed of sound. 

\subsection{The case $c\neq0, \mu\neq0$}\label{sect2.3}
\begin{figure}[H]\hspace{0.4cm}
\centering
\subfigure[$\nu=4.5,l=1,\alpha=0.1,c=\pm0.5$]{\includegraphics[width=0.4\columnwidth]{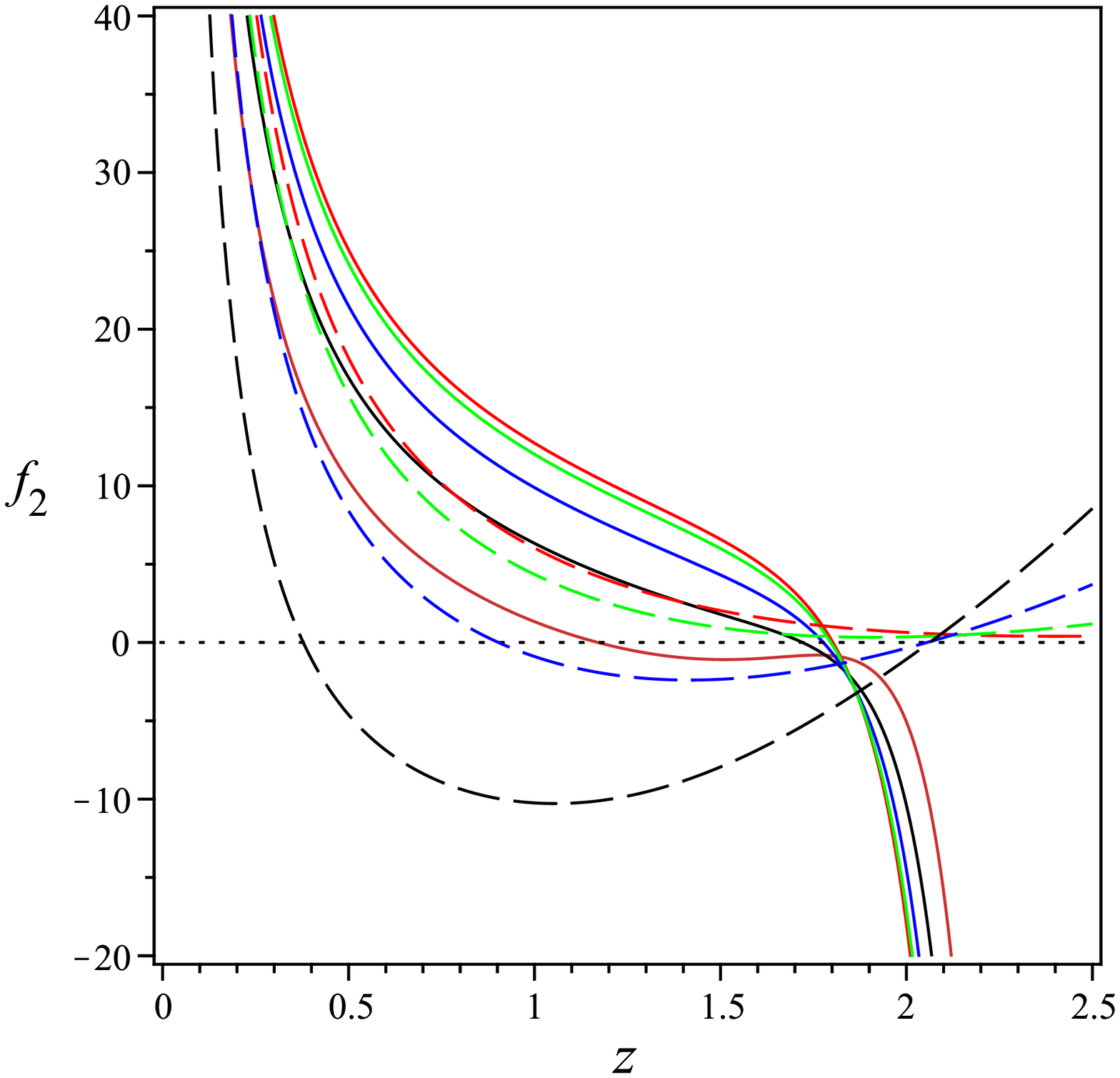}}
\subfigure[$\nu=4.5,l=1,\alpha=0.1,c=\pm0.5$]{\includegraphics[width=0.4\columnwidth]{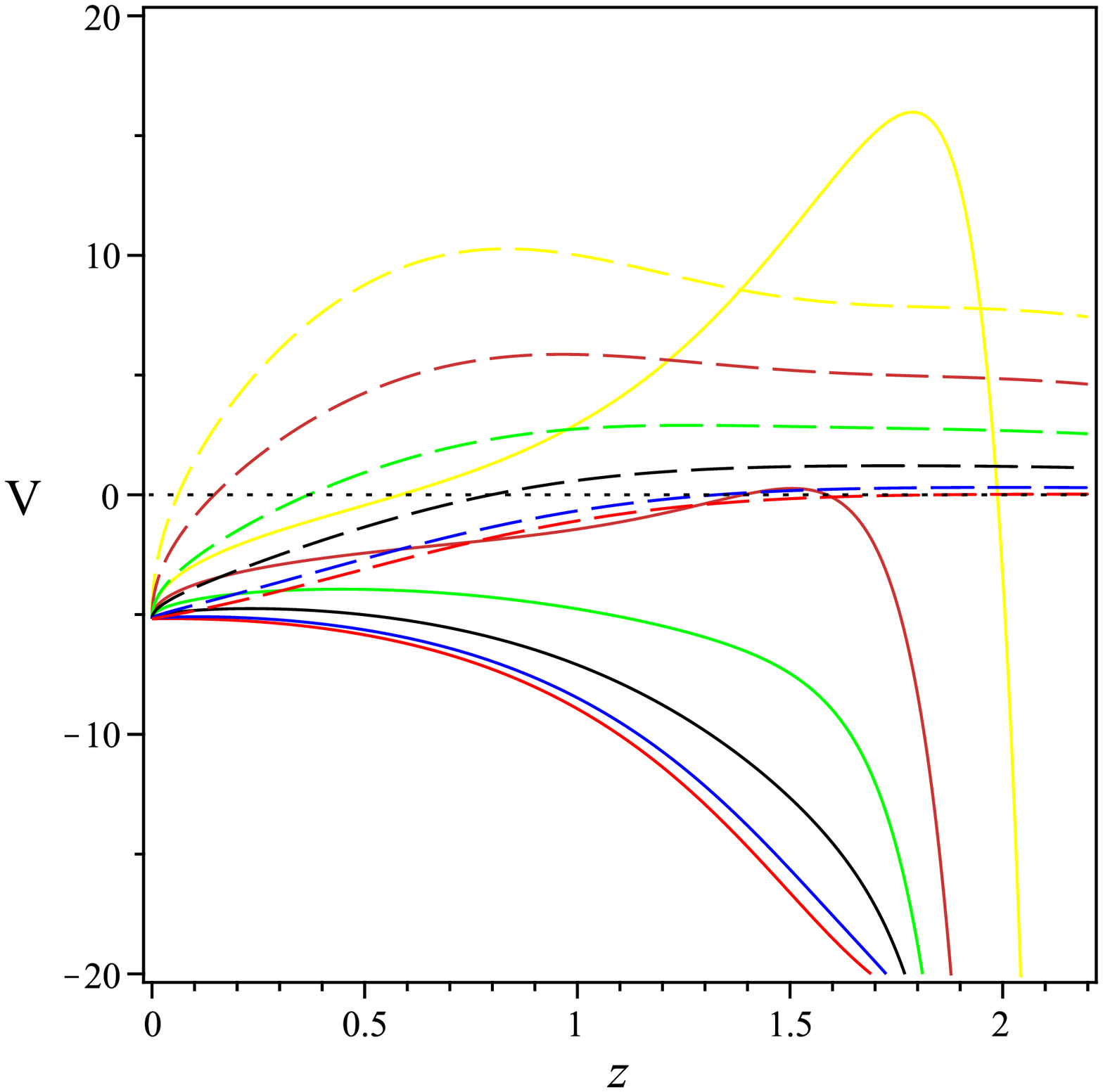}}
\caption{{Plots of $f_{2}$ and $V$ in terms of $z$ for $\mu=\textcolor{red}{0},\textcolor{blue}{0.05},\textcolor{black}{0.1},\textcolor{green}{0.15},\textcolor{orange}{0.2},\textcolor{yellow}{0.25}$}.} 
\label{fvplotmu}
\end{figure}

In this case the differential equation \eqref{eqqtotal}, becomes
\begin{align}
&4\nu lz^{2}e^{-\frac{cz^{2}}{4}}(l^{2}\nu^{2}e^{-\frac{cz^{2}}{2}}-\alpha(2+\nu cz^{2})^{2}g(z))g^{\prime\prime}-2zle^{-\frac{cz^{2}}{4}}(\nu^{2}l^{2}(4+(2+3z^{2})\nu)-\nonumber\\
&\alpha(2+\nu c z^{2})(8+4\nu+6\nu c z^{2}-6c\nu^{2}z^{2}+c^{2}\nu^{2}z^{4})g)g^{\prime}-4\alpha \nu z^{2}le^{-\frac{cz^{2}}{4}}(2+\nu cz^{2})^2g^{\prime 2}\nonumber\\
&-c^{2}\nu^{3}l \mathfrak{c}_{2}^{2}z^{\frac{4\nu+2}{\nu}}e^{\frac{cz^{2}}{4}}=0
\end{align}
in order to solve the above differential equation, we assume
\begin{equation}
g(z)=1+\epsilon g_{1}(z)+\mathcal{O}(\epsilon^{2}),
\end{equation}
by inserting it, one can achieve a non-homogeneous differential equation as
\begin{equation}\label{eqq36}
g_{1}^{\prime\prime}+\dfrac{[-l^{2}\nu^{3}(2+3cz^{2})-4\nu^{2}(l^{2}+4\alpha cz^{2})+4\alpha \nu (2+cz^{2})+16\alpha]g_{1}^{\prime}}{\nu z(l^{2}\nu^{2}-4\alpha)}+\dfrac{c\nu \mathfrak{c}_{2}^{2}z^{\frac{2}{\nu}}e^{\frac{cz^{2}}{2}}}{4\alpha (4+c\nu z^{2})}=0.
\end{equation}
In order to solve the non-homogeneous differential equation \eqref{eqq36} in the case of $c<0, \mathfrak{c}_{2}\neq0$, we consider the following particular solution:
\begin{equation}
g_{2}(z)=e^{\frac{cz^{2}}{2}}z^{-\frac{2(\nu-1)}{\nu}}\sum_{i}h_{i}z^{i},
\end{equation}
where $h_{i}$ are coefficients of expansions. By inserting the above solution into the differential equation \eqref{eqq36}, and solving order by order for coefficients, one can get
\begin{align}\label{eqqfi52}
&\hspace{5cm}h_{0}=h_{1}=h_{3}=h_{5}=0,\nonumber\\
h_{2}&=\dfrac{\mathfrak{c}_{2}^{2}\nu^2}{8\alpha (\nu-2)},\;\;\;h_{4}=-\dfrac{c\nu^{3}\mathfrak{c}_{2}^{2}(-20\alpha\nu+ l^{2}\nu^{3}+l^{2}\nu^{2}-12\alpha)}{16\alpha(\nu-2)(\nu^{4}l^2+3l^{2}\nu^{3}-4\alpha \nu^{2}+2l^{2}\nu^{2}-12\alpha \nu-8\alpha)},\nonumber\\
h_{6}&=\dfrac{\nu^{4}c^{2}\mathfrak{c}_{2}^{2}}{32\alpha (2+3\nu)(l^2\nu^2-4\alpha)^{2}(1+2\nu)(1+\nu)(\nu^2-4)}(3l^{4}\nu^{6}+6l^{5}\nu^{5}-40\nu^{4}l^{2}\alpha+3\nu^{4}l^{4}+256\alpha^{2}\nu^{3}
\nonumber\\
&-56\nu^{3}l^{2}\alpha+624\nu^{2}\alpha^{2}-24\nu^{2}l^{2}\alpha+448\alpha^{2}
\nu+112\alpha^{2}),
\end{align} 
where $\mathfrak{c}_{2}$ used from \eqref{eqqcond10}.
Therefore, the metric function by considering a particular solution becomes
\begin{equation}
g(z)=1-\dfrac{\text{erf}\left(\dfrac{1}{2}\sqrt{\dfrac{c(-6l^{2}\nu^{2}-32\nu\alpha+8\alpha)}{(l^{2}\nu^{2}-4\alpha)}}z\right)}{\text{erf}\left(\dfrac{1}{2}\sqrt{\dfrac{c(-6l^{2}\nu^{2}-32\nu\alpha+8\alpha)}{(l^{2}\nu^{2}-4\alpha)}}z_{h}\right)}+e^{\frac{cz^{2}}{2}}z^{-\frac{2(\nu-1)}{\nu}}\sum_{i}h_{i}z^{i},
\end{equation}
with $h_{i}$ are provided in \eqref{eqqfi52}. In figure \ref{fvplotmu}, the effect of $\mu$ on the behavior of $f_{2}$ and $V$ have been shown. As can be seen in left panel by increasing chemical potential, $f_{2}$ decreases and in right panel $V$ increases.

\begin{figure}[H]\hspace{0.4cm}
\centering
\subfigure[$\nu=4.5,l=1,c=-0.5,\alpha=0.1$]{\includegraphics[width=0.3\columnwidth]{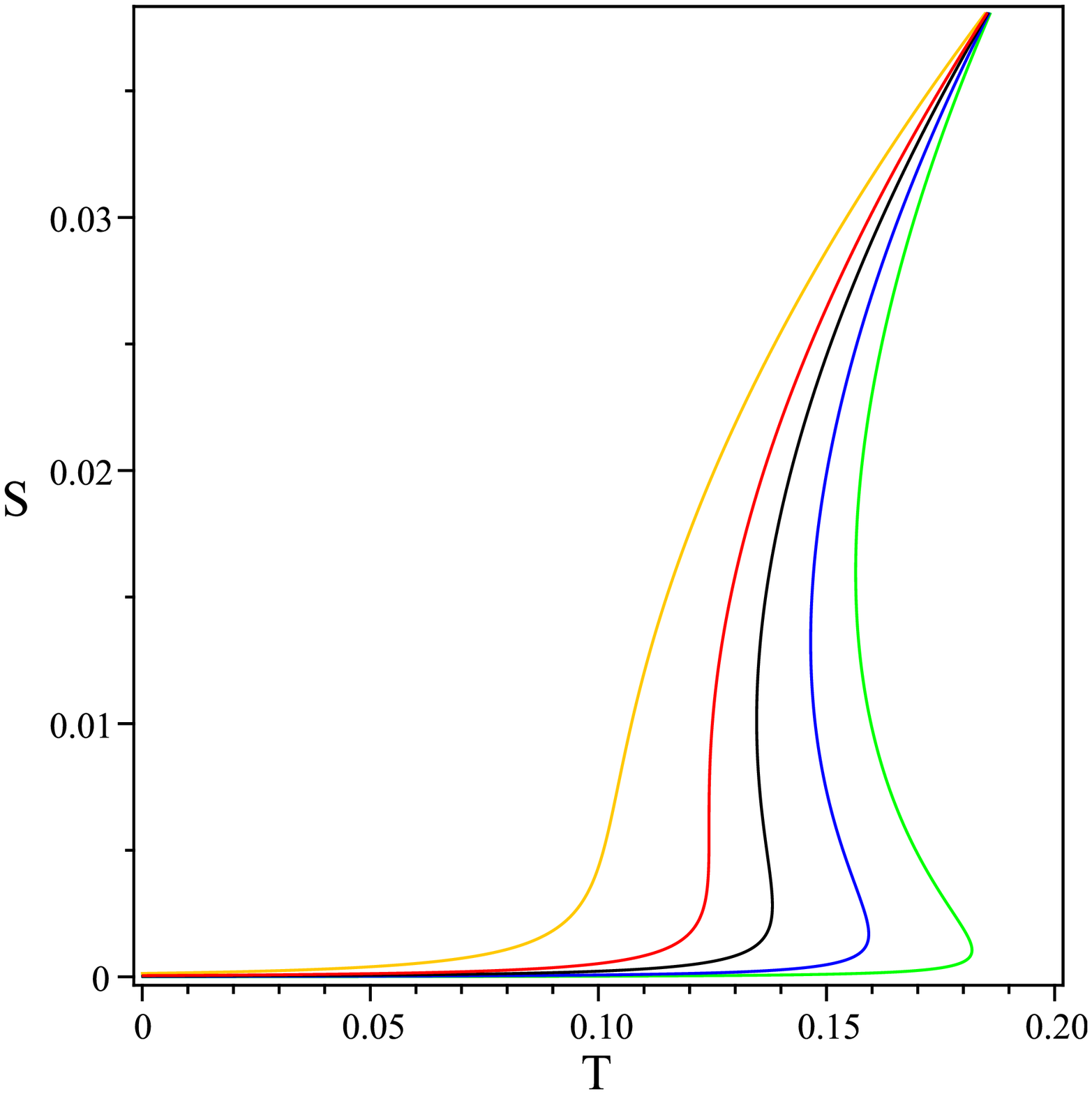}}
\subfigure[$\nu=4.5,l=1,c=-0.5,\alpha=0.1$]{\includegraphics[width=0.3\columnwidth]{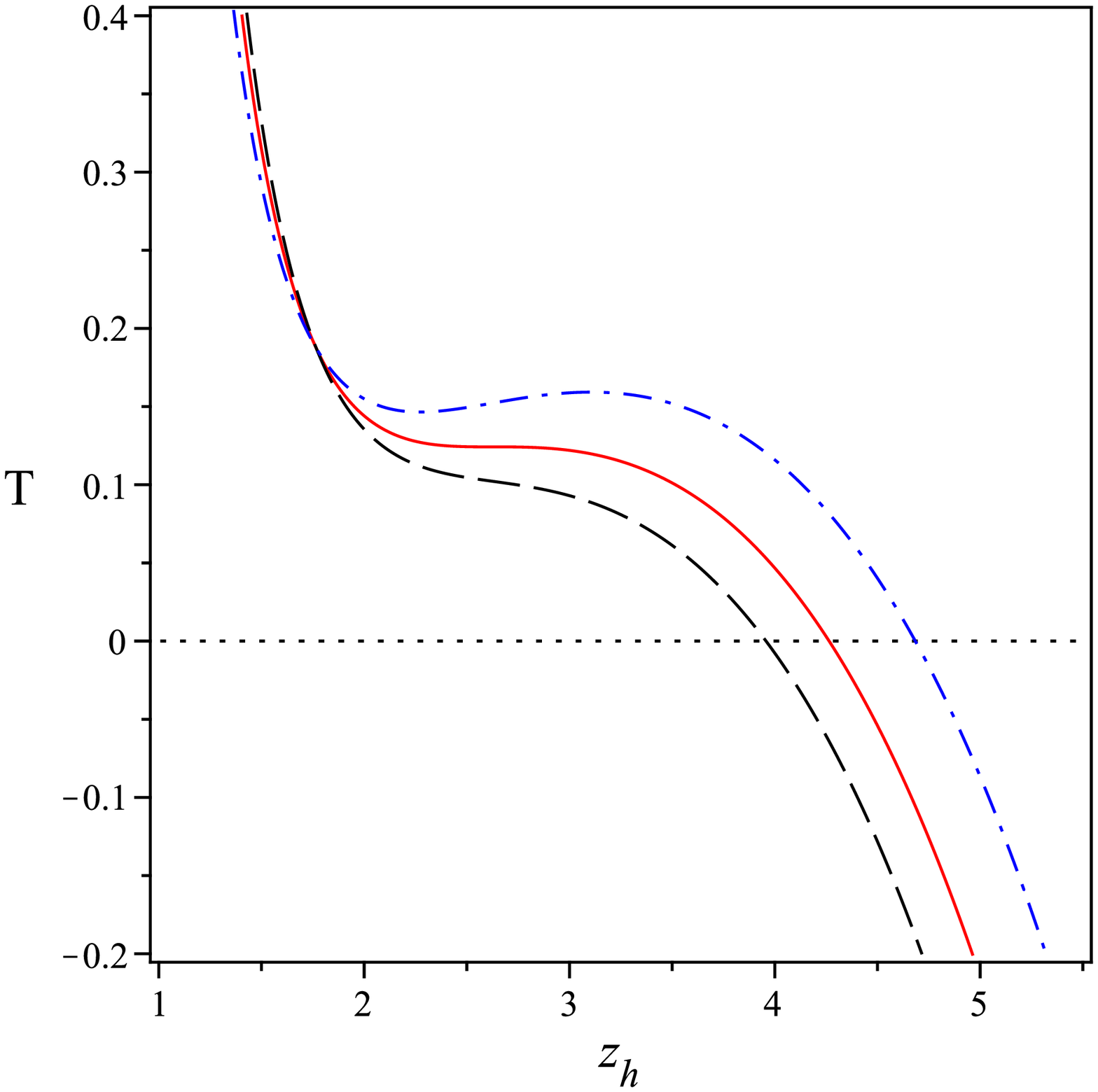}}
\subfigure[$\nu=4.5,l=1,c=-0.5,\alpha=0.1$]{\includegraphics[width=0.3\columnwidth]{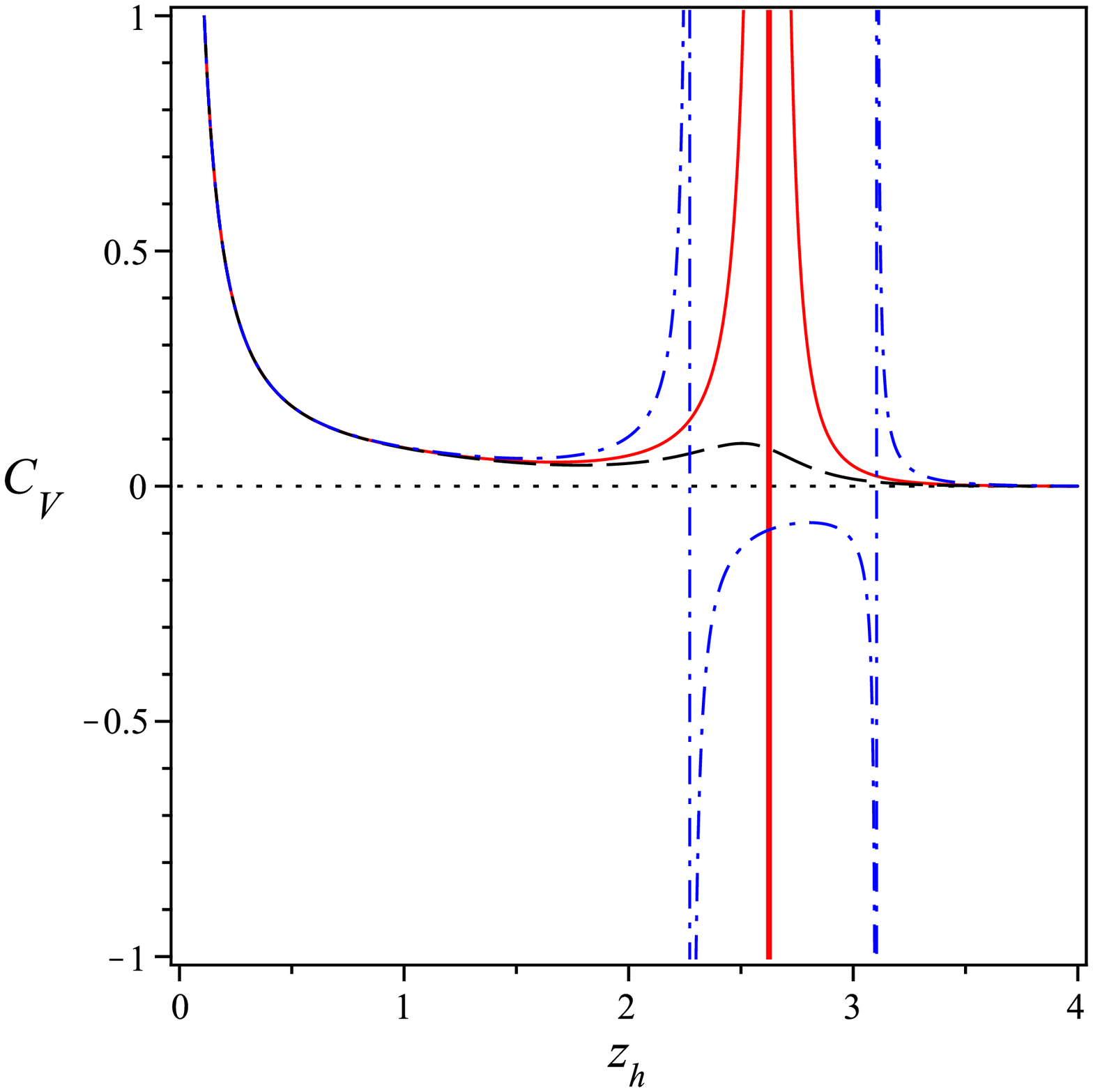}}
\caption{Plots of $s, T$ and $C_{V}$ in terms of $z_{h}$ for $\mu=\textcolor{green}{0.1},\textcolor{blue}{0.15},\textcolor{black}{0.2},\textcolor{red}{0.22728},\textcolor{orange}{0.3}$ (left) and $\mu=\textcolor{blue}{0.2},\textcolor{red}{0.22728},0.3$ (middle and right).} 
\label{stcvplot}
\end{figure}
\begin{figure}[H]\hspace{0.4cm}
\centering
\subfigure[$\nu=4.5,l=1,c=-0.5$]{\includegraphics[width=0.5\columnwidth]{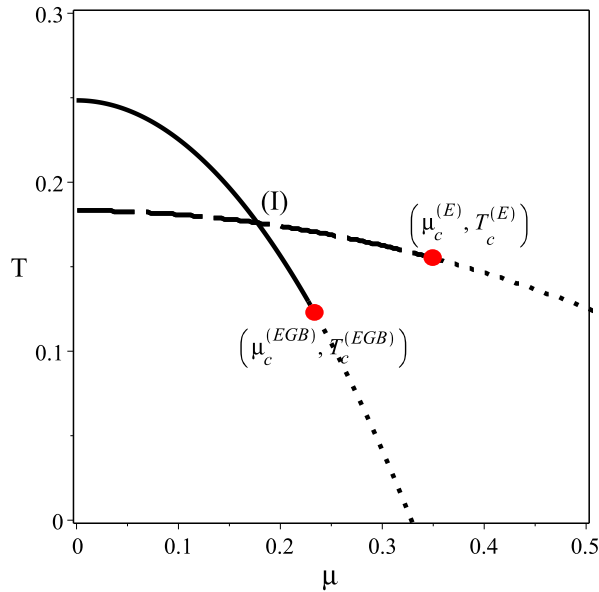}}
\caption{The phase diagram in $T$ and $\mu$ plane for anisotropic background. At small $\mu$, the system undergoes a first order phase transition at finite $T$. The first order phase transition stops at the critical point $(\mu_{c}, T_{c}) \sim (0.22728, 0.1242)$, where the phase transition becomes second order. The solid line is for $\alpha=0.1$ and long dashed line is for $\alpha=0$.} 
\label{TMUplote}
\end{figure}

\noindent In fig. \ref{stcvplot}a, the behavior of entropy is shown in terms of temperature. The figure shows the minimum and maximum temperature. As $\mu$ increases, the minimum temperature decreases. This plot also shows that for $0<\mu<\mu_{c}$, there are minimal $T_{min}$ and maximal $T_{max}$ temperatures, between which the entropy
is a function of $T$ with three branches. When we decrease the temperature, the entropy decreases along the first branch ($T_{min}< T < \infty$). Then the entropy decreases along the second branch with an increase of temperature from $T_{min}$ to $T_{max}$, i.e. here the black holes are unstable. Finally the entropy increases along the third branch with an increase of temperature for $0<T < T_{max}$. Such a behavior in terms of event horizon, one can see in the \ref{stcvplot}b and c. In each panel, the critical point has been shown in red color curve. Upon varying the Hawking temperature, a phase transition from the large black hole phase to the thermal AdS phase takes place at a critical temperature $T_{c}=0.1242$. This is the famous black hole-thermal AdS Hawking-Page phase transition which occurs in the presence of chemical potential $\mu_{c}=0.22728$. {In fig. \ref{TMUplote} by using the approximate analysis, the phase diagram of the holographic QCD model for anisotropic background and for Einstien gravity (long dashed line), Einstien-Gauss-Bonnet gravity (solid line) has been shown. As one can see at $\mu=0$ the system undergoes a black hole to thermal gas first order phase transition so that $T^{(EGB)}(\mu=0)>T^{(E)}(\mu=0)$. For $0<\mu<\mu_{c}$ (in the transition lines), the system undergoes a large black hole to a small black hole first-order phase transition. For $0<\mu<\mu_{I}$, the temperature of the black hole to black hole transition of Einstien gravity ($T^{(E)}(\mu)$) is less than Einstien-Gauss-Bonnet gravity ($T^{(EGB)}(\mu)$) and for $\mu_{I}<\mu<\mu_{c}$ vice versa. The first order phase transition stops at the critical point ($\mu_{c}, T_{c}$), where the phase transition becomes second order, herewith $T^{(EGB)}_{c}<T^{(E)}_{c}$, $\mu_{c}^{(EGB)}<\mu_{c}^{(E)}$. For $\mu>\mu_{c}$, the system has a sharp but smooth crossover.} These thermal AdS and black hole phases in the usual language of gauge-gravity duality are dual to the confinement and deconfinement phases in the dual boundary theory.

\section{Conclusion}\label{conclud}
In this work, we extended the AdS/QCD model to quadratic gravity to gain insight into the influence of gravity on QCD. To do so, we considered an anisotropic black hole metric as a solution to a system of $5D$ Einstein-quadratic-two Maxwell-dilaton fields. The anisotropic background is specified by an arbitrary exponent, a non-zero dilaton field, a non-zero time component of the first Maxwell field, and a non-zero longitudinal magnetic component of the second Maxwell field. The field equations for the considered theory are coupled and bulky differential equations for six unknown functions. Therefore, obtaining the solution to such field equations is too hard, this is why we considered the special case of field equation, i.e. $\gamma=\alpha, \beta=-4\alpha$ (Einstien-Gauss-Bonnet gravity). The differential equation for the metric function in EGB gravity is a nonlinear second-order equation that has been solved in  special cases. At the first, we obtained the exact solutions for the differential equations with zero warp functions. In this case, it doesn't occur any thermodynamical phase transitions to the black brane. The second case that we have considered is the case with zero chemical potential. The blackening function in this case supports the Van der Waals-like phase transition between small and large black holes for suitable values of parameters. The third case that has been considered is nonzero warp function and chemical potential. In this case, in addition to the small/large phase transition, the blackening function supported the phase transition from the large black hole phase to the thermal AdS phase at a critical temperature. Holographically, this phase transition corresponds to the confinement-deconfinement phase transition in QCD. In each case, we investigated the anisotropy influence and the effect of parameters of theory on the thermodynamic properties of our background, in particular, on the small/large black holes phase transition diagram. {In fig. \eqref{TMUplote}, the effect of the Gauss-Bonnet term on the phase transition has been shown. This figure shows that before/after the intersection point for constant chemical potential, the temperature of a black hole in EGB gravity is more/less than that of Einstien gravity. Clearly, before the intersection point, $\alpha$ has a dominant impact on the temperature (compared to the effect of $\mu$ on the temperature). In the isotropic case corresponding to $\nu=1$ (zero magnetic fields) and $, \alpha \to 0$ reproduces previously known results \cite{Arefeva:2018hyo}.}\\
For future work, one can consider the Weyl-squared term by using the combination $\gamma=6\alpha$ and $\beta=-4/3\gamma$ for the parameters of the theory. But since in this case, the field equation for the metric has a 4th-order derivative, the differential equations should be solved numerically. Also, following the paper \cite{Bohra:2019ebj}, one can study the effect of the magnetic field on the system in the framework of EGB.

\section*{Acknowledgements}
I would like to thank the referee for her/his fruitful comments. I also would like to thank the School of Physics of the Institute for Research in Fundamental Sciences (IPM).

\appendix

\section*{Constants}
The constants related to equation\eqref{eqphi18}:
\begin{small}
\begin{align}\label{appeq48}
\mathcal{E}&=(\nu+1)((\nu+1)(\nu^4l^{4}-8\alpha c_{2})-4\alpha \nu c_{1}z^{\frac{2\nu+2}{\nu}}),\nonumber\\
\mathcal{D}&=\nu l^2-\nu^{2}l^{2}-\sqrt{\nu^{4}l^{4}-8\alpha c_{2}},\nonumber\\
\mathcal{F}&=\nu l^2-\nu^{2}l^{2}+\sqrt{\nu^{4}l^{4}-8\alpha c_{2}},\nonumber\\
\mathcal{G}&=\dfrac{\nu^{3}l^{2}-\nu l^{2}-\sqrt{\mathcal{E}}}{\nu+1},\nonumber\\
\mathcal{H}&=-l^{2}\nu(\nu-1)\sqrt{\nu^{4}l^{4}-8\alpha c_{2}}+\nu^{4}l^{4}-8\alpha c_{2},\nonumber\\
\mathcal{K}&=-l^{2}\nu\left(\nu-\frac{1}{2}\right)\sqrt{\mathcal{E}}-2\alpha c_{1}\nu z^{\frac{2\nu+2}{\nu}}+(\nu+1)\left(\nu^{4}l^{4}-4\alpha c_{2}-\dfrac{l^{4}\nu^{3}}{2}\right),\nonumber\\
\mathcal{L}&=l^{2}\nu (\nu-1)\sqrt{\nu^{4}l^{4}-8\alpha c_{2}}+\nu^{4}l^{4}-8\alpha c_{2}.
\end{align}
\end{small}The constants related to equation\eqref{eqf221}:
\begin{small}
\begin{align}\label{appeq49}
\bar{\mathcal{A}}&=(\nu+1)(-4\alpha c_{1}\nu z^{\frac{2\nu+2}{\nu}}+(\nu+1)(\nu^{4}l^{4}-8\alpha c_{2})),\nonumber\\
\bar{\mathcal{B}}&=-4\alpha \nu^{2}c_{1}z^{\frac{2\nu+2}{\nu}}+(\nu+1)(\nu+\frac{3}{2})(\nu^{4}l^{4}-8\alpha c_{2}),\nonumber\\
\bar{\mathcal{C}}&=-4\alpha \nu c_{1}z^{\frac{2\nu+2}{\nu}}+(\nu+1)(\nu^{4}l^{4}-8\alpha c_{2}),\nonumber\\
\bar{\mathcal{E}}&=-4c_{1}^{2}\alpha^{2}\nu^{2}(3\nu^{3}-4\nu^{2}+3\nu+2)z^{\frac{2(2\nu+2)}{\nu}}+\nonumber\\
&2c_{1}\nu\alpha(\nu^{2}-1)(\nu^{4}l^{4}+2l^{4}\nu^{3}-16\alpha \nu c_{2}-8\alpha c_{2})z^{\frac{2\nu+2}{\nu}}+\nonumber\\
&(\nu-1)(\nu+1)^{2}(\nu^{4}l^{4}-8\alpha c_{2})(4\alpha c_{2}+\nu^{5}l^{5}-l^{4}\nu^{3}-\frac{\nu^{4}l^{4}}{2}),\nonumber\\
\bar{\mathcal{F}}&=\nu^{7}l^{4}-\frac{7}{2}\nu^{6}l^{4}+2\nu^{5}l^{4}+\frac{5}{2}\nu^{4}l^{4}+4\nu^{3}\alpha c_{2}+24\alpha \nu^{2}c_{2}-36\alpha \nu c_{2}-8\alpha c_{2},\nonumber\\
\bar{\mathcal{H}}&=4c_{1}^{2}\alpha^{2}(5\nu+2-2\nu^{2}+\nu^{3})z^{\frac{2(2\nu+2)}{\nu}}-4c_{1}\alpha(\nu^{2}-1)(l^{4}\nu^{3}-8\alpha \nu c_{2}+4\alpha c_{2})z^{\frac{2\nu+2}{\nu}}\nonumber\\
&-(\nu+1)^{2}(\nu-1)\nu^{2}l^{4}(\nu^{2}-\frac{1}{2}\nu-1)(\nu^{4}l^{4}-8\alpha c_{2}),\nonumber\\
\bar{\mathcal{G}}&=\nu^{5}l^{4}+2\alpha c_{2}+\frac{1}{2}\nu^{4}l^{4}-\frac{1}{2}\nu^{3}l^{4},\;\;\;\;\;
\bar{\mathcal{K}}=\nu^{5}l^{4}+\frac{1}{3}\nu^{4}l^{4}-\frac{1}{2}l^{4}\nu^{3}+4\alpha c_{2},\nonumber\\
\bar{\mathcal{L}}&=576\alpha^{2}c_{2}^{2}\bar{\mathcal{G}}\bar{\mathcal{C}}+\frac{1}{2}\bar{\mathcal{A}}^{2}\bar{\mathcal{H}},\;\;\;\;\bar{\mathcal{J}}=320\alpha^{2}c_{2}^{2}\bar{\mathcal{G}}\bar{\mathcal{C}}-\frac{1}{2}(\bar{\mathcal{A}}^{2}(\bar{\mathcal{A}}\bar{\mathcal{B}}-\bar{\mathcal{H}})).
\end{align}
\end{small}


\begin{thebibliography}{99}

\bibitem{Maldacena:1997re}
J.~M.~Maldacena,
Adv. Theor. Math. Phys. \textbf{2}, 231-252 (1998)
doi:10.1023/A:1026654312961
[arXiv:hep-th/9711200 [hep-th]].

\bibitem{Witten:1998qj}
E.~Witten,
Adv. Theor. Math. Phys. \textbf{2}, 253-291 (1998)
doi:10.4310/ATMP.1998.v2.n2.a2
[arXiv:hep-th/9802150 [hep-th]].

\bibitem{Casalderrey-Solana:2011dxg}
J.~Casalderrey-Solana, H.~Liu, D.~Mateos, K.~Rajagopal and U.~A.~Wiedemann,
Cambridge University Press, 2014,
ISBN 978-1-139-13674-7
doi:10.1017/CBO9781139136747
[arXiv:1101.0618 [hep-th]].

\bibitem{Arefeva:2014kyw}
I.~Y.~Aref'eva,
Phys. Usp. \textbf{57}, 527-555 (2014)
doi:10.3367/UFNe.0184.201406a.0569

\bibitem{Strickland:2013uga}
M.~Strickland,
Pramana \textbf{84}, no.5, 671-684 (2015)
doi:10.1007/s12043-015-0972-1
[arXiv:1312.2285 [hep-ph]].

\bibitem{Andreev:2006ct}
O.~Andreev and V.~I.~Zakharov,
Phys. Rev. D \textbf{74}, 025023 (2006)
doi:10.1103/PhysRevD.74.025023
[arXiv:hep-ph/0604204 [hep-ph]].


\bibitem{He:2013qq}
S.~He, S.~Y.~Wu, Y.~Yang and P.~H.~Yuan,
JHEP \textbf{04}, 093 (2013)
doi:10.1007/JHEP04(2013)093
[arXiv:1301.0385 [hep-th]].


\bibitem{Arefeva:2018hyo}
I.~Aref'eva and K.~Rannu,
JHEP \textbf{05}, 206 (2018)
doi:10.1007/JHEP05(2018)206
[arXiv:1802.05652 [hep-th]].


\bibitem{Bohra:2019ebj}
H.~Bohra, D.~Dudal, A.~Hajilou and S.~Mahapatra,
Phys. Lett. B \textbf{801}, 135184 (2020)
doi:10.1016/j.physletb.2019.135184
[arXiv:1907.01852 [hep-th]].

\bibitem{Fan:2014ala}
Z.~Y.~Fan and H.~Lu,
Phys. Rev. D \textbf{91}, no.6, 064009 (2015)
doi:10.1103/PhysRevD.91.064009
[arXiv:1501.00006 [hep-th]].

\bibitem{Sajadi:2022ybs}
S.~N.~Sajadi and S.~H.~Hendi,
Eur. Phys. J. C \textbf{82}, no.8, 675 (2022)
doi:10.1140/epjc/s10052-022-10647-9
[arXiv:2207.13435 [gr-qc]].

\bibitem{Sajadi:2020axg}
S.~N.~Sajadi, R.~B.~Mann, N.~Riazi and S.~Fakhry,
doi:10.1103/PhysRevD.102.124026
[arXiv:2010.15039 [gr-qc]].

\bibitem{Sajadi:2022tgi}
S.~N.~Sajadi, A.~Hajilou and S.~H.~Hendi,
[arXiv:2208.09892 [gr-qc]].

\bibitem{Sajadi:2022pcz}
S.~N.~Sajadi, L.~Shahkarami, F.~Charmchi and S.~H.~Hendi,
[arXiv:2207.07374 [gr-qc]].


\bibitem{Ferrara:1996hh}
S.~Ferrara, R.~R.~Khuri and R.~Minasian,
Phys. Lett. B \textbf{375}, 81-88 (1996)
doi:10.1016/0370-2693(96)00270-5
[arXiv:hep-th/9602102 [hep-th]].

 
\bibitem{Hofman:2009ug}
D.~M.~Hofman,
Nucl. Phys. B \textbf{823}, 174-194 (2009)
doi:10.1016/j.nuclphysb.2009.08.001
[arXiv:0907.1625 [hep-th]].

\bibitem{Brigante:2007nu}
M.~Brigante, H.~Liu, R.~C.~Myers, S.~Shenker and S.~Yaida,
Phys. Rev. D \textbf{77}, 126006 (2008)
doi:10.1103/PhysRevD.77.126006
[arXiv:0712.0805 [hep-th]].

\bibitem{deBoer:2009pn}
J.~de Boer, M.~Kulaxizi and A.~Parnachev,
JHEP \textbf{03}, 087 (2010)
doi:10.1007/JHEP03(2010)087
[arXiv:0910.5347 [hep-th]].

\bibitem{Camanho:2009hu}
X.~O.~Camanho and J.~D.~Edelstein,
JHEP \textbf{06}, 099 (2010)
doi:10.1007/JHEP06(2010)099
[arXiv:0912.1944 [hep-th]].


\bibitem{Garraffo:2008hu}
C.~Garraffo and G.~Giribet,
Mod. Phys. Lett. A \textbf{23}, 1801-1818 (2008)
doi:10.1142/S0217732308027497
[arXiv:0805.3575 [gr-qc]].

\bibitem{Charmousis:2008kc}
C.~Charmousis,
Lect. Notes Phys. \textbf{769}, 299-346 (2009)
doi:10.1007/978-3-540-88460-6\_8
[arXiv:0805.0568 [gr-qc]].

\bibitem{Canfora:2021ttl}
F.~Canfora, A.~Cisterna, S.~Fuenzalida, C.~Henriquez-Baez and J.~Oliva,
Phys. Rev. D \textbf{104}, no.4, 044026 (2021)
doi:10.1103/PhysRevD.104.044026
[arXiv:2103.09110 [hep-th]].



\bibitem{Deruelle:2003ck}
N.~Deruelle and J.~Madore,
[arXiv:gr-qc/0305004 [gr-qc]].

\bibitem{Zwiebach:1985uq}
B.~Zwiebach,
Phys. Lett. B \textbf{156}, 315-317 (1985)
doi:10.1016/0370-2693(85)91616-8

\bibitem{Kanti:1995cp}
P.~Kanti and K.~Tamvakis,
Phys. Rev. D \textbf{52}, 3506-3511 (1995)
doi:10.1103/PhysRevD.52.3506
[arXiv:hep-th/9504031 [hep-th]].


\bibitem{Alexeev:1996vs}
S.~O.~Alexeev and M.~V.~Pomazanov,
Phys. Rev. D \textbf{55}, 2110-2118 (1997)
doi:10.1103/PhysRevD.55.2110
[arXiv:hep-th/9605106 [hep-th]].

\bibitem{Kanti:1997br}
P.~Kanti, N.~E.~Mavromatos, J.~Rizos, K.~Tamvakis and E.~Winstanley,
Phys. Rev. D \textbf{57}, 6255-6264 (1998)
doi:10.1103/PhysRevD.57.6255
[arXiv:hep-th/9703192 [hep-th]].


\bibitem{Padmanabhan:2013xyr}
T.~Padmanabhan and D.~Kothawala,
Phys. Rept. \textbf{531}, 115-171 (2013)
doi:10.1016/j.physrep.2013.05.007
[arXiv:1302.2151 [gr-qc]].

\bibitem{Horndeski:1974wa}
G.~W.~Horndeski,
Int. J. Theor. Phys. \textbf{10}, 363-384 (1974)
doi:10.1007/BF01807638


\bibitem{Kobayashi:2019hrl}
T.~Kobayashi,
Rept. Prog. Phys. \textbf{82}, no.8, 086901 (2019)
doi:10.1088/1361-6633/ab2429
[arXiv:1901.07183 [gr-qc]].


\bibitem{Fernandes:2022zrq}
P.~G.~S.~Fernandes, P.~Carrilho, T.~Clifton and D.~J.~Mulryne,
Class. Quant. Grav. \textbf{39}, no.6, 063001 (2022)
doi:10.1088/1361-6382/ac500a
[arXiv:2202.13908 [gr-qc]].

  \bibitem{Wald1}
  R.~M.~Wald,
  Phys.\ Rev.\ D {\bf 48}, no. 8, R3427 (1993)
  doi:10.1103/PhysRevD.48.R3427
  [gr-qc/9307038].

\bibitem{Wald2}
  V.~Iyer and R.~M.~Wald,
  Phys.\ Rev.\ D {\bf 50}, 846 (1994)
  doi:10.1103/PhysRevD.50.846
  [gr-qc/9403028].

\bibitem{Li:2011hp}
D.~Li, S.~He, M.~Huang and Q.~S.~Yan,
JHEP \textbf{09}, 041 (2011)
doi:10.1007/JHEP09(2011)041
[arXiv:1103.5389 [hep-th]].

\end{thebibliography}
\end{document}